\documentclass[mnsc]{informs3}

\OneAndAHalfSpacedXI

\usepackage{amsmath,amssymb}
\usepackage{mathtools}
\usepackage{bm}

\usepackage{graphicx}
\usepackage{float}       
\usepackage{placeins}    

\usepackage{enumitem}
\usepackage{booktabs}

\usepackage{xcolor}

\usepackage{url,xurl}
\usepackage[pageanchor]{hyperref}

\usepackage{natbib}

\bibpunct[, ]{(}{)}{,}{a}{}{,}%
\def\bibfont{\small}%
\def\bibsep{\smallskipamount}%
\def\bibhang{24pt}%
\TheoremsNumberedThrough

\ECRepeatTheorems

\EquationsNumberedThrough

\newcommand{\E}{\mathbb{E}}
\newcommand{\Var}{\mathrm{Var}}
\newcommand{\Cov}{\mathrm{Cov}}
\newcommand{\R}{\mathbb{R}}

\newcommand{\MSE}{\mathrm{MSE}}
\newcommand{\diag}{\mathrm{diag}}

\newcommand{\vone}{\mathbf{1}}
\newcommand{\vw}{\bm{w}}
\newcommand{\vwhat}{\hat{\bm{w}}}
\newcommand{\veps}{\bm{\epsilon}}
\newcommand{\vepshat}{\hat{\bm{\epsilon}}}

\newcommand{\vBbarte}{\bar{\bm B}_{\mathrm{te}}}

\renewcommand{\footnoterule}{%
  \kern -3pt
  \hrule width 0.4\columnwidth
  \kern 2.6pt
}

\begin{document}

\RUNAUTHOR{Ham, Jasin, and Zhang}

\RUNTITLE{Combining Past Experimental Data for Efficient Estimation}


\TITLE{Combining Past Experimental Data for Efficient Estimation:
A Shrinkage Approach}


\ARTICLEAUTHORS{
\AUTHOR{Qianxi Zhang}
\AFF{Stephen M. Ross School of Business,
University of Michigan, Ann Arbor, Michigan, USA}

\AUTHOR{Dae Woong Ham}
\AFF{Stephen M. Ross School of Business,
University of Michigan, Ann Arbor, Michigan, USA}

\AUTHOR{Stefanus Jasin}
\AFF{Stephen M. Ross School of Business,
University of Michigan, Ann Arbor, Michigan, USA}
}


\ABSTRACT{
Firms routinely conduct randomized experiments, creating an archive of past experiments that could be leveraged to improve the estimation of treatment effects (in terms of mean squared error (MSE)) in new experiments. However, past experiments evaluate related but different interventions, so their estimates may be biased for the current target. We study when and how borrowing from multiple auxiliary experiments can reduce the MSE of a single treatment-effect estimate. In the oracle case when biases and variances are known, we show that borrowing from even one biased experiment strictly improves the MSE. In the non-oracle case, we propose three practical shrinkage estimators and characterize their asymptotic MSE as the number of auxiliary experiments grows. Our analysis shows that the average bias of past experiments and the variances of the bias affects the relative performance of all proposed estimators. As expected, we find all proposed estimators perform better when the bias is low. Surprisingly, however, we find that the variance of the bias has both benefits and harm to our first proposed estimator - the general plug in estimator. On one hand, a more dispersed bias allows flexibility in cancelling the bias. On the other hand, the bias cancellation is imperfect as we only have an  estimate of the bias, leading to overfitting. Motivated by these tradeoffs, we propose a pooled plug-in estimator that restricts study-specific reweighting to limit this source of overfitting. Lastly, we find both estimators suffer when bias is large. To address this, we also develop a fixed-weight estimator with a hypothesis-testing gate that only borrows past experiments only when the bias is statistically significantly low. Simulations illustrate the relative strengths of these estimators across bias environments and the relevance of the limiting analysis for finite collections of experiments. We also provide managerial insights on how to choose different proposed shrinkage estimators depending on both the average bias and bias dispersion.
}




\maketitle


\section{Introduction}
\label{sec:introduction}

Randomized experiments have become a routine part of decision making at large technology firms. Organizations such as Airbnb, Amazon, Booking.com, Google, Microsoft, and Netflix now run experiments at massive scale, collectively testing more than one hundred thousand treatments per year \citep{gupta2019top}. As experimentation programs mature, firms accumulate large archives of past A/B tests. At Booking.com, for example, every product change is tested through an experiment, and the results of past experiments are stored in a searchable central repository organized by product area, team, and visitor segment \citep{kaufman2017democratizing}. These repositories help teams generate hypotheses and disseminate organizational learning. However, they are typically used only qualitatively: past experiments inform intuition about what may work, but they are rarely incorporated directly into the statistical estimation of a new treatment effect. This raises a natural question: can historical experiments be used as a statistical resource to improve the accuracy of estimation in a new experiment?

When a firm launches a new intervention on a familiar product surface, past experiments on related interventions may contain useful information about the likely magnitude of the new treatment effect. This magnitude is often directly relevant for managerial decisions: firms use estimated effects to compare product changes, prioritize launches, and assess business impact. In such settings, noisy estimates can lead to suboptimal decisions with meaningful operational or revenue consequences \citep{xie2016improving}. We therefore focus on improving point-estimation accuracy, measured by mean squared error (MSE), when estimating a new treatment effect using both the current experiment and related historical experiments. This objective differs from improving confidence-interval efficiency or optimizing hypothesis-testing procedures; see, for example, \citet{fiksel2024adjustedCI} and \citet{Peng2026bvalue}.

Constructing an MSE-efficient estimator that leverages historical data is challenging because past experiments typically evaluate related but non-identical interventions. Their treatment-effect estimates may therefore be biased for the estimand of the current experiment. Moreover, both the magnitude and the direction of these biases can vary across auxiliary experiments and are not known in advance. As a result, naively averaging all available estimates can be counterproductive: auxiliary data may reduce variance, but sufficiently large biases can more than offset this gain.

At a high level, the value of a historical experiment archive depends on two distinct features of its mismatch with the current target. The first is \emph{systematic misalignment}: whether the auxiliary treatment effects are, on average, shifted away from the target effect. The second is \emph{cross-experiment heterogeneity}: how much the study-specific deviations from the target vary across historical experiments. We show these two dimensions create different challenges for borrowing. A large systematic shift can make borrowing harmful even when the auxiliary experiments are relatively homogeneous, whereas we find surprising results that cross-experiment heterogeneity can be both beneficial and harmful to borrowing past experiments (detailed below).
The central question is therefore whether, and under what conditions, borrowing from multiple biased auxiliary experiments can improve point estimation of the target treatment effect.

\subsection{Literature Review and Contributions}
The literature of shrinkage estimators is relevant and provides a natural framework for this problem. The basic idea is to combine a noisy but unbiased estimator with an informative but potentially biased reference point, accepting some bias in exchange for a reduction in variance. Classical James--Stein results show that shrinkage can strictly reduce risk when estimating a vector of three or more parameters simultaneously \citet{james1961estimation}. In a related causal-inference setting, \citet{rosenman2020combining} leverages the Stein-method to similarly combine experimental and observational data by shrinking a vector of unbiased experimental estimates toward a vector of potentially biased observational estimates across subgroups.  Related work in operations management also
uses shrinkage to improve decisions across multiple problems.
\citet{li2023cluster} develop a cluster-based Shrunken-SAA approach that
exploits similarities among decision problems to guide data pooling.
They show that clustering can improve aggregate performance over pooling
all problems together, particularly when the problems are heterogeneous.

Our setting differs in both the source of dimensionality and the managerial decision being studied. In Stein-based shrinkage methods like \citet{rosenman2020combining}, the estimand is a vector with dimensions related to the number of target causal effects, such as subgroup-level effects or treatment-arm-specific effects, and the objective is to reduce the aggregated risk across each dimension, which is greater than one. Results in these works show a reduction in MSE when aggregating across a multi-vector estimand. However, the reduction is generally not guaranteed for any specific univariate estimand, as the typical James-Stein requires dimension to be at least 3 to show MSE dominance.
In contrast, our target, a single treatment effect of one new experiment, is a univariate estimand \(\theta\). Therefore, typical James--Stein shrinkage results for vector-based estimands do not apply. Regardless, the motivation and idea behind our paper is most closely related to the shrinkage literature, thus for the rest of our paper we label our estimator that borrow past experiments also a ``shrinkage'' estimator with the understanding that our guarantee and univariate estimand setting is different than that of the typical shrinkage estimator literature (detailed further in Section~\ref{sec:oracle}).

We highlight our main contributions. First, we show in an oracle setting that borrowing from even one arbitrarily biased experiment strictly reduces MSE, provided that the bias is finite and the target variance is positive. Unfortunately this oracle case is infeasible in practice because the optimal weights require the true biases and variances of every experiment to be known, hence named the oracle setting. We, therefore, present three different non-oracle shrinkage estimators, each with different strengths and weakness. The first proposed one is 
a practical plug-in shrinkage estimator that estimates the oracle weights through sample splitting. For all shrinkage estimators, we characterize the limiting MSE as the number of past experiments grows. 
This characterization yields two insights, which are governed by two quantities: the average bias, which captures the systematic misalignment between the auxiliary experiments and the target, and the cross-experiment bias dispersion, which captures the heterogeneity of the biases. As expected, a smaller magnitude of  average auxiliary bias leads to a larger MSE reduction. However, we surprisingly find a non-monotonic relationship between bias dispersion and MSE improvement. 
On one hand a
larger bias dispersion allows greater flexibility to cancel the bias. 
However, this larger dispersion does not offset the bias perfectly as it offsets the estimated bias, creating a harm of overfitting to an imperfect target. 
We fix this overfitting issue with a pooled plug-in shrinkage estimator. Both these proposed estimators work well in small bias regimes. However, as expected, they may perform poorly in large bias regimes. Therefore, we lastly propose a fixed-hypothesis shrinkage-based estimator that only borrows if we believe the bias is small, giving a robustness property desirable in large bias regimes. 
In summary, each proposed estimators are preferred in differing contexts. We mathematically characterize these different contexts and summarize the managerial insights in Table~\ref{tab:estimator_recommendation}.


The remainder of the paper is organized as follows. Section~\ref{sec:setup} introduces the model and basic setup. Section~\ref{sec:oracle} studies the oracle shrinkage problem and characterizes the MSE improvement that is achievable when the bias and variance parameters are known. Section~\ref{sec:practical} analyzes practical plug-in shrinkage with estimated biases and variances, and studies its asymptotic behavior as the number of auxiliary experiments grows. Section~\ref{sec:pooled} develops the pooled-bias plug-in shrinkage estimator and compares it with the general plug-in estimator. Section~\ref{sec:fixed} proposes a fixed-weight shrinkage estimator with a hypothesis-testing gate. Section~\ref{subsec:comparison} compares all proposed estimators and gives managerial insights when each estimator is preferable. Section~\ref{sec:inference} presents inference results for constructing confidence intervals for our proposed shrinkage estimators. Section~\ref{sec:simulation} presents simulations demonstrating our theoretical properties, and Section~\ref{sec:discussions} concludes with a discussion.

\section{Problem Formulation and Oracle Benchmark}\label{sec:oracle}

\subsection{Sampling Model and Benchmark}\label{sec:setup}

We study the problem of estimating a univariate target treatment effect by combining data
from a current randomized experiment with data from \(M\) past auxiliary randomized
experiments. Let \(\theta\) denote the treatment effect of interest in the current experiment.
The target experiment directly evaluates the intervention whose effect we wish to estimate and
therefore provides an unbiased benchmark. The auxiliary experiments could be historical experiments that evaluate related but
non-identical interventions. They may therefore contain useful information about \(\theta\), but
their treatment-effect estimators are generally biased for the current target estimand.

We impose only weak distributional assumptions. Within each experiment, observations are
independent and identically distributed with finite second moments. The distributions may
differ across experiments and need not be Gaussian.

\begin{assumption}[Independent sampling model]\label{ass:model}
Let $A_1,\dots,A_{N_A}$ denote observations from the target experiment. We assume
$
A_1,\dots,A_{N_A} \overset{\mathrm{i.i.d.}}{\sim} F_A, \,
\E[A_j]=\theta, \mbox{ and }
0<\Var(A_j)=\sigma_A^2<\infty .
$
For each auxiliary study $i=1,\dots,M$, let $B_{i,1},\dots,B_{i,N_{B,i}}$ denote observations from auxiliary dataset $i$. We assume
$
B_{i,1},\dots,B_{i,N_{B,i}} \overset{\mathrm{i.i.d.}}{\sim} F_{B,i},
\,
\E[B_{i,j}]=\theta+\epsilon_i,
\mbox{ and }
0<\Var(B_{i,j})=\sigma_{B,i}^2<\infty .
$
Here $\epsilon_i$ denotes the (fixed constant) bias of auxiliary study $i$ relative to the target effect $\theta$. 

Finally, the target dataset and all auxiliary datasets are mutually independent:
$
\{A_j:1\le j\le N_A\}
\perp\!\!\!\perp
\{B_{i,j}:1\le j\le N_{B,i}\},
\, i=1,\dots,M,
$
and
$
\{B_{i,j}:1\le j\le N_{B,i}\}
\perp\!\!\!\perp
\{B_{k,j}:1\le j\le N_{B,k}\},
\, i\neq k .
$
\end{assumption}
\noindent Assumption~\ref{ass:model} is deliberately general. We do not make any homoscedastic variance assumptions, i.e., every experiment $i$ can have a different variance $\sigma^2_{B, i}$. 

Let \(\bar A\) and \(\bar B_i\) denote the sample means of the target experiment and the \(i\)th auxiliary experiment, respectively. Under Assumption~\ref{ass:model},
\[
\E[\bar A]=\theta,\quad V_A:=\Var(\bar A)=\frac{\sigma_A^2}{N_A},\qquad \E[\bar B_i]=\theta+\epsilon_i,\quad V_{B,i}:=\Var(\bar B_i)=\frac{\sigma_{B,i}^2}{N_{B,i}},\quad i=1,\ldots,M.
\]
Throughout the paper, we use ``A-side'' to refer to the target experiment and ``B-side'' to refer to the auxiliary experiments, or past experiments, interchangeably. The no-borrowing or no shrinkage-estimator benchmark is when the analyst does not utilize any of the $B_{i, j}$ samples from previous experiments and only utilizes the $A_i$ samples. Formally, the no-borrowing benchmark utilizes 
\(\bar A\) to estimate $\theta$ (the uniformly most powerful unbiased estimator).
Estimator performance is evaluated by mean squared error (MSE). For any estimator
\(\widehat\theta\), define
\[
    \MSE(\widehat\theta):=\E\left[(\widehat\theta-\theta)^2\right],
    \qquad
    \MSE(\bar A)=V_A.
\]
Accordingly, a borrowing estimator improves upon the no-borrowing benchmark whenever
\(\MSE(\widehat\theta)<V_A\).

We begin with an idealized oracle setting in which all bias and variance quantities are known:
the target variance \(V_A\), the auxiliary variances \(V_{B,1},\ldots,V_{B,M}\), and the
auxiliary biases \(\epsilon_1,\ldots,\epsilon_M\). This oracle setting evaluates the potential value
of shrinkage before accounting for the statistical cost of estimating the shrinkage
weights. As we show below, when these quantities are known, optimally weighted shrinkage
strictly improves upon the no-borrowing benchmark.

\subsection{Oracle Shrinkage with One Auxiliary Estimator}
\label{subsection:univariate_oracle}

We first consider the case of a single auxiliary estimator (\(M=1\))
to make the bias--variance tradeoff transparent. Following the classic linear
shrinkage formulation of combining an initial estimator with an auxiliary
reference in
\citep{rosenman2020combining,li2023cluster},
for a fixed weight \(w\in\R\), define

\begin{equation}\label{eq:oracle-theta1}
\hat\theta(w)=(1-w)\bar A+w\bar B_1.
\end{equation}
As a reminder, a no-shrinkage baseline estimator simply uses the unbiased estimator $\bar{A}$ to estimate the treatment effect $\theta$. Under Assumption~\ref{ass:model}, its MSE is
\[
\MSE\bigl(\hat\theta(w)\bigr)
=
(1-w)^2V_A+w^2\bigl(V_{B,1}+\epsilon_1^2\bigr).
\]
The first term reflects the remaining sampling variance from the target estimator, whereas the
second captures the variance and squared-bias cost introduced by borrowing from the auxiliary
estimator. The MSE-minimizing oracle weight is
\begin{equation}\label{eq:oracle-w1}
w_1^*
=
\frac{V_A}{\epsilon_1^2+V_A+V_{B,1}}.
\end{equation}
Equation~\eqref{eq:oracle-w1} reflects the shrinkage tradeoff explicit. A noisier target estimator
increases the shrinkage weight, whereas a larger auxiliary bias or auxiliary variance reduces
the weight placed on the auxiliary estimator. As expected, a larger auxiliary bias magnitude \(|\epsilon_1|\) or auxiliary variance \(V_{B,1}\) reduces the oracle weight assigned to the auxiliary estimator. We explain below why a larger $V_A$ receives larger shrinkage weight when extending this result to $M$ auxiliary experiments. 

\begin{theorem}\label{thm:oracle_one}
Under Assumption~\ref{ass:model}, the oracle shrinkage estimator \(\hat\theta(w_1^*)\) satisfies
\[
\MSE\bigl(\hat\theta(w_1^*)\bigr)-\MSE(\bar A)
=
-\frac{V_A^2}{\epsilon_1^2+V_A+V_{B,1}}
<0.
\]
\end{theorem}

Theorem~\ref{thm:oracle_one} shows that, when the bias and variance components are known,
borrowing strictly improves upon the unbiased benchmark for any finite auxiliary bias and
variance. A similar finding appears in \citet{oberst2023}, who study the risks and rewards of combining an unbiased and a biased estimator in the context of causal inference.

\subsection{Oracle Shrinkage with \(M\) Auxiliary Estimators}

We now extend the oracle problem to \(M\) auxiliary estimators. Let
\[
\vw=(w_1,\dots,w_M)^\top\in\R^M,
\qquad
\veps=(\epsilon_1,\dots,\epsilon_M)^\top\in\R^M,
\qquad
\Sigma:=\diag(V_{B,1},\dots,V_{B,M}),
\]
where \(\vw\) is the vector of auxiliary weights, \(\veps\) is the vector of auxiliary biases,
and \(\Sigma\) is the covariance matrix of the auxiliary sample means. 

Define
\begin{equation}\label{eq:oracle-thetaM}
\hat\theta(\vw)
=
\bigl(1-\vone^\top\vw\bigr)\bar A
+
\sum_{i=1}^M w_i\bar B_i.
\end{equation}
Under Assumption~\ref{ass:model},
\begin{equation}\label{eq:MSE-expression}
    \MSE\bigl(\hat\theta(\vw)\bigr)
=
(\vw^\top\veps)^2
+
\bigl(1-\mathbf 1^\top\vw\bigr)^2V_A
+
\vw^\top\Sigma\vw.
\end{equation}
Let
\[
Q:=\veps\veps^\top+\Sigma+V_A\mathbf 1\mathbf 1^\top.
\]
The MSE-minimizing oracle weight vector is
\begin{equation}\label{eq:oracle-wM}
\vw_M^*
=
V_AQ^{-1}\mathbf 1.
\end{equation}
\noindent We remark that Equation~\eqref{eq:oracle-wM} is just the multivariate extension of Equation~\eqref{eq:oracle-w1}. 

We highlight two insights from Equation~\eqref{eq:oracle-wM}. First, lower auxiliary variance is preferred. To see why, we can examine both how the oracle rule assigns shrinkage weights and how auxiliary variance enters the MSE. When two auxiliary estimators have the same bias, the more precise estimator receives more weight. In particular, when \(M=2\) and \(\epsilon_1=\epsilon_2\),
\[
    \frac{w_1^*}{w_2^*}
    =
    \frac{V_{B,2}}{V_{B,1}}.
\]
The reason follows directly from the MSE decomposition in Equation~\eqref{eq:MSE-expression}: when two auxiliary estimators have the same bias, shifting weight between them while holding their total auxiliary weight fixed only changes the auxiliary variance contribution \(\vw^\top\Sigma\vw\). Placing more weight on the estimator with smaller \(V_{B,i}\) therefore reduces MSE. Consistent with this mechanism, Theorem~\ref{thm:oracle_M} shows that auxiliary-side data with a lower general variance leads to a larger oracle MSE reduction.

Furthermore, when \(M\ge2\), shrinkage allows bias cancellation through the term of weighted auxiliary biases $(\vw^\top\veps)$ in Equation~\eqref{eq:MSE-expression}. Intuitively, When the auxiliary biases differ in sign, the oracle weights can combine positive and negative biases so that they partially  offset with each other.

We now present the main oracle result that holds for any number of auxiliary experiments.

\begin{theorem}\label{thm:oracle_M}
Under Assumption~\ref{ass:model}, the oracle shrinkage estimator \(\hat\theta(\vw_M^*)\) satisfies
\[
\MSE\bigl(\hat\theta(\vw_M^*)\bigr)-\MSE(\bar A)
=
-\,V_A^2\,\mathbf 1^\top Q^{-1}\mathbf 1
<0,
\]
where \(Q=\veps\veps^\top+\Sigma+V_A\mathbf 1\mathbf 1^\top\). Thus, for every
\(M\ge1\), the oracle shrinkage estimator has strictly smaller MSE than \(\bar A\).
\end{theorem}

Theorem~\ref{thm:oracle_M} extends the univariate result: with perfect knowledge of the
bias and variance structure, shrinkage can always help. The magnitude of the oracle improvement is
\(V_A^2\mathbf 1^\top Q^{-1}\mathbf 1\). Holding the auxiliary biases and variances fixed, the oracle improvement increases with \(V_A\). Thus, a noisier target estimator creates more room for auxiliary information to improve estimation. This monotonicity is consistent with the classical shrinkage intuition that a noisier unbiased
estimator creates more room for risk reduction through borrowing. For example, after rescaling
the identity-variance formulation in \citet{james1961estimation}, the James--Stein
estimator can be written as
\[
\hat{\bm\theta}^{JS}
=
\left(
1-\frac{(p-2)\sigma^2}{\|\bm X\|^2}
\right)\bm X,
\qquad
\bm X\sim N_p(\bm\theta,\sigma^2 I_p).
\]
In this expression, a larger noise variance \(\sigma^2\) leads to stronger shrinkage as there is more variance in $\bar{X}$, providing a greater opportunity for shrinkage to reduce this ``larger'' variance. More closely
related to our setting, \citet{rosenman2020combining} use Stein-type shrinkage to
combine unbiased experimental estimates with potentially biased observational estimates across
multiple treatment effects. Their analysis similarly shows that borrowing from a biased but
informative source is more valuable when the unbiased experimental estimator is noisier. However, the connection is conceptual rather than mechanical. In James--Stein shrinkage and in     
\citet{rosenman2020combining}, the dimension comes from the number of target effects being
estimated. In our setting, the target remains the single current-experiment effect \(\theta\), and
the dimension comes from the number of auxiliary experiments available to inform that
target. 

To summarize the main takeaways, borrowing from past experiments can always feasibly reduce the MSE in the oracle setting. 
A noisier target creates more
room for variance reduction, while smaller auxiliary biases and
variances reduce the cost of borrowing. Therefore, borrowing is most attractive when the target estimator is noisy
and the available auxiliary experiments have relatively small
biases and low sampling variances.

\subsection{Improvement from Adding an Additional Auxiliary Estimator}
Theorem~\ref{thm:oracle_M} shows that optimally borrowing from \(M\)
auxiliary experiments strictly reduces MSE relative to using
\(\bar A\) alone. However, this does not establish whether every
additional auxiliary experiment provides a further MSE reduction,
or whether some experiments offer no additional benefit given the
existing collection. We therefore examine how the oracle MSE
changes when one more auxiliary experiment becomes available,
and characterize when the additional experiment yields a strict
improvement.

Let \(\MSE_M^*\) denote the oracle MSE when the first \(M\) auxiliary
estimators are available, with corresponding bias vector
\(\veps=(\epsilon_1,\ldots,\epsilon_M)^\top\), variance matrix
\(\Sigma=\diag(V_{B,1},\ldots,V_{B,M})\), and
\[
Q_M=\veps\veps^\top+\Sigma+V_A\mathbf 1_M\mathbf 1_M^\top.
\]

\begin{proposition}\label{prop:oracle_mse_monotone}
For every \(M\ge1\), adding the \((M+1)\)-st auxiliary estimator always leads to a weak reduction in MSE, i.e.,:
\[
\MSE_{M+1}^*\le \MSE_M^*.
\]
Equality holds if and only if the \((M+1)\)-st auxiliary estimator receives zero oracle weight,
equivalently,
\[
\epsilon_{M+1}\sum_{i=1}^M\frac{\epsilon_i}{V_{B,i}}
=
1+\sum_{i=1}^M\frac{\epsilon_i^2}{V_{B,i}}.
\]
Otherwise, adding the new auxiliary estimator strictly reduces the oracle MSE.
\end{proposition}

\footnotetext{Specifically,
\[
\MSE_M^*-\MSE_{M+1}^*
=
V_A^2
\frac{
\left[1-(\epsilon_{M+1}\veps+V_A\mathbf 1_M)^\top Q_M^{-1}\mathbf 1_M\right]^2
}{
\epsilon_{M+1}^2+V_{B,M+1}+V_A
-(\epsilon_{M+1}\veps+V_A\mathbf 1_M)^\top Q_M^{-1}
(\epsilon_{M+1}\veps+V_A\mathbf 1_M)
}.
\]
}
The weak improvement follows from the fact that: after a new auxiliary estimator becomes available, the previous \(M\)-study solution remains feasible by assigning zero weight to the new estimator. The substantive content of
Proposition~\ref{prop:oracle_mse_monotone} is therefore more about the equality condition, which
characterizes exactly when a newly available historical experiment has no marginal value.
The equality condition depends on the bias of the new experiment relative to the precision-weighted bias structure already present in the archive. When the condition holds, it implies that the new auxiliary estimator does not provide additional
\textit{bias} information beyond what is already captured by the existing oracle shrinkage structure. We see here, and will continue to see later, that bias plays a more dominant role than auxiliary variance in our shrinkage structure: whether an additional auxiliary estimator is informative depends only on its bias, whereas its variance determines how much weight it receives once it is used.

We also remark that the equality condition is similar to an ``edge-case'', since it requires the bias of the new estimator to be one precise value. In fact, when the auxiliary-side bias is viewed as an independent draw from a continuous distribution, such as a normal or uniform distribution, the equality event occurs with probability zero, and hence
\[
    P(\MSE_{M+1}^*<\MSE_M^*)=1.
\]
That is, adding one more auxiliary estimator almost surely improves the oracle MSE.

In this sense, the result mirrors the James--Stein phenomenon, where higher dimension creates more room for risk reduction. The notion of dimension, however, is different here. In the James--Stein setting, dimension refers to the number of quantities being estimated jointly. In our setting, the relevant dimension is the number of related past experiments available to inform the current target experiment. Another difference is that James--Stein dominance arises only when the parameter dimension is at least three, whereas here the weak improvement from shrinkage appears as soon as a single auxiliary experiment is added.

\section{Practical Plug-in Shrinkage Estimator}\label{sec:practical}

As shown in Equation~\eqref{eq:oracle-wM}, the oracle weights assume that the bias and
variance components are known, which is practically infeasible. Therefore, for the rest of the paper, we propose different practical shrinkage estimators that bypasses this issue. The first estimator we propose is one that simply uses a plug-in approach to the oracle estimator after sample splitting.
This makes the shrinkage weights random, which introduces an additional source of estimation
error in the MSE computation. Consequently, the oracle improvement result in Theorem~\ref{thm:oracle_M} is not
guaranteed to carry over directly to feasible shrinkage rules. A poorly estimated weight may
introduce more noise or bias, potentially making the combined estimator worse than
the unbiased benchmark \(\bar A\). This motivates the current section: does the shrinkage estimator after learning the weights still perform better or worse (in terms of MSE) than not shrinking?

To obtain tractable expressions for the MSE under random weights, we use a pilot/testing
sample split. The pilot sample is used only to estimate the shrinkage weights, while the testing
sample is reserved for the construction of final estimator. This separation ensures that the estimated shrinkage weights are independent
of the sample means to which they are applied, allowing analytical
characterization of the MSE and subsequent confidence interval construction. We then impose additional
working approximations and regularity conditions on the sample summaries that enter the
plug-in shrinkage weights. 

We formally introduce the sample-splitting scheme: each dataset is randomly partitioned into independent pilot and testing subsets, with a split ratio $\rho$ determining the proportion of observations allocated to each subset. The pilot samples estimate the shrinkage rule, and the testing samples form the final estimator.
 For the target dataset, the split sizes are
\[
    n_{A,\mathrm{pi}}=(1-\rho_A)N_A,
    \qquad
    n_{A,\mathrm{te}}=\rho_A N_A,
    \qquad
    \rho_A\in(0,1),
\]
yielding pilot and testing sample means \(\bar A_{\mathrm{pi}}\) and
\(\bar A_{\mathrm{te}}\), with true variances
\[
    V_{A,\mathrm{pi}}
    :=
    \frac{\sigma_A^2}{n_{A,\mathrm{pi}}},
    \qquad
    V_{A,\mathrm{te}}
    :=
    \frac{\sigma_A^2}{n_{A,\mathrm{te}}}.
\]
For each auxiliary dataset \(i=1,\ldots,M\), the split sizes are
\[
    n_{B,i,\mathrm{pi}}=(1-\rho_{B,i})N_{B,i},
    \qquad
    n_{B,i,\mathrm{te}}=\rho_{B,i}N_{B,i},
    \qquad
    \rho_{B,i}\in(0,1),
\]
yielding pilot and testing sample means \(\bar B_{i,\mathrm{pi}}\) and
\(\bar B_{i,\mathrm{te}}\), with true variances
\[
    V_{B,i,\mathrm{pi}}
    :=
    \frac{\sigma_{B,i}^2}{n_{B,i,\mathrm{pi}}},
    \qquad
    V_{B,i,\mathrm{te}}
    :=
    \frac{\sigma_{B,i}^2}{n_{B,i,\mathrm{te}}}.
\]
All pilot and testing subsets are mutually independent.  The test sample, as shown formally later, will be used to estimate the true effects of the target experiment $A$ and auxiliary experiments $B$. The independent pilot sample will be used to estimate the oracle weights $w$. Before we state this formally, we make some regularity and distributional assumptions on the mean and variances.

\begin{assumption}[Working approximation for sample summaries]\label{ass:working}
In addition to Assumption~\ref{ass:model}, suppose the following hold.

\begin{enumerate}[label=\textup{(A\arabic*)},leftmargin=2.5em]

    \item
    All distributions have finite fourth moments and zero third central
    moments:
    \[
        \E[(A_j-\theta)^4]<\infty,
        \qquad
        \E[(A_j-\theta)^3]=0,
        \qquad
        \E[(B_{i,j}-(\theta+\epsilon_i))^4]<\infty,
        \qquad
        \E[(B_{i,j}-(\theta+\epsilon_i))^3]=0.
    \]

    \item
    We use Gaussian working approximations for the pilot and testing sample means:
    \[
        \bar A_{\mathrm{pi}}\sim N(\theta,V_{A,\mathrm{pi}}),
        \qquad
        \bar A_{\mathrm{te}}\sim N(\theta,V_{A,\mathrm{te}}),
    \]
    and, for each \(i=1,\ldots,M\),
    \[
        \bar B_{i,\mathrm{pi}}\sim N(\theta+\epsilon_i,V_{B,i,\mathrm{pi}}),
        \qquad
        \bar B_{i,\mathrm{te}}\sim N(\theta+\epsilon_i,V_{B,i,\mathrm{te}}).
    \]
    For the pilot-sample variance estimators, we use the scaled chi-squared working
    approximations
    \[
        \frac{S_{A,\mathrm{pi}}^2}{\sigma_A^2}
        \sim
        \frac{\chi^2_{\nu_{A,\mathrm{pi}}}}{\nu_{A,\mathrm{pi}}},
        \qquad
        \frac{S_{B_i,\mathrm{pi}}^2}{\sigma_{B,i}^2}
        \sim
        \frac{\chi^2_{\nu_{B_i,\mathrm{pi}}}}{\nu_{B_i,\mathrm{pi}}}.
    \]
    and treat each pilot-sample mean as independent of its corresponding pilot-sample
    variance estimator:
    \[
        \bar A_{\mathrm{pi}}\perp\!\!\!\perp S_{A,\mathrm{pi}}^2,
        \qquad
        \bar B_{i,\mathrm{pi}}\perp\!\!\!\perp S_{B_i,\mathrm{pi}}^2.
    \]

\end{enumerate}

The effective degrees of freedom \(\nu_{A,\mathrm{pi}}\), \(\nu_{B_i,\mathrm{pi}}\)
are determined by moment matching. Their exact formulas,
together with the corresponding kurtosis definitions and inverse-moment conditions used in
the theoretical analysis, are provided in the Appendix~\ref{app:working-assumption}. 
\end{assumption}

The finite-fourth-moment condition in Assumption~\ref{ass:working}(A1) is for
regularity. The zero-third-central-moment condition is satisfied by distributions
that are symmetric about their means. This condition implies that the sample
mean and sample variance are uncorrelated in finite samples and asymptotically
independent as sample sizes grow, supporting the independence approximation
in Assumption~\ref{ass:working}(A2). The working distributional approximations
in (A2) are exact under Gaussian sampling, while the Gaussian approximations
for the sample means are asymptotically supported by the Central Limit Theorem
as \(N_A\) and \(N_{B,i}\) grow. However, the scaled chi-squared approximation
for the sample variance remains a working approximation outside the Gaussian
case. For expositional simplicity, we impose (A2) directly rather than state
all subsequent results asymptotically as \(N_A,N_{B,i}\to\infty\).

\subsection{Shrinkage Estimator with Plug-in Estimation of Shrinkage Weights}

Moving from the oracle problem to a practically feasible estimator, the final shrinkage estimator combines
sample mean obtained from the test samples
\[
    \bar A_{\mathrm{te}},
    \ \bar B_{1,\mathrm{te}},
    \ \ldots,
    \ \bar B_{M,\mathrm{te}},
\]
while the pilot samples are used only to estimate the quantities needed for the shrinkage weights. Under this test split, we aim to estimate the optimal weights for just the test sample. 
Therefore the corresponding optimal quantities in  Section~\ref{sec:oracle} are:
\[
    Q_{\mathrm{te}}
    =
    \veps\veps^\top
    +
    \Sigma_{\mathrm{te}}
    +
    V_{A,\mathrm{te}}\vone\vone^\top,
    \qquad
    \Sigma_{\mathrm{te}}
    :=
    \diag\!\left(V_{B,1,\mathrm{te}},\ldots,V_{B,M,\mathrm{te}}\right).
\]
If these testing-side bias and variance quantities were known, the optimal oracle weight for combining
the testing-sample means would be
\( 
    \vw_{\mathrm{te}}^*
    =
    V_{A,\mathrm{te}}Q_{\mathrm{te}}^{-1}\vone.
\)
In practice, however, the bias vector \(\veps\) and the testing-sample variances
\(V_{A,\mathrm{te}},V_{B,1,\mathrm{te}},\ldots,V_{B,M,\mathrm{te}}\) are unknown. We therefore estimate
these quantities using the pilot data only, and then plug the resulting estimates into the oracle
weight formula.

The bias of auxiliary dataset \(i\) is estimated by the difference between the pilot means:
\[
    \hat\epsilon_i
    :=
    \bar B_{i,\mathrm{pi}}-\bar A_{\mathrm{pi}},
    \qquad i=1,\ldots,M.
\]
Let \(S_{A,\mathrm{pi}}^2\) and \(S_{B_i,\mathrm{pi}}^2\) denote the usual unbiased
pilot-sample observation level variance estimators. The testing-sample variances is estimated by
\[
    \widehat V_{A,\mathrm{te}}
    :=
    \frac{S_{A,\mathrm{pi}}^2}{n_{A,\mathrm{te}}},
    \qquad
    \widehat V_{B,i,\mathrm{te}}
    :=
    \frac{S_{B_i,\mathrm{pi}}^2}{n_{B,i,\mathrm{te}}},
    \qquad i=1,\ldots,M.
\]
The notation \(\widehat V_{A,\mathrm{te}}\) and \(\widehat V_{B,i,\mathrm{te}}\) emphasizes that these are estimators of the variances of testing-datasets, but they are computed
entirely from pilot-sample variance estimators (as shown by $S_{A,\mathrm{pi}}^2, S_{B_i,\mathrm{pi}}^2$). Finally, define
\[
    \widehat\Sigma_{\mathrm{te}}
    :=
    \diag\!\left(
        \widehat V_{B,1,\mathrm{te}},
        \ldots,
        \widehat V_{B,M,\mathrm{te}}
    \right),
    \qquad
    \widehat Q
    :=
    \vepshat\vepshat^\top
    +
    \widehat\Sigma_{\mathrm{te}}
    +
    \widehat V_{A,\mathrm{te}}\vone\vone^\top.
\]
Replacing the unknown oracle quantities in \(\vw_{\mathrm{te}}^*\) by these pilot-based
estimators yields a plug-in weight
\begin{equation}\label{eq:feasible-wM}
    \vwhat
    =
    \widehat V_{A,\mathrm{te}}\widehat Q^{-1}\vone.
\end{equation}
Denote
\[
\vepshat:=(\hat\epsilon_1,\ldots,\hat\epsilon_M)^\top,\qquad
\bar{\mathbf B}_{\mathrm{te}} :=
(\bar B_{1,\mathrm{te}},\ldots,\bar B_{M,\mathrm{te}})^\top.
\]
Applying the estimated weight vector to the independent testing-sample means gives the
final plug-in shrinkage estimator
\begin{equation}\label{eq:feasible-thetaM}
    \hat\theta(\vwhat)
    =
    \left(1-\vone^\top\vwhat\right)\bar A_{\mathrm{te}}
    +
    \vwhat^\top\bar{\mathbf B}_{\mathrm{te}}.
\end{equation}
We also write \(\widehat\theta_M:=\widehat\theta(\widehat{\mathbf w})\) for the estimator based on the first \(M\) auxiliary studies. Throughout the remainder of the paper, we refer to
\eqref{eq:feasible-thetaM} as the practical plug-in shrinkage estimator,
since it can be computed entirely from observed data without oracle quantities.
We also call it the general plug-in estimator to distinguish it from the
pooled plug-in estimator introduced later.

For an in-depth analysis, we include the analysis of the plug-in estimator in
the univariate case with one auxiliary estimator in Appendix~\ref{app:univariate}, which provides intuition for the bias–variance tradeoffs. Our main analysis focuses on the \(M\)-asymptotic regime, which corresponds to the organizational setting in which a firm utilizes a large number of prior experiments.

\subsection{\(M\)-Asymptotic Analysis}
\label{sec: general plug-in}
Unlike the oracle setting, in which the shrinkage weights \(\mathbf{w}\)
are fixed deterministic constants, analyzing the plug-in shrinkage estimator
in \eqref{eq:feasible-thetaM} is more challenging because the estimated weights
\(\vwhat\) depend nonlinearly on the pilot-sample means and variance estimators
from the target and all \(M\) auxiliary datasets. Under
Assumption~\ref{ass:working}, the resulting MSE involves expectations of
nonlinear functions of Gaussian and scaled chi-squared random variables,
which generally do not admit simple closed-form expressions. To obtain a tractable
characterization, we study the asymptotic regime in which
\[
    M\to\infty,
\]
i.e., while the number of auxillary experiment increases, while the target experiment is still univarate and held fixed. Formally, 
$
    N_A, n_{A,\mathrm{pi}},n_{A,\mathrm{te}},\sigma_A^2
$
do not vary with \(M\), and our focus is still estimating the univariate treatment effect $\theta$ corresponding to one target experiment with samples $A_i$.

Although this asymptotic analysis permits analytical tractability, it is also motivated by practice. In many firms, experimentation is part of
routine daily operations, and over time the firm accumulates a large archive of past A/B tests on
interventions that are similar, though not identical, to the current target intervention
\citep{gupta2019top,kaufman2017democratizing}. The regime \(M\to\infty\) is therefore a
useful approximation for studying how the practical plug-in estimator behaves as the number of
available auxiliary experiments becomes large. Simulation shows that the large-\(M\) approximation can be informative for moderate numbers of auxiliary studies, particularly when bias dispersion is limited, where the asymptotics stabilize quickly even for moderate $M$ ($M \approx20$). 
Under substantial heterogeneity, however, finite-\(M\) performance can converge more slowly and depend strongly on the realized composition of the auxiliary archive, as illustrated in Section~\ref{sec:simulation}.

To proceed with asymptotic analysis, we impose asymptotic regularity conditions:
\begin{assumption}[\(M\)-asymptotic regularity]\label{ass:largeM}
\leavevmode
\begin{enumerate}[label=\textup{(A\arabic*)},leftmargin=2.5em]

    \item\label{ass:largeM:reg} \textbf{Regularity.}
    There exist constants \(0<\underline v<\overline v<\infty\) such that, for all \(i\ge1\),
    \[
        \underline v\le V_{B,i,\mathrm{pi}},V_{B,i,\mathrm{te}}\le\overline v,
        \qquad
        \sup_{i\ge1}|\epsilon_i|<\infty,
        \qquad
        \inf_{i\ge1}\nu_{B,i,\mathrm{pi}}>4.
    \]

    \footnotetext{This condition ensures that the required inverse moments of
    \(\widehat V_{B,i,\mathrm{te}}\) exist. Together with
    Assumption~\ref{ass:working}, it implies
    \[
    \sup_{i\ge1}\E\!\left[\widehat V_{B,i,\mathrm{te}}^{-1}\right]<\infty,
    \qquad
    \sup_{i\ge1}\E\!\left[\widehat V_{B,i,\mathrm{te}}^{-2}\right]<\infty.
    \]
    }
    
    \item\label{ass:largeM:avg} \textbf{\(M\)-asymptotic averaging.}
    As \(M\to\infty\), the following weighted empirical averages converge in probability to
    finite constants:
    \[
        \frac{1}{M}\sum_{i=1}^M
        \frac{1}{\widehat V_{B,i,\mathrm{te}}}
        \xrightarrow[M\to\infty]{p}
        \tilde\mu_0,
        \qquad
        \frac{1}{M}\sum_{i=1}^M
        \frac{\epsilon_i}{\widehat V_{B,i,\mathrm{te}}}
        \xrightarrow[M\to\infty]{p}
        \tilde\mu_1,
    \]
    \[
        \frac{1}{M}\sum_{i=1}^M
        \frac{\epsilon_i^2}{\widehat V_{B,i,\mathrm{te}}}
        \xrightarrow[M\to\infty]{p}
        \tilde\mu_2,
        \qquad
        \frac{1}{M}\sum_{i=1}^M
        \frac{V_{B,i,\mathrm{pi}}}{\widehat V_{B,i,\mathrm{te}}}
        \xrightarrow[M\to\infty]{p}
        \tilde\nu.
    \]

    \item\label{ass:largeM:UI} \textbf{Uniform integrability.}
    Let \(\mathcal T_M\) denote the \(\sigma\)-field generated by all pilot-sample quantities.
    The families
    \(\{\Var(\widehat\theta_M\mid\mathcal T_M):M\ge1\}\) and
    \(\{[\E(\widehat\theta_M-\theta\mid\mathcal T_M)]^2:M\ge1\}\)
    are uniformly integrable, i.e.,
    \[
    \begin{aligned}
    &\lim_{K\to\infty}\sup_{M\ge1}
    \E\!\left[
    \Var(\widehat\theta_M\mid\mathcal T_M)
    \mathbf 1\!\left\{
    \Var(\widehat\theta_M\mid\mathcal T_M)>K
    \right\}
    \right]=0,\\
    &\lim_{K\to\infty}\sup_{M\ge1}
    \E\!\left[
    \{\E(\widehat\theta_M-\theta\mid\mathcal T_M)\}^2
    \mathbf 1\!\left\{
    \{\E(\widehat\theta_M-\theta\mid\mathcal T_M)\}^2>K
    \right\}
    \right]=0.
    \end{aligned}
    \]
    \end{enumerate}
\end{assumption}

The boundedness conditions in Assumption~\ref{ass:largeM}\ref{ass:largeM:reg} rule out
auxiliary experiments whose variances are either vanishingly small or arbitrarily large. The lower and upper limits $(\underline v, \overline v)$ can be arbitrarily small and large, respectively, and are not assumed to be known. The requirement on for degree of freedom \(\nu_{B_i,\mathrm{pi}}\) is mild: under Gaussian
sampling, it requires only \(\inf_{i\ge 1}n_{B_i,\mathrm{pi}}>5\); more generally, it holds for sufficiently large pilot sample sizes relative to the kurtosis of the
underlying distributions. 

Assumption~\ref{ass:largeM}\ref{ass:largeM:avg} is a law-of-large-numbers-type
condition. It allows the auxiliary studies to
be heterogeneous while requiring their aggregate precision and bias structure
to stabilize as the number of studies grows. For simplicity, one may alternatively adopt a Bayesian interpretation to justify these assumptions. Formally, suppose
that each auxiliary bias \(\epsilon_i\) is viewed as an i.i.d. draw from a common bias
distribution, while the corresponding study design characteristics---including sample size,
outcome variance---are i.i.d. draws from another distribution,
independently of the bias draws. Under the required finite-moment conditions, Assumption~\ref{ass:largeM}\ref{ass:largeM:avg} then directly follows from the law of
large numbers. 
 
 For notational convenience, define
\[
    \tilde A
    :=
    \tilde\mu_0\tilde\mu_2-\tilde\mu_1^2,
    \qquad
    \tilde\Delta
    :=
    \tilde A+\tilde\mu_0\tilde\nu.
\]

 Under these conditions, the practical plug-in multivariate shrinkage estimator has a well-defined limiting MSE.

\begin{theorem}\label{thm:MSE_infty1}
Under Assumption~\ref{ass:model}, Assumption~\ref{ass:working}, and Assumption~\ref{ass:largeM},
\[
    \MSE_M
    :=
    \E\!\left[(\hat\theta(\vwhat)-\theta)^2\right]
    \xrightarrow[M\to\infty]{}
    \MSE_\infty
    \qquad
    \text{as } M\to\infty,
\]
where
\[
    \MSE_\infty
    =
    \frac{
        (\tilde\mu_1\tilde\nu)^2
        +
        V_{A,\mathrm{pi}}\tilde A^2
    }{
        \tilde\Delta^2
    }.
\]
\end{theorem}

\subsection{A Simplified Limiting Expression and Managerial Insights}\label{sec:simplified}

To interpret Theorem~\ref{thm:MSE_infty1},  we further simplify this result by imposing two additional assumptions on the
bias sequence.

\begin{assumption}[Bias moment convergence and no alignment]\label{ass:noalign}
\leavevmode
\begin{enumerate}[label=\textup{(B\arabic*)},leftmargin=2.5em]
    \item\label{ass:noalign:moments} \textbf{Moment convergence.}
    The empirical first and second moments of the bias sequence converge:
    \[
        e
        :=
        \lim_{M\to\infty}
        \frac{1}{M}\sum_{i=1}^M\epsilon_i,
        \qquad
        \tau^2
        :=
        \lim_{M\to\infty}
        \frac{1}{M}\sum_{i=1}^M\epsilon_i^2,
        \qquad
        \delta
        :=
        \tau^2-e^2
        \ge 0.
    \]

    \item\label{ass:noalign:align} \textbf{No systematic alignment.}
    The bias sequence is not systematically aligned with the \(B\)-side precision terms:
    \[
        \frac{1}{M}\sum_{i=1}^M
        \left(\epsilon_i-\frac{1}{M}\sum_{k=1}^M\epsilon_k\right)
        \left(
            \E\!\left[\frac{1}{\widehat V_{B,i,\mathrm{te}}}\right]
            -
            \frac{1}{M}\sum_{j=1}^M
            \E\!\left[\frac{1}{\widehat V_{B,j,\mathrm{te}}}\right]
        \right)
        \to 0,
    \]
    and
    \[
        \frac{1}{M}\sum_{i=1}^M
        \left(\epsilon_i^2-\frac{1}{M}\sum_{k=1}^M\epsilon_k^2\right)
        \left(
            \E\!\left[\frac{1}{\widehat V_{B,i,\mathrm{te}}}\right]
            -
            \frac{1}{M}\sum_{j=1}^M
            \E\!\left[\frac{1}{\widehat V_{B,j,\mathrm{te}}}\right]
        \right)
        \to 0.
    \]
\end{enumerate}
\end{assumption}

In Assumption~\ref{ass:noalign}, \(e\) is the limiting average bias and \(\delta\) is the limiting
dispersion of the bias sequence around that average. The no-alignment condition rules out systematic relationships between auxiliary-study bias and auxiliary-
study precision, and is used only to express the large-\(M\) limit in terms of \(e\) and \(\delta\); without it, the limit in Theorem~\ref{thm:MSE_infty1} still holds but involves precision-weighted bias moments, which obfuscates the main insight.
If the bias vectors were generated i.i.d from any distribution that is independent with the auxiliary variance and has finite second moment, Assumption~\ref{ass:noalign} is immediately satisfied. The no systematic alignment condition can fail in cases when auxiliary-study bias and precision are correlated. For example, older auxiliary experiments may have accumulated larger sample sizes and therefore receive greater precision weight, while also being more biased for the current target because the product, user base, or policy environment has changed over time. In such settings, the general expression in Theorem~\ref{thm:MSE_infty1} is more appropriate.

Under Assumption~\ref{ass:noalign}, the limiting MSE in
Theorem~\ref{thm:MSE_infty1} simplifies as follows.

\begin{corollary}\label{cor:simplified}
Under the assumptions of Theorem~\ref{thm:MSE_infty1}, together with
Assumption~\ref{ass:noalign}, let
\[
    \tilde v_B'
    :=
    \frac{\tilde\nu}{\tilde\mu_0}
    >0.
\]
Then
\begin{equation}\label{eq:asymp_theta}
    \hat\theta(\vwhat) - \theta \xrightarrow[M\to\infty]{d} \hat\theta_{\infty} - \theta = \underbrace{\frac{\tilde v_B'
    }{\delta+\tilde v_B'
    }e}_{\text{Bias Contribution}} + \underbrace{\frac{ \delta 
    }{\delta+\tilde v_B'
    }( \bar{A}_{\mathrm{pi}} - \theta)}_{\text{A-side pilot error}} , 
\end{equation}
\begin{equation}\label{eq:MSE-plug-in}
    \MSE_\infty
    =
    \underbrace{\frac{
        \tilde v_B'^2
    }{
        (\delta+\tilde v_B')^2
    }e^2}_{\text{Bias contribution}} + \underbrace{\frac{
       \delta^2
    }{
        (\delta+\tilde v_B')^2
    }V_{A, \mathrm{pi}}}_{\text{A-side pilot error}}.
\end{equation}

\end{corollary}

Here \(\tilde v_B'\) summarizes the sampling noise from B-side pilot data entering the estimated auxiliary biases, incorporating both the average auxiliary variance and the effect of the
pilot/test split. As a reminder, $e$ represents the average bias and $\delta$ represents the variance of the bias  or how dispersed they are. 


The main driving forces behind the aforementioned corollaries are the average bias level $e$  and cross-study bias dispersion $\delta$. First, as expected, a larger bias magnitude makes the performance of our practical shrinkage estimator worse. This is immediate from the first term of Equations~\eqref{eq:asymp_theta}-\eqref{eq:MSE-plug-in} in which a larger bias $|e|$ monotonically worsens our MSE. Hence, unsurprisingly, managers should aim to borrow from previous experiments that are most alike the current experiment's treatment effect. 

Understanding the role of $\delta$ is more nuanced. One may expect a similar monotonic relationship between $\delta$ and MSE, i.e., a smaller bias and smaller bias dispersion should always benefit the MSE. Equation~\eqref{eq:MSE-plug-in} shows this is not the case. We first explain the benefit that more dispersed bias could bring: in a large $M$-regime, when bias are more dispersed, the weights $\hat w_i$ have a greater flexibility to offset the bias through bias cancellation. This is theoretically supported by the first term in Equation~\eqref{eq:MSE-plug-in}. We see as $\delta \to \infty$, i.e., bias is more dispersed, we offset the bias contribution to zero by being able to flexibly and perfectly offset $e$.

However,  since we are in a non-oracle plug-in case, this correction is not perfect as it is guided by the noisy target pilot mean $\bar A_{\mathrm{pi}}$ rather than the unknown true target effect. As dispersion increases, the reduction in systematic bias comes at the cost of transmitting more target-pilot randomness into the final estimator as shown in Equation~\eqref{eq:MSE-plug-in}. This is because, asymptotically, what our estimated shrinkage weights do is to match all the B-side weighted means not with $\theta$ ($\theta$ is unfortunately not available to match to) but our best guess of $\theta$: $\bar{A}_{\mathrm{pi}}$. This is shown in the following expression
\begin{equation}
\label{eq:overfitting}
\sum_{i=1}^M
    \widehat w_i\bar B_{i,\mathrm{pi}}
    \xrightarrow[M\to\infty]{p}
    \bar A_{\mathrm{pi}},    
\end{equation}
where we see the estimated shrinkage weights are selected to align the weighted auxiliary pilot mean with \(\bar A_{\mathrm{pi}}\), rather than the unknown perfect target \(\theta\). Although Equation~\eqref{eq:overfitting} does not directly characterize
the limiting error of $\hat\theta(\vwhat)$, the same estimated weights
are subsequently applied to independent auxiliary testing samples to form the $\sum_{i=1}^M
    \widehat w_i\bar B_{i,\mathrm{te}}$ in $\hat\theta(\vwhat)$.
Consequently, the overfitting to  $\bar A_{\mathrm{pi}}$ can persist in
the final estimator $\hat\theta(\vwhat)$ through the weighted true auxiliary biases
(see Appendix~\ref{app:pilot-alignment-general} for the formal mathematical connection). In particular, the second term in Equation~\eqref{eq:asymp_theta} shows that the limiting estimation error inherits randomness from
the target pilot mean through \(\bar A_{\mathrm{pi}}-\theta\),
with its influence increasing as bias dispersion \(\delta\) grows.
Therefore, on one hand a larger $\delta$ allows us to overfit more perfectly as shown in Equation~\eqref{eq:asymp_theta} to offset the bias (benefit). 
However, if $\bar{A}_{\mathrm{pi}}$ is far from $\theta$ due to sampling noise, a larger $\delta$ magnifies this bias correction as a larger $\delta$ helps further overfit to the imperfect target (harm).



Unlike the bias $e$, $\delta$ has a non-monotonic optimal $\delta$ that is neither 0 nor $\infty$ that minimizes the MSE. The limiting MSE is minimized at
\begin{equation}\label{eq:optimal_delta*}
    \delta^\star
    =
    \frac{e^2\widetilde v'_B}
    {V_{A,\mathrm{pi}}}.
\end{equation}
Unless bias is exactly zero or infinite, even for small bias regimes ($e$ close to zero), managers should attempt to find previous experiments that still have some heterogeneity. Equation~\eqref{eq:optimal_delta*} gives managerial guidance on how to choose previous experiments' bias dispersion. If they believe the borrowed experiments have high bias, they should consequently choose more heterogeneous experiments proportional to the bias squared $e^2$ for bias cancellations. On the other hand, if managers have chosen experiments that on average have low bias, they should aim to choose experiments that are not as heterogeneous in bias. In summary, we recommend managers borrow from experiments with the lowest bias (potentially zero) and smallest (not necessarily zero) bias heterogeneity.

\section{Pooled Plug-in Shrinkage Estimator}\label{sec:pooled}
The general plug-in shrinkage estimator in Section~\ref{sec: general plug-in}
estimates each auxiliary study's bias separately as \(\widehat\epsilon_i\),
using only the pilot mean from study \(i\) and the A-side pilot mean. We see that such flexibility allows our B-side weights to overfit to the noisy pilot sample estimate as shown in Equation~\eqref{eq:overfitting}. Motivated by this overfitting issue and also the common practice in meta-analysis of combining multiple
study-level estimates into a single precision-weighted summary estimate
\citep{cochran1954combination},
we consider replacing the \(M\) study-specific bias estimates
\(\widehat\epsilon_1,\ldots,\widehat\epsilon_M\) with a single pooled bias
estimate \(\widehat\epsilon_{\mathrm{pool}}\).
We show that this pooled approach can offer substantial advantages over
the general plug-in estimator in certain settings, although not uniformly.

\subsection{Construction of Practical Pooled-bias Plug-in Shrinkage Estimator}

To construct the practical ``pooled-bias'' shrinkage estimator, we replace
the full vector of study-specific bias estimates
\(\vepshat=(\hat\epsilon_1,\dots,\hat\epsilon_M)^\top\)
with a single pooled bias estimate \(\widehat\epsilon_{\mathrm{pool}}\).
Specifically, we replace the term
\(\vepshat\vepshat^\top\) in the general plug-in matrix by
\(\widehat\epsilon_{\mathrm{pool}}^{\,2}\mathbf 1\mathbf 1^\top\).
Under this restriction, the plug-in matrix and estimated shrinkage weights
change from \eqref{eq:feasible-wM} to
\[
\widehat Q_{\mathrm{pool}}
=
\widehat\Sigma_{\mathrm{te}}
+
\widehat\epsilon_{\mathrm{pool}}^{\,2}\mathbf 1\mathbf 1^\top
+
\widehat V_{A,\mathrm{te}}\mathbf 1\mathbf 1^\top,
\qquad
\vwhat_{\mathrm{pool}}
=
\widehat V_{A,\mathrm{te}}
\widehat Q_{\mathrm{pool}}^{-1}\vone.
\]

It remains to specify how \(\widehat\epsilon_{\mathrm{pool}}\) should be
estimated. A naive choice would be the unweighted average of the
study-specific bias estimates
\(\widehat\epsilon_1,\ldots,\widehat\epsilon_M\).
This choice treats all auxiliary experiments symmetrically, but ignores
the fact that some auxiliary estimates are more precise than others.
Since the pooled-bias restriction is introduced to stabilize the weight
construction, we still want the pooled bias itself to reflect the relative
reliability of the auxiliary experiments.

We therefore choose \(\widehat\epsilon_{\mathrm{pool}}\) by following the
same oracle-to-feasible logic used above. First, within the pooled-bias
shrinkage class, we identify the scalar pooled bias
\(\epsilon_{\mathrm{pool},\mathrm{opt}}\) that minimizes MSE if all the parameters were known. This oracle
calculation determines the appropriate precision-weighted form of the
pooled bias. We then obtain the feasible pooled-bias estimate
\(\widehat\epsilon_{\mathrm{pool}}\) by replacing the unknown biases and
variances in that oracle expression with their pilot-based estimates.

We define the oracle pooled bias as the value that minimizes the MSE of
the corresponding pooled estimator.
For any \(\epsilon_{\mathrm{pool}}\in\mathbb R\), let
\[
Q_{\mathrm{pool}}(\epsilon_{\mathrm{pool}})
=
\Sigma_{\mathrm{te}}
+
\epsilon_{\mathrm{pool}}^2\mathbf 1\mathbf 1^\top
+
V_{A,\mathrm{te}}\mathbf 1\mathbf 1^\top,
\qquad
\bm w_{\mathrm{pool}}(\epsilon_{\mathrm{pool}})
=
V_{A,\mathrm{te}}
Q_{\mathrm{pool}}(\epsilon_{\mathrm{pool}})^{-1}\mathbf 1.
\]
The corresponding estimator is
\(
\hat\theta_{\mathrm{pool}}(\epsilon_{\mathrm{pool}})
=
\bigl(
1-\mathbf 1^\top
\bm w_{\mathrm{pool}}(\epsilon_{\mathrm{pool}})
\bigr)\bar A_{\mathrm{te}}
+
\bm w_{\mathrm{pool}}(\epsilon_{\mathrm{pool}})^\top
\bar{\mathbf B}_{\mathrm{te}},
\)

\begin{proposition}[Oracle-optimal pooled bias]
\label{prop:pooled_bias_opt}
Under Assumption~\ref{ass:model}, an MSE-minimizing oracle pooled bias
within the pooled-bias shrinkage class is
\[
\epsilon_{\mathrm{pool},\mathrm{opt}}
=
\frac{
\sum_{i=1}^M \epsilon_i/V_{B,i,\mathrm{te}}
}{
\sum_{i=1}^M 1/V_{B,i,\mathrm{te}}
}.
\]
\end{proposition}

Proposition~\ref{prop:pooled_bias_opt} shows that the oracle pooled bias
is a precision-weighted average of the auxiliary biases, with weights
proportional to \(1/V_{B,i,\mathrm{te}}\). Accordingly, its feasible
pilot-based analogue is
\[
\widehat\epsilon_{\mathrm{pool}}
=
\frac{
\sum_{i=1}^M
\widehat\epsilon_i/\widehat V_{B,i,\mathrm{te}}
}{
\sum_{i=1}^M
1/\widehat V_{B,i,\mathrm{te}}
},
\]
where \(\widehat\epsilon_i\) and \(\widehat V_{B,i,\mathrm{te}}\)
are computed from the pilot samples in the same way as in the
general plug-in estimator. Substituting \(\widehat\epsilon_{\mathrm{pool}}\) into
\(\widehat Q_{\mathrm{pool}}\) and \(\vwhat_{\mathrm{pool}}\) gives our
final practical pooled-bias plug-in estimator:
\begin{equation}\label{eq:pooled-thetaM}
\hat\theta_{\mathrm{pool}}
=
\bigl(
1-\vone^\top\vwhat_{\mathrm{pool}}
\bigr)\bar A_{\mathrm{te}}
+
\vwhat_{\mathrm{pool}}^\top\vBbarte.
\end{equation}
Throughout the remainder of the paper, we refer to
\eqref{eq:pooled-thetaM} as the pooled plug-in shrinkage estimator, or
simply the pooled shrinkage estimator.
This pooled structure has an immediate implication for how weight is
allocated across auxiliary studies. In particular, for any auxiliary
studies \(i\) and \(j\),
\[
    \frac{\widehat w_{\mathrm{pool},i}}
         {\widehat w_{\mathrm{pool},j}}
    =
    \frac{\widehat V_{B,j,\mathrm{te}}}
         {\widehat V_{B,i,\mathrm{te}}}.
\]
Thus, after pooling, the relative allocation of weight within the B side
depends only on the estimated variances. This differs from the general
plug-in estimator in \eqref{eq:feasible-wM}, where the study-specific bias
estimates \(\widehat\epsilon_i\) enter the weight construction and can
therefore affect which auxiliary studies receive more weight. We study the theoretical and managerial consequences of this difference below. 


\subsection{\(M\)-Asymptotic Analysis of the Pooled Plug-in Estimator}

We again rely on the \(M\)-asymptotic regime in which the number of auxiliary experiments
grows while the target-side design remains fixed.

\begin{theorem}\label{thm:pooled_bias_limit}
Under Assumptions~\ref{ass:model}--\ref{ass:noalign}, the pooled plug-in estimator satisfies
\[
\hat\theta_{\mathrm{pool}}
\xrightarrow[M\to\infty]{p}
\hat\theta_{\mathrm{pool},\infty}
=
\bigl(1-\hat T_\infty\bigr)\bar A_{\mathrm{te}}
+
\hat T_\infty(\theta+e),
\qquad
\hat T_\infty
=
\frac{
    \widehat V_{A,\mathrm{te}}
}{
    \widehat V_{A,\mathrm{te}}
    +
    [e-(\bar A_{\mathrm{pi}}-\theta)]^2
}.
\]
The corresponding limiting MSE is
\begin{equation}
\label{eq:pooled_MSE_nonclosed}
\MSE_{\mathrm{pool},\infty}
=
\mathbb E_{\bar A_{\mathrm{pi}},\widehat V_{A,\mathrm{te}}}
\Bigg[
\frac{
e^2\widehat V_{A,\mathrm{te}}^{\,2}
}{
\left\{
\widehat V_{A,\mathrm{te}}
+
[e-(\bar A_{\mathrm{pi}}-\theta)]^2
\right\}^{2}
}
+
V_{A,\mathrm{te}}
\frac{
[e-(\bar A_{\mathrm{pi}}-\theta)]^4
}{
\left\{
\widehat V_{A,\mathrm{te}}
+
[e-(\bar A_{\mathrm{pi}}-\theta)]^2
\right\}^{2}
}
\Bigg].
\end{equation}
\end{theorem}
A key implication of Theorem~\ref{thm:pooled_bias_limit} is that the information from B-side experiments enter
the limiting MSE only through their average bias \(e\); and unlike the limiting MSE of general plug-in estimator in  \eqref{eq:MSE-plug-in}, \(\MSE_{\mathrm{pool},\infty}\)
does not depend on the cross-study bias dispersion
\(\delta\). This follows from the pooled weighting structure described
above. Because the pooled estimator uses a single common bias estimate,
the relative B-side weights no longer depend on the individual
auxiliary biases. Consequently, the weighting rule cannot exploit
bias dispersion to offset the average auxiliary bias as shown by the lack of $\delta$ term in Theorem~\ref{thm:pooled_bias_limit}, but it also
avoids the corresponding overfitting of the relative weights to
the  $\bar{A}_{\mathrm{pi}}$ in \eqref{eq:overfitting}.
We remark that the issue of a noisy A-side estimate still affects the total amount of
borrowing in \(\widehat T_\infty\) through $[e-(\bar A_{\mathrm{pi}}-\theta)]$ in
Theorem~\ref{thm:pooled_bias_limit} . 

Since the pooled plug-in estimator is an extension of the general plug-in estimator, we focus the discussion on when one is more preferable than the other. First, the main difference between the two estimators is the role (or lack of role) of $\delta$ on the limiting MSE. As we noted in Section~\ref{sec:simplified}, bias dispersion is neither strictly harmful or beneficial. Hence, intuitively, the pooled estimator, that rids the role of $\delta$, is more preferable when the harm of $\delta$ is the largest, i.e., when the pilot estimate of $\bar{A}_{\mathrm{pi}}$ is very noisy and we overfit to this noise. 
Similarly, the general plug-in estimator is more preferable when the benefit of $\delta$ is the largest, i.e., when the bias is very large that we offset this bias. Consequently, the general plug-in estimator is particularly
attractive when both the average bias magnitude and bias dispersion
are small, as the benefits of its flexible study-specific reweighting
can outweigh the cost of overfitting to target-pilot noise.
By contrast, the pooled estimator restricts relative auxiliary
weights to depend only on estimated precision, limiting its ability
to exploit this flexibility.
However, when $e$ is small but $\delta$ is large, the general plug-in would unnecessarily overfit to a noisy pilot A-side estimate $\bar{A}_{\mathrm{pi}}$, and the pool-plug in approach becomes more attractive because
it restricts the potentially harmful bias-driven reweighting.
Initially, this may appear counterintuitive, as it suggests pooling, which uses a common bias parameter, can outperform the general plug-in estimator when the true biases are more heterogeneous. In this case, however, the pooled estimator acts as a regularization method to avoid the overfitting issue, i.e., when the harm of $\delta$ outweighs the benefit (summarized in row one of Table~\ref{tab:estimator_recommendation}). If we only compare these two estimators, then we generally would prefer the general plug-in estimator when $e$ is large and $\delta$ is large as it can utilize the bias dispersion for bias cancellation. However, we present another robust method in Section~\ref{sec:fixed} that generally outperforms both estimators when bias is large. 

Before presenting this alternative robust estimator, we provide a tractable approximation based on a second-order Delta expansion for cases in which
managers wishes to analytically approximate \(\MSE_{\mathrm{pool},\infty}\).

\begin{proposition}[Second-order Delta approximation for the pooled-bias risk]
\label{prop:pooled_bias_delta}
Approximate the expectation in
\eqref{eq:pooled_MSE_nonclosed} by applying a second-order Taylor expansion to the
integrand around
\[
(\bar A_{\mathrm{pi}},\widehat V_{A,\mathrm{te}})
=
(\theta,V_{A,\mathrm{te}}),
\]
and evaluate the resulting quadratic approximation using
\[
\Var(\bar A_{\mathrm{pi}})=V_{A,\mathrm{pi}},
\qquad
\Var(\widehat V_{A,\mathrm{te}})
\approx
\frac{2V_{A,\mathrm{te}}^2}{\nu_{A,\mathrm{pi}}},
\qquad
\Cov(\bar A_{\mathrm{pi}},\widehat V_{A,\mathrm{te}})\approx 0.
\]
The resulting second-order Delta approximation is
\[
\MSE_{\mathrm{pool},\infty}
\approx
\frac{V_{A,\mathrm{te}}e^2}{V_{A,\mathrm{te}}+e^2}
+
\frac{4V_{A,\mathrm{pi}}\,V_{A,\mathrm{te}}^2 e^2}{(V_{A,\mathrm{te}}+e^2)^3}
+
\frac{2V_{A,\mathrm{te}}^2}{\nu_{A,\mathrm{pi}}}
\frac{e^4}{(V_{A,\mathrm{te}}+e^2)^3},
\]
where
\[
\nu_{A,\mathrm{pi}}
=
\frac{2n_{A,\mathrm{pi}}}
{\kappa_A-\frac{n_{A,\mathrm{pi}}-3}{n_{A,\mathrm{pi}}-1}}.
\]
\end{proposition}

\section{Fixed-Weight Shrinkage with Hypothesis Testing}\label{sec:fixed}

The plug-in shrinkage rules in Sections~\ref{sec:practical}--\ref{sec:pooled}
estimate optimal weights from data, at the cost of sample splitting. This creates
a sample size inefficiency. Furthermore, as discussed above, both estimators suffer when bias is very large, which suggests a robust estimator that ``fallbacks'' to a non-shrinkage estimator may be preferable. Motivated by this, we consider a simpler alternative that uses all observations in the final estimator. Specifically, we assign deterministic weights proportional to sample size, eliminating the need for a separate pilot sample to estimate the weights, and directly pool the target and auxiliary sample means. Because this fixed-weight rule does not adjust for auxiliary
bias, we pair it with a hypothesis-testing gate: the estimator borrows from the
auxiliary datasets only when the aggregated auxiliary mean is not significantly
different from the target mean, and otherwise reverts to the unbiased benchmark
\(\bar A\). This construction is closely related with pretest, or test-and-pool,
estimators in the data-integration literature~\citep{gao2023pretest}.

\subsection{Construction of Fixed-weight Shrinkage with Hypothesis Testing}

For each \(M\), define the corresponding sample-size weights  \(u_{i,M}\),
and the aggregated auxiliary mean by sample size \(\widetilde B_M\) as
\[
    \widetilde N_{B,M}:=\sum_{i=1}^M N_{B,i},
    \qquad
    u_{i,M}:=\frac{N_{B,i}}{\widetilde N_{B,M}},
    \qquad
    \widetilde B_M:=\sum_{i=1}^M u_{i,M}\bar B_i
    \qquad i=1,\ldots,M.
\]
Using the full target and auxiliary samples, define the deterministic fixed-weight estimator as
\[
    \hat\theta_M^{\mathrm{fix}}
    =
    (1-\lambda_M)\bar A+\lambda_M\widetilde B_M
    =
    \frac{N_A\bar A+\sum_{i=1}^M N_{B,i}\bar B_i}
    {N_A+\widetilde N_{B,M}},
    \qquad
    \lambda_M:=\frac{\widetilde N_{B,M}}{N_A+\widetilde N_{B,M}}.
\]
In words, the fixed-weight rule simply pools all observations and assigns weights according to
sample size. 
Although this removes the need for sample splitting, $\hat\theta_M^{\mathrm{fix}}$ always shrinks to the auxiliary estimates, which may be undesirable especially when bias is high. As shown by Section~\ref{sec:oracle}, we know the optimal weights should be close to zero if the bias is very high.
Consequently, we pair it with a
hypothesis-testing gate to guard against large auxiliary bias.
Let \(S_A^2\) and \(S_{B_i}^2\) denote the usual unbiased \textit{full-sample}
variance estimators, and define
\[
    \widehat V_A:=\frac{S_A^2}{N_A},
    \qquad
    \widehat V_{B,i}:=\frac{S_{B_i}^2}{N_{B,i}},
    \qquad i=1,\ldots,M.
\]
These variance estimators are constructed from the full samples, in contrast to the pilot-based
variance estimators used in Section~\ref{sec:practical}.

We first define
\[
    D_M:=\widetilde B_M-\bar A,
    \qquad
    \widehat U_M:=\sum_{i=1}^M u_{i,M}^2\widehat V_{B,i}
    \]
We care about whether the bias of the auxiliary experiments are large, i.e., how large $D_M$ may be. Thus, we create a hypothesis testing-based ``gate'' to assess whether we shrink or not:
\[
    \widehat I_M
    =
    \mathbf 1\!\left\{
        |D_M|
        \le
        t_{1-\alpha/2,\nu_A}
        \sqrt{\widehat V_A+\widehat U_M}
    \right\},
\]
where \(t_{1-\alpha/2,\nu_A}\) is the \((1-\alpha/2)\)-quantile of the Student \(t\)
distribution with \(\nu_A\) degrees of freedom, defined in
Assumption~\ref{ass:working_full}. 
The final fixed-weight estimator with Hypothesis testing is
\begin{equation}\label{eq:theta-FH}
    \hat\theta_M^{\mathrm{FH}}
    =
    \widehat I_M\hat\theta_M^{\mathrm{fix}}
    +(1-\widehat I_M)\bar A.
\end{equation}
We refer to \(\hat\theta_M^{\mathrm{FH}}\) as the FH (Fixed-Hypothesis) estimator.  
It uses the fixed sample-size-weighted auxiliary aggregate, solving the need for sample splitting, only when its
discrepancy from the target estimate is not statistically significant
at the user-specified type-1 error level \(\alpha\);
otherwise, it reverts to the unbiased full-sample benchmark \(\bar A\), giving robustness to large bias regimes. 

\subsection{\(M\)-Asymptotic Analysis of FH Estimator}

We study the same asymptotic regime as in
Sections~\ref{sec:practical}--\ref{sec:pooled}: \(M\to\infty\). Because the FH estimator uses the full target and auxiliary samples and does not employ a pilot/testing split,
Assumption~\ref{ass:working}, which is formulated for pilot- and testing-sample
summaries, is not directly applicable. We therefore introduce the following
full-sample counterpart of that working approximation.

\begin{assumption}[Working approximation for full-sample summaries]
\label{ass:working_full}

In addition to Assumption~\ref{ass:model}, we use Gaussian working approximations for the
full-sample means:
\[
    \bar A\sim N(\theta,V_A),
    \qquad
    \bar B_i\sim N(\theta+\epsilon_i,V_{B,i}),
    \qquad i=1,\ldots,M.
\]
For the target variance estimator, we use the scaled chi-squared working approximation
\[
R_A:=\frac{\widehat V_A}{V_A}
=\frac{S_A^2}{\sigma_A^2}
\sim\frac{\chi^2_{\nu_A}}{\nu_A},
\]
and treat the full-sample means as independent of the corresponding full-sample variance
estimators:
\[
    (\bar A,\bar B_1,\ldots,\bar B_M)
    \perp\!\!\!\perp
    (S_A^2,S_{B,1}^2,\ldots,S_{B,M}^2).
\]
The exact moment conditions and the moment-matching definition of degree of freedom \(\nu_A\) are provided in
Appendix~\ref{app:working-FH}.
\end{assumption}

In addition, we impose the following regularity conditions on the auxiliary
datasets.

\begin{assumption}[Fixed-weight regularity]\label{ass:fw}
\leavevmode
\begin{enumerate}[label=\textup{(FW\arabic*)},leftmargin=2.5em]
    \item\label{ass:fw1}
    \textbf{No dominant auxiliary study.}
    As \(M\to\infty\), the fixed weights satisfy
    \[
        \max_{1\le i\le M}u_{i,M}\to0,
        \qquad
        \sup_{i\ge1}V_{B,i}<\infty.
    \]

    \item\label{ass:fw2}
    \textbf{No sample-size alignment with bias.}
    As \(M\to\infty\), the sample-size-weighted average auxiliary bias
    converges to the same limit \(e\) as the simple average:
    \[
        \bar\epsilon_M^{\,w}
        :=
        \sum_{i=1}^M u_{i,M}\epsilon_i
        =
        \frac{
            \sum_{i=1}^M N_{B,i}\epsilon_i
        }{
            \sum_{i=1}^M N_{B,i}
        }
        \xrightarrow[M\to\infty]{}
        e,
        \qquad
        e
        :=
        \lim_{M\to\infty}
        \frac{1}{M}
        \sum_{i=1}^M\epsilon_i.
    \]
\end{enumerate}
\end{assumption}

Condition~\ref{ass:fw1} implies that the variance of the aggregated auxiliary mean vanishes
    \[
    \Var(\widetilde B_M)
    =
    U_M
    \xrightarrow[M\to\infty]{} 0.
    \]

Condition~\ref{ass:fw2} is the fixed-weight analogue of the
no-systematic-alignment condition in
Assumption~\ref{ass:noalign}\ref{ass:noalign:align}. Whereas
Assumption~\ref{ass:noalign}\ref{ass:noalign:align} rules out systematic
alignment between biases and precision weights
\(1/\widehat V_{B,i,\mathrm{te}}\), Condition~\ref{ass:fw2} rules out systematic alignment between biases and
sample-size weights \(u_{i,M}=N_{B,i}/\widetilde N_{B,M}\). Intuitively, auxiliary
studies with systematically larger or smaller sample sizes should not also
have treatment effects that are systematically farther above or below the
target effect \(\theta\). It ensures that weighting the auxiliary studies by
sample size does not shift the limiting average bias away from the simple
average \(e\). Condition~\ref{ass:fw2} is plausible when auxiliary sample sizes are driven primarily by operational scale, such as traffic volume, market size, or experiment duration, rather than by the direction of the treatment effect. The condition may fail, however, when experiment size is itself related to treatment effectiveness; for example, if promising interventions are allowed to run longer or are expanded to more users, or if older, larger experiments are systematically more biased for the current target because the operating environment has changed.

Condition~\ref{ass:fw1} together with Condition~\ref{ass:fw2}  imply that the aggregated auxiliary mean behaves asymptotically like a signal
centered at \(\theta+e\), with vanishing B-side sampling variance.
Let \(\widehat I_\infty\) denote the limiting borrowing indicator:
\[
\widehat I_\infty
:=
\mathbf1\!\left\{
|e-(\bar A-\theta)|
\le t_{1-\alpha/2,\nu_A}\sqrt{\widehat V_A}
\right\},
\]

\begin{theorem}\label{thm:fixed_weight}
Under Assumptions~\ref{ass:model}, \ref{ass:working_full}, and~\ref{ass:fw},
\[
\widehat\theta_M^{\mathrm{FH}}
\xrightarrow[M\to\infty]{p}
(\theta+e)\widehat I_\infty
+\bar A(1-\widehat I_\infty),
\qquad
    \MSE_M^{\mathrm{FH}}
    :=
    \E\!\left[
        (\hat\theta_M^{\mathrm{FH}}-\theta)^2
    \right]
    \xrightarrow[M\to\infty]{}
    \MSE_\infty^{\mathrm{FH}},
\]
where
\[
\begin{aligned}
    \MSE_\infty^{\mathrm{FH}}
    =
    \E_{R_A}\Big[
        &
        e^2p_F(R_A) +
        V_A
        \big\{
            1-p_F(R_A)
            +
            b_F(R_A)\phi(b_F(R_A))
            -
            a_F(R_A)\phi(a_F(R_A))
        \big\}
    \Big],
\end{aligned}
\]
and, for \(r>0\),
\[
a_F(r):=\frac{e}{\sqrt{V_A}}-t_{1-\alpha/2}\sqrt r,
\qquad
b_F(r):=\frac{e}{\sqrt{V_A}}+t_{1-\alpha/2}\sqrt r,
\qquad
p_F(r):=\Phi\!\bigl(b_F(r)\bigr)-\Phi\!\bigl(a_F(r)\bigr).
\]
\end{theorem}


We remark that all distributional equalities and limiting MSE formulas in the FH analysis are evaluated under the summary-level working model in Assumption~\ref{ass:working_full}. They are exact under Gaussian sampling and otherwise should be interpreted as working-model approximations.  

To see the effect of the hypothesis-testing "gate", we further
let the target sample size \(N_A\to\infty\), so that
\(\nu_A\to\infty\) (after taking \(M\to\infty\) as in
Theorem~\ref{thm:fixed_weight}). Then,
\[
R_A\xrightarrow[N_A\to\infty]{p}1,
    \qquad
    t_{1-\alpha/2,\nu_A}
    \xrightarrow[N_A\to\infty]{}
    z_{1-\alpha/2}.
    \]
Using these limits, we obtain the following closed-form
approximation to the limiting MSE for large \(N_A\):
\begin{equation}
\label{eq:simplified_FH}
\MSE_\infty^{\mathrm{FH}} \approx e^2\, p_F + V_A\bigl(1 - p_F + b_F\,\phi(b_F) - a_F\,\phi(a_F)\bigr),
\end{equation}
with
\[
a_F := \frac{e}{\sqrt{V_A}} - z_{1-\alpha/2}, \qquad
b_F := \frac{e}{\sqrt{V_A}} + z_{1-\alpha/2}, \qquad
p_F := \Phi(b_F) - \Phi(a_F).\]

Equation~\eqref{eq:simplified_FH} shows how bias $e^2$ gets reduced not by bias correction through bias dispersion as in Equation~\eqref{eq:MSE-plug-in}, instead, FH controls
this contribution by deciding whether to borrow.
In particular, increasing
the testing level \(\alpha\) lowers the critical value of the
testing gate and therefore reduces the probability of borrowing.
The two extreme cases illustrate this effect:
\[
    \lim_{\alpha\to 0}\MSE_\infty^{\mathrm{FH}}=e^2,
    \qquad
    \lim_{\alpha\to 1}\MSE_\infty^{\mathrm{FH}}=V_A.
\]
Thus, managers seeking greater protection against substantial
auxiliary bias can choose a larger \(\alpha\), e.g., $\alpha = 0.10$, to make the FH
estimator less likely to borrow (at the cost of more frequently
rejecting potentially useful auxiliary information).
We detail this argument and summarize all these comparison and managerial insights in the following section.

\section{Comparison of Estimators with Managerial Insights}
\label{subsec:comparison}
By now, we have proposed three different shrinkage-based estimators, all with their strengths and weakness in different regimes. We now compare these formally and characterize when each estimator should be preferred and summarize this as managerial insights.

We begin by comparing when the FH estimator works best through Theorem~\ref{thm:fixed_weight}.  Theorem~\ref{thm:fixed_weight} shows that, like the pooled plug-in estimator,
the limiting MSE of the FH estimator does not depend directly on the
cross-study bias dispersion \(\delta\). The key structural advantage of FH,
however, is that it can shut down borrowing entirely. As a result, even under
severe systematic misalignment of the auxiliary archive, FH estimator can protect
against large auxiliary bias by reverting to the full-sample no-borrowing
estimator, as indicated in Equation~\eqref{eq:simplified_FH}
\[
\lim_{|e|\to\infty}\widehat I_\infty = 0
, \qquad
\lim_{|e|\to\infty}
\MSE_\infty^{\mathrm{FH}}=V_A.
\]
However, this protection is not costless and we characterize it mathematically in Proposition~\eqref{prop:fh-endpoint-comparison}.


\begin{proposition}[Endpoint comparison for the FH estimator]
\label{prop:fh-endpoint-comparison}
Under the conditions of Theorem~\ref{thm:fixed_weight}, for any fixed
\(\alpha\in(0,1)\) and every finite \(e\in\mathbb R\),
\[
    \MSE_\infty^{\mathrm{FH}}(e)
    >
    \min\{e^2,V_A\}.
\]
\end{proposition}
The two quantities, \(e^2\) and \(V_A\), correspond to the limiting
MSEs of always borrowing from the auxiliary aggregate and never
borrowing, respectively. Proposition~\ref{prop:fh-endpoint-comparison}
shows that the FH estimator has strictly higher MSE than the better
of two extreme decisions. In particular, when \(e^2\) is small,
FH cannot fully realize the MSE reduction available from always
borrowing, because its testing gate may reject useful auxiliary
information based on the observed discrepancy rather than the
unknown true bias.
Thus, FH's protection against large auxiliary bias comes at the
cost of potentially sacrificing MSE improvements when the average
bias is small. In such favorable borrowing environments, the general
and pooled plug-in estimators can achieve lower MSE, as illustrated
by the simulations in Section~\ref{sec:simulation}. By contrast,
FH's distinctive advantage emerges when the average bias is large,
as its testing gate allows the estimator to revert to the
full-sample benchmark.




We now summarize the comparisons of all three proposed estimators. First, as expected, shrinking is most effective when the average bias of auxiliary experiments are small ($e$ close to zero). As shown in Section~\ref{sec:simulation}, this is the regime where all proposed shrinkage estimators have better performances than the non-shrinkage estimators. Second, within a small bias regime and moderate bias dispersion, the plug-in based estimators (both pooled and non-pooled) are generally preferred over the FH estimator as the FH estimator does not optimize the shrinkage weights. Third, in the same small bias regime, the general plug-in based estimator is preferred over the pooled version when $\delta$, the bias dispersion, is small so that bias correction benefit outweighs the overfitting harm detailed in Section~\ref{sec:simplified}. Lastly, the FH estimator is preferred over both plug-in based estimators when bias is large (regardless of $\delta$) as it allows us to revert back to the non-shrinkage estimator, protecting from potentially harmful shrinkage. Table~\ref{tab:estimator_recommendation} summarizes these insights visually.


\begin{table}[h!]
\centering
\begin{tabular}{|c|c|c|}
\hline
 & \textbf{Small \(\delta\)} & \textbf{Large \(\delta\)} \\
\hline
\textbf{Small \(e\)}
& General plug-in
& Pooled plug-in  \\
\hline
\textbf{Large \(e\)}
& FH
& FH \\
\hline
\end{tabular}
\caption{Recommended estimator by bias environment.}
\label{tab:estimator_recommendation}
\end{table}

In practice, the top-left quadrant arises when the auxiliary experiments
evaluate interventions that are close variants of the target. For example,
successive iterations of the same promotional campaign tested on the same
customer segment. We remark that many large companies with long-established platforms are in this quadrant. For example, 80-90\% of experiments in Google and Bing have close to zero treatment effects \citep{exp_unsuccessful}. In many of these established companies, performing an experiment with a large treatment effect (both positive or negative) is generally difficult as the platform is already largely optimized. In these cases, treatment effects are usually close to each other, e.g., within 1-2\%, thus both $e$ and $\delta$ are small in this case. As shown in Section~\ref{sec:simulation}, this quadrant is where the shrinkage estimator have the best performance.

The top-right quadrant arises when a firm draws on a broad portfolio of past
A/B tests across different product lines or markets. The interventions differ
enough that individual biases relative to the target vary widely, producing
large \(\delta\). However, if the auxiliary interventions are not
systematically more or less effective than the target---for instance, because
the firm repeatedly tests variants of the same type of intervention and
deviations from the target effect are roughly symmetric---then the average
bias \(e\) may remain small even though individual biases are dispersed.
This is a plausible scenario when the auxiliary experiments share the same causal
mechanism as the target but differ in implementation details that shift
effects in both directions.

The bottom row corresponds to settings in which the auxiliary experiments
are systematically misaligned with the target, so that the average bias
magnitude is large regardless of cross-study heterogeneity. For example,
this may arise when borrowing from a different therapeutic indication in a
clinical trial or from a different geographic market with structurally
different consumer behavior. In this regime, the shrinkage estimators, unsurprisingly, performs the worst, i.e., borrowing from unrelated interventions does not really help inform the current intervention.

\section{Inference for the Shrinkage Estimators}\label{sec:inference}

Although the main focus of the paper is point estimation and MSE reduction,
practitioners typically require uncertainty quantification as opposed to only point estimation after applying the
proposed shrinkage estimators. For completeness, we present (asymptotically) valid confidence intervals, beginning with the oracle shrinkage estimator.

\begin{proposition}[Asymptotic confidence interval for the oracle shrinkage estimator]
\label{prop:oracle-ci}
Under Assumption~\ref{ass:model}, fix \(M\) and consider an asymptotic sequence
in which
\[
    N_A\to\infty,
    \qquad
    N_{B,i}\to\infty,
    \quad
    i=1,\ldots,M.
\]
Then the oracle shrinkage estimator \eqref{eq:oracle-thetaM}, with oracle weight
vector \eqref{eq:oracle-wM}, satisfies
\begin{equation}\label{eq:oracle-asymptotic-normality}
    \frac{
        \hat\theta(w_M^*)
        -
        \theta
        -
        (w_M^*)^\top\epsilon
    }{
        \sqrt{
            \bigl(1-\mathbf 1^\top w_M^*\bigr)^2V_A
            +
            (w_M^*)^\top\Sigma w_M^*
        }
    }
    \xrightarrow{d}
    N(0,1).
\end{equation}
Consequently, an asymptotic \((1-\xi)\) confidence interval for \(\theta\) is
\begin{equation}\label{eq:oracle-asymptotic-ci}
\begin{aligned}
\Biggl[
&
\hat\theta(w_M^*)-(w_M^*)^\top\epsilon
-
z_{1-\xi/2}
\sqrt{
    \bigl(1-\mathbf 1^\top w_M^*\bigr)^2V_A
    +
    (w_M^*)^\top\Sigma w_M^*
},
\\
&
\hat\theta(w_M^*)-(w_M^*)^\top\epsilon
+
z_{1-\xi/2}
\sqrt{
    \bigl(1-\mathbf 1^\top w_M^*\bigr)^2V_A
    +
    (w_M^*)^\top\Sigma w_M^*
}
\Biggr].
\end{aligned}
\end{equation}
\end{proposition}

We next move to inference on the general plug-in shrinkage estimator that utilizes $M$ asymptotics instead:
\begin{proposition}[Large-\(M\) confidence interval for the general plug-in shrinkage estimator]
\label{prop:adaptive-ci}
Under Assumptions~\ref{ass:model}, \ref{ass:working}, and~\ref{ass:largeM}, when $\widetilde A>0$, the
general plug-in shrinkage estimator
\eqref{eq:feasible-thetaM} satisfies
\[
    \hat\theta_M-\theta
    \xrightarrow[M\to\infty]{d}
    N\!\left(
        \frac{\tilde\mu_1\tilde\nu}{\tilde\Delta},
        \frac{\tilde A^2}{\tilde\Delta^2}V_{A,\mathrm{pi}}
    \right).
\]
Accordingly, an \(M\)-asymptotic \((1-\xi)\) confidence interval for
\(\theta\) is
\[
\left[
    \hat\theta_M
    -
    \frac{\tilde\mu_1\tilde\nu}{\tilde\Delta}
    -
    z_{1-\xi/2}
    \sqrt{V_{A,\mathrm{pi}}}
    \frac{\tilde A}{\tilde\Delta},
    \;
    \hat\theta_M
    -
    \frac{\tilde\mu_1\tilde\nu}{\tilde\Delta}
    +
    z_{1-\xi/2}
    \sqrt{V_{A,\mathrm{pi}}}
    \frac{\tilde A}{\tilde\Delta}
\right].
\]
Under Assumption~\ref{ass:noalign}, this interval can be simplified as
\[
\left[
    \hat\theta_M
    -
    \frac{e\tilde v_B'}{\delta+\tilde v_B'}
    -
    z_{1-\xi/2}
    \sqrt{V_{A,\mathrm{pi}}}
    \frac{\delta}{\delta+\tilde v_B'},
    \;
    \hat\theta_M
    -
    \frac{e\tilde v_B'}{\delta+\tilde v_B'}
    +
    z_{1-\xi/2}
    \sqrt{V_{A,\mathrm{pi}}}
    \frac{\delta}{\delta+\tilde v_B'}
\right].
\]
\end{proposition}

The pooled plug-in estimator has a generally non-Gaussian
unconditional limiting distribution because its conditional
mean and variance depend on the random pilot-based weight.
Conditioning on the pilot data fixes the estimated weights
and allows us to exploit their independence from the
testing-sample means.

\begin{proposition}[Large-\(M\) conditional confidence interval
for the pooled plug-in estimator]
\label{prop:pooled-CI}

Under Assumptions~\ref{ass:model}, \ref{ass:working},
\ref{ass:largeM}, and~\ref{ass:noalign}, and based on
Theorem~\ref{thm:pooled_bias_limit},
conditional on
\[
    \widehat T_\infty=t,
\]
the limiting pooled plug-in estimator satisfies
\[
    \widehat\theta_{\mathrm{pool},\infty}-\theta
    \,\Bigm|\,
    \widehat T_\infty=t
    \sim
    N\!\left(
        et,\,
        V_{A,\mathrm{te}}(1-t)^2
    \right).
\]

Therefore, a conditional \((1-\xi)\) confidence interval for
\(\theta\), given \(\widehat T_\infty=t\), is
\[
\left[
    \widehat\theta_{\mathrm{pool},\infty}
    -
    et
    -
    z_{1-\xi/2}(1-t)\sqrt{V_{A,\mathrm{te}}},
    \;
    \widehat\theta_{\mathrm{pool},\infty}
    -
    et
    +
    z_{1-\xi/2}(1-t)\sqrt{V_{A,\mathrm{te}}}
\right].
\]
\end{proposition}

For the fixed-weight estimator with hypothesis testing
\(\hat\theta_M^{\mathrm{FH}}\) in \eqref{eq:theta-FH}, the unconditional
distribution is also a non-Gaussian mixture because the estimator selected by the
gate depends on the observed data. The practical gate is completely determined
once the discrepancy \(D_M\) and its estimated variance
\(\widehat V_A+\widehat U_M\) are observed. We therefore condition on these two
quantities.

\begin{proposition}[Conditional confidence interval for the FH estimator]
\label{prop:fh-conditional-ci}

Under Assumptions~\ref{ass:model} and~\ref{ass:working_full},
\[
    \hat\theta_M^{\mathrm{FH}}-\theta
    \mid
    D_M=d,\,
    \widehat V_A+\widehat U_M=q
    \sim
    N\!\left(m_M(d,q),s_M^2\right),
\]
where
\[
    m_M(d,q)
    :=
    -\frac{V_A}{V_A+U_M}\left(d-\bar\epsilon_M^{\,w}\right)
    +
    \lambda_M d\,
    \mathbf 1\!\left\{|d|\le t_{1-\alpha/2}\sqrt q\right\},
    \qquad
    s_M^2:=\frac{V_AU_M}{V_A+U_M},
\]
with
\[
    \bar\epsilon_M^{\,w}:=\sum_{i=1}^M u_{i,M}\epsilon_i,
    \qquad
    U_M:=\sum_{i=1}^M u_{i,M}^2V_{B,i}.
\]

Therefore, a conditional \((1-\xi)\) confidence interval for \(\theta\) is
\[
    \left[
    \bar A+\frac{V_A}{V_A+U_M}\left(d-\bar\epsilon_M^{\,w}\right)
    -z_{1-\xi/2}\sqrt{\frac{V_AU_M}{V_A+U_M}},
    \;
    \bar A+\frac{V_A}{V_A+U_M}\left(d-\bar\epsilon_M^{\,w}\right)
    +z_{1-\xi/2}\sqrt{\frac{V_AU_M}{V_A+U_M}}
    \right].
\]
\end{proposition}
The \(t\)-quantile \(t_{1-\alpha/2}\) is used only for the hypothesis-testing
gate, because it determines whether the estimator borrows auxiliary
information. The confidence interval uses \(z_{1-\xi/2}\) because, conditional
on the quantities determining the gate, the estimator has a Gaussian
distribution with conditional variance \(s_M^2\).

The resulting intervals depend on bias and variance quantities that may not be
known in practice. Since confidence intervals is not the focus of the paper, we do not delve into any plug-in approach and merely state them for completeness. We, therefore, view these results primarily as distributional
characterizations and recommend complementing them with sensitivity analyses
over plausible bias specifications when implementing the procedures.
See \citet{Peng2026bvalue}, who study inference for combined unbiased and
biased estimators by constructing confidence intervals indexed by the
magnitude of the unknown bias.

\section{Simulation Study}\label{sec:simulation}

We examine how the proposed shrinkage estimators perform with
a finite number of auxiliary experiments and whether their empirical
 performance reflects the expected theoretical large-$M$ results. The
comparison focuses on two features of the auxiliary archive:
the average bias $e$ and the cross-study bias dispersion
$\delta$. These quantities govern the theoretical comparisons
in Corollary~\ref{cor:simplified},
Theorem~\ref{thm:pooled_bias_limit}, and
Theorem~\ref{thm:fixed_weight}. We compare the
general plug-in, pooled plug-in, and FH estimators across
four illustrative bias environments.

\paragraph{Simulation design.}
We generate independent Gaussian observations according to
\[
    A_j\sim N(\theta,\sigma_A^2),
    \qquad
    B_{i,j}\sim N(\theta+\epsilon_i,\sigma_{B,i}^2).
\]
Gaussian sampling makes the working distributions of the
sample means and variance estimators in Assumptions~\ref{ass:working}
and ~\ref{ass:working_full} exact. We set $\theta=10$, $N_A=100$, and $\sigma_A^2=64$,
so the full-sample no-borrowing benchmark has
$V_A=\sigma_A^2/N_A=0.64$. The A-side testing fraction
is $\rho_A=0.20$, yielding $n_{A,\mathrm{pi}}=80$ and
$n_{A,\mathrm{te}}=20$.
For each auxiliary study, we independently generate
\[
    N_{B,i}\sim U_d\{100,\ldots,150\},
    \qquad
    \rho_{B,i}\sim U[0.30,0.70],
    \qquad
    \sigma_{B,i}^2\sim U[81,100].
\]
The B-side testing count is obtained by rounding
$\rho_{B,i}N_{B,i}$ to the nearest integer, with the remaining
observations assigned to the pilot sample. We use the FH gate
level $\alpha=0.10$.

Independently of these study characteristics, the auxiliary
biases follow
\[
    \epsilon_i
    \stackrel{\mathrm{i.i.d.}}{\sim}
    U\!\left[
        e-\sqrt{3\delta},
        e+\sqrt{3\delta}
    \right],
    \qquad
    \E(\epsilon_i)=e,
    \quad
    \Var(\epsilon_i)=\delta.
\]
This construction satisfies the bounded-bias and no-alignment
conditions in Assumption~\ref{ass:largeM} and Assumption~\ref{ass:noalign}
underlying the simplified large-$M$ results.
We consider $(e,\delta)=(0.15,0.05)$, $(0.15,12)$,
$(5,2)$, and $(5,36)$, representing archives with different
degrees of systematic misalignment and bias heterogeneity.
The labels ``small'' and ``large'' for $\delta$ describe
the heterogeneity relative to the corresponding level of $e$. To facilitate comparisons across the four environments, we
generate a common sequence of auxiliary study designs
$(N_{B,i},\rho_{B,i},\sigma_{B,i}^2)$ and standardized
bias draws $U_i\sim U[-\sqrt{3},\sqrt{3}]$.
For each environment $(e,\delta)$, we construct the
study-specific biases as
$\epsilon_i=e+\sqrt{\delta}U_i$.
Thus, the four environments have distinct realized
bias sequences but share the same auxiliary study
designs and underlying standardized bias draws.

For each bias environment, we construct an auxiliary archive
containing $M_{\max}=80$ studies, using the common study designs
and standardized bias draws described above. Within each
archive, we consider nested collections consisting of the
first $M=1,\ldots,80$ auxiliary studies. Conditional on the
realized archive in each environment, we conduct $R=50{,}000$
Monte Carlo replications to evaluate estimator performance.
 For computational efficiency, pilot and testing
means and unbiased variance estimators are generated directly
from their exact Gaussian and scaled chi-squared sampling
distributions. The general and pooled plug-in estimators use
pilot summaries to estimate shrinkage weights and independent
testing means to form the final estimates; the FH estimator
uses the reconstructed full-sample summaries.

\begin{figure}[!htbp]
    \centering
    \includegraphics[width=0.95\textwidth]
    {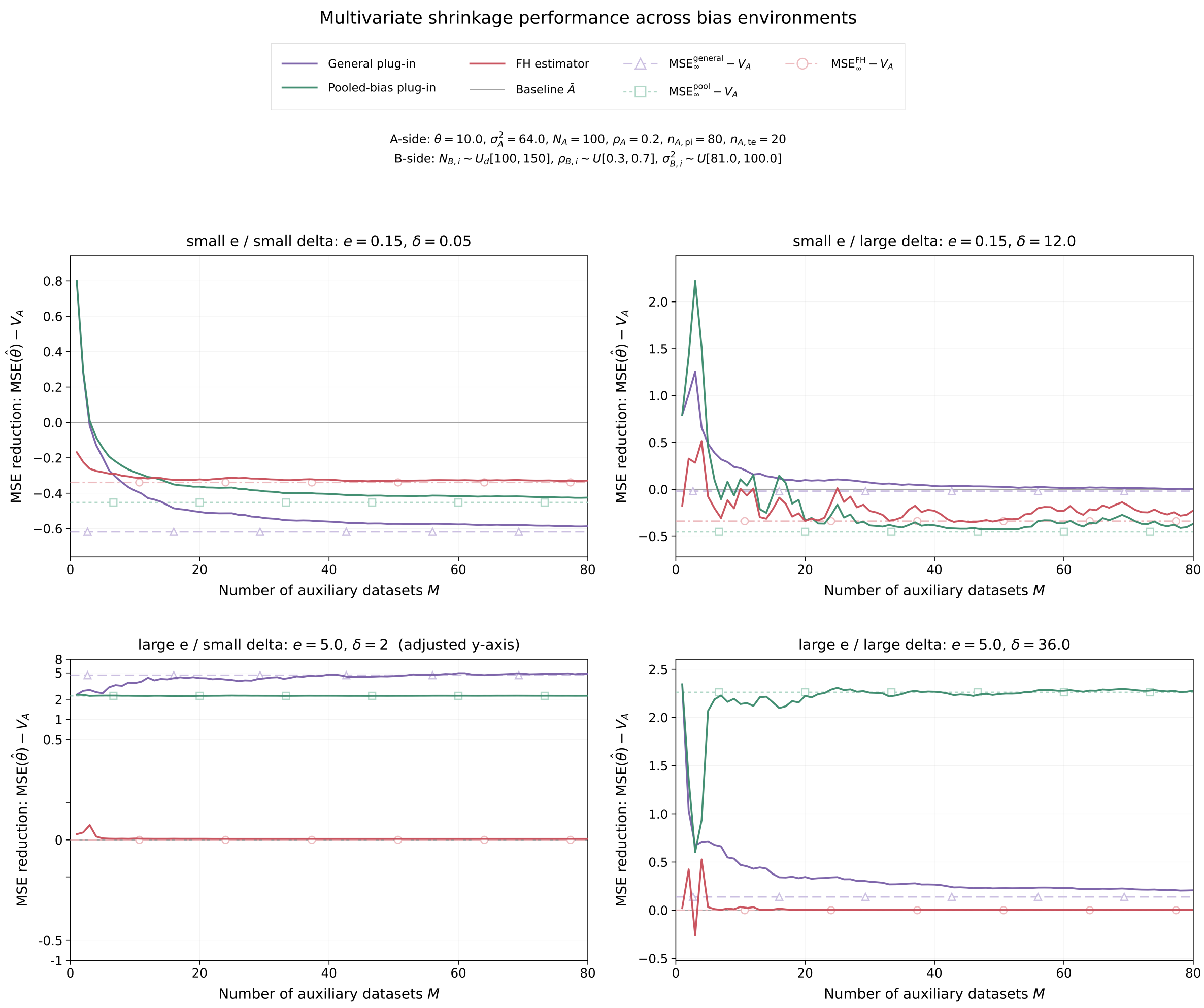}
    \caption{
    \textbf{Finite-$M$ shrinkage performance across auxiliary
    bias environments.}
    Solid curves show empirical MSE differences relative to
    the full-sample target estimator $\bar A$. The zero line represents no
    borrowing; negative values indicate MSE improvement.
    Horizontal dashed lines show the population-level $M$-asymptotic
    MSE differences for the three practical estimators.
    Finite-$M$ curves
    reflect the realized archive composition, whereas the
    dashed limits use the population bias parameters
    $(e,\delta)$.
    }
    \label{fig:multi_compare}
\end{figure}

We estimate the MSE by
\[
    \widehat{\MSE}(\hat\theta_M)
    =
    \frac{1}{R}
    \sum_{r=1}^{R}
    \bigl(\hat\theta_M^{(r)}-\theta\bigr)^2.
\]
Figure~\ref{fig:multi_compare} reports
$\widehat{\MSE}(\hat\theta_M)-V_A$, so negative values indicate
improvement over no borrowing. The horizontal dashed lines
are the corresponding population-level large-$M$ MSE
differences: the general plug-in limit is calculated from
Corollary~\ref{cor:simplified}, the pooled limit from
Theorem~\ref{thm:pooled_bias_limit}, and the FH limit from
Theorem~\ref{thm:fixed_weight}. The pooled expectation is
evaluated using $500{,}000$ Monte Carlo draws over A-side
pilot randomness, while the FH expectation is evaluated by
numerical integration over the target variance-estimation
distribution. We further show the asymptotic theoretical expressions using respective dash lines.

Figure~\ref{fig:multi_compare} highlights four main findings. First, the finite-$M$ MSEs generally move toward their
corresponding large-$M$ limits as more auxiliary experiments
become available, confirming our $M$-asymptotic theory. However, convergence can be slower and more
volatile when bias dispersion is large, as the realized
composition of auxiliary biases may fluctuate substantially
with the addition of new studies. Since each simulation examines a particular realized
sequence of auxiliary experiments, other sequences may stabilize
faster or slower. Thus, the large-$M$ results capture the overall
performance patterns, while their finite-$M$ accuracy also depends
on the realized auxiliary archive.

Second, the upper-left panel illustrates the advantage of the
general plug-in estimator when both $e$ and $\delta$ are small.
Although both plug-in estimators follow the same weighting framework,
the general plug-in can flexibly reallocate weights across auxiliary
studies using their study-specific bias estimates. When auxiliary
biases are small and similar, such reweighting incurs little
additional true-bias cost, allowing the general plug-in estimator
to benefit from its greater flexibility. The pooled estimator
cannot fully exploit this opportunity because its relative
auxiliary weights depend only on estimated precision. In this
setting, all three practical shrinkage estimators also achieve
lower MSE than the no-borrowing benchmark once sufficiently many
auxiliary studies are available.

Third, when the average bias remains small but bias dispersion
becomes large (upper-right panel), the pooled plug-in outperforms
the general plug-in. This confirms our earlier theoretical
comparison: pooling provides greater stability under substantial
bias heterogeneity, whereas the general plug-in's flexible
study-specific reweighting becomes less advantageous.

Fourth, the bottom panels highlight two findings when the average
auxiliary bias is large. As a reminder, this is a region where bias is large, i.e., the region where borrowing is undesirable and may create more harm than good.
Comparing the bottom-left and bottom-right
panels, increasing \(\delta\) reduces the MSE of the general plug-in
estimator. This is consistent with our theoretical finding that,
when \(e\) is large, increasing bias dispersion toward the
MSE-minimizing level \(\delta^\star\) can improve performance by
allowing study-specific reweighting to partially offset the large
average auxiliary bias. Nevertheless, both plug-in estimators still
have substantially higher MSE than the no-borrowing benchmark in
these settings. In contrast, the FH estimator detects the large discrepancy and largely
reverts to the target-only estimator $\bar A$, keeping its MSE
close to $V_A$, showing the desired robustness property.

\section{Concluding Remark}
\label{sec:discussions}
This paper studies how to combine a target randomized experiment with multiple auxiliary
experiments when the auxiliary treatment effects may be systematically biased relative to the
target. We develop an oracle shrinkage benchmark and three non-oracle practical estimators that differ in how
much structure they impose on cross-study bias. The analysis shows that the value of auxiliary
information is governed by two distinct features: the average auxiliary bias, which captures
systematic misalignment with the target, and cross-study bias dispersion, which determines how
much can be gained from study-specific reweighting. We find that, as expected, having lower average bias monotonically leads to better performance for shrinkage-based estimators. On the other hand, bias dispersion or the variance of the bias, does not have a monotonic relationship with the shrinkage performance. A larger dispersion benefits by allowing greater flexibility of bias cancellation to offset the bias. However, this also exacerbates an overfitting issue to an imperfectly estimated target. 

All three proposed  practical (non-oracle) shrinkage-based estimators have different strengths and weaknesses. The general plug-in estimator is most useful when average bias and bias dispersion are small such that the benefit of bias cancellation outweighs the overfitting issue detailed in Section~\ref{sec:simplified}. Similarly, the pooled plug-in estimator is preferred when the harm of overfitting outweighs the bias cancellation (large $\delta$ small $e$). Lastly, the FH estimator
provides protection against substantial aggregate mismatch by reverting to the target estimator
when the auxiliary signal is inconsistent with the target data. The simulations show that these
asymptotic comparisons also provide useful guidance for finite collections of auxiliary
experiments.


More broadly, the results highlight that the usefulness of past experimental data depends not
only on how much auxiliary information is available, but also on how its discrepancies from the
current target are structured. This distinction is central to designing borrowing procedures that
can improve estimation while remaining robust to imperfect comparability across experiments.

\newpage

\bibliographystyle{ormsv080}
\bibliography{ref}

@misc{oberst2023,
      title={Understanding the Risks and Rewards of Combining Unbiased and Possibly Biased Estimators, with Applications to Causal Inference}, 
      author={Michael Oberst and Alexander D'Amour and Minmin Chen and Yuyan Wang and David Sontag and Steve Yadlowsky},
      year={2023},
      eprint={2205.10467},
      archivePrefix={arXiv},
      primaryClass={stat.ME},
      url={https://arxiv.org/abs/2205.10467} 
}

@book{exp_unsuccessful,
author = {Kohavi, Ron and Tang, Diane and Xu, Ya},
year = {2020},
month = {03},
pages = {},
title = {Trustworthy Online Controlled Experiments: A Practical Guide to A/B Testing},
isbn = {9781108724265},
doi = {10.1017/9781108653985}
}

@misc{rosenman2020combining,
      title={Combining Observational and Experimental Datasets Using Shrinkage Estimators}, 
      author={Evan Rosenman and Guillaume Basse and Art Owen and Michael Baiocchi},
      year={2020},
      eprint={2002.06708},
      archivePrefix={arXiv},
      primaryClass={stat.ME},
      url={https://arxiv.org/abs/2002.06708} 
}

@misc{li2023cluster,
      title={Mostly Beneficial Clustering: Aggregating Data for Operational Decision Making}, 
      author={Chengzhang Li and Zhenkang Peng and Ying Rong},
      year={2023},
      eprint={2311.17326},
      archivePrefix={arXiv},
      primaryClass={cs.LG},
      url={https://arxiv.org/abs/2311.17326}, 
}

@article{gupta2019top,
  title={Top Challenges from the First Practical Online Controlled Experiments Summit},
  author={Gupta, Somit and Kohavi, Ronny and Tang, Diane and Xu, Ya and Andersen, Reid and Bakshy, Eytan and Cardin, Niall and Chandran, Sumita and Chen, Nanyu and Coey, Dominic and Curtis, Mike and Deng, Alex and Duan, Weitao and Forbes, Peter and Frasca, Brian and Guy, Tommy and Imbens, Guido W. and Saint Jacques, Guillaume and Kantawala, Pranav and Katsev, Ilya and Katzwer, Moshe and Konutgan, Mikael and Kunakova, Elena and Lee, Minyong and Lee, MJ and Liu, Joseph and McQueen, James and Najmi, Amir and Smith, Brent and Trehan, Vivek and Vermeer, Lukas and Walker, Toby and Wong, Jeffrey and Yashkov, Igor},
  journal={ACM SIGKDD Explorations Newsletter},
  volume={21},
  number={1},
  pages={20--35},
  year={2019},
  publisher={Association for Computing Machinery},
  doi={10.1145/3331651.3331655}
}

@misc{kaufman2017democratizing,
      title={Democratizing online controlled experiments at Booking.com}, 
      author={Raphael Lopez Kaufman and Jegar Pitchforth and Lukas Vermeer},
      year={2017},
      eprint={1710.08217},
      archivePrefix={arXiv},
      primaryClass={cs.HC},
      url={https://arxiv.org/abs/1710.08217}
}

@inproceedings{xie2016improving,
  title={Improving the Sensitivity of Online Controlled Experiments: Case Studies at Netflix},
author = {Xie, Huizhi and Aurisset, Juliette},
  year={2016},
  month = {08},
    pages = {645-654},
    doi = {10.1145/2939672.2939733}
}

@article{fiksel2024adjustedCI,
  author  = {Fiksel, Jacob},
  title   = {On exact randomization-based covariate-adjusted confidence intervals},
  journal = {Biometrics},
  volume  = {80},
  number  = {2},
  pages    = {ujae051},
  year    = {2024},
  month   = jun,
  doi     = {10.1093/biomtc/ujae051},
  url     = {https://doi.org/10.1093/biomtc/ujae051}
}

@misc{Peng2026bvalue,
      title={Introducing the b-value: combining unbiased and biased estimators from a sensitivity analysis perspective}, 
      author={Zhexiao Lin and Peter J. Bickel and Peng Ding},
      year={2026},
      eprint={2602.16310},
      archivePrefix={arXiv},
      primaryClass={stat.ME},
      url={https://arxiv.org/abs/2602.16310}
}

@inproceedings{james1961estimation,
  author    = {James, Willard and Stein, Charles},
  title     = {Estimation with Quadratic Loss},
  booktitle = {Proceedings of the Fourth Berkeley Symposium on Mathematical Statistics and Probability},
  volume    = {1},
  pages     = {361--379},
  year      = {1961},
  publisher = {University of California Press},
  address   = {Berkeley, CA}
}

@article{cochran1954combination,
  author    = {William G. Cochran},
  title     = {The Combination of Estimates from Different Experiments},
  journal   = {Biometrics},
  volume    = {10},
  number    = {1},
  pages     = {101--129},
  year      = {1954},
  publisher = {Wiley},
  issn      = {0006-341X},
  url       = {http://www.jstor.org/stable/3001666}
}

@misc{gao2023pretest,
      title={Pretest estimation in combining probability and non-probability samples}, 
      author={Chenyin Gao and Shu Yang},
      year={2023},
      eprint={2305.17801},
      archivePrefix={arXiv},
      primaryClass={stat.ME},
      url={https://arxiv.org/abs/2305.17801}, 
}

\newpage

\begin{APPENDICES}

\section{Exact statement of Assumptions}

\subsection{Exact statement of Assumption~\ref{ass:working}}
\label{app:working-assumption}

Here we present the precise moment conditions and moment-matching formulas underlying
the working approximation in Assumption~\ref{ass:working}. 

In addition to Assumption~\ref{ass:model}, suppose the following hold.

\begin{enumerate}[label=\textup{(A\arabic*)},leftmargin=2.5em]
    \item
    The target distribution and each auxiliary distribution have finite fourth moments and zero
    third central moments:
    \[
        \E\!\left[(A_j-\theta)^4\right]<\infty,
        \qquad
        \E\!\left[(A_j-\theta)^3\right]=0,
    \]
    and, for each \(i=1,\ldots,M\),
    \[
        \E\!\left[(B_{i,j}-(\theta+\epsilon_i))^4\right]<\infty,
        \qquad
        \E\!\left[(B_{i,j}-(\theta+\epsilon_i))^3\right]=0.
    \]
    Define
    \[
        \kappa_A
        :=
        \frac{\E[(A_j-\theta)^4]}{\sigma_A^4},
        \qquad
        \kappa_{B,i}
        :=
        \frac{\E[(B_{i,j}-(\theta+\epsilon_i))^4]}{\sigma_{B,i}^4}.
    \]

    \item 
    For each pilot/testing split, the sample means are approximated by Gaussian with the
    corresponding population means and variances:
    \[
        \bar A_{\mathrm{pi}}
        \sim
        N\!\left(\theta,\frac{\sigma_A^2}{n_{A,\mathrm{pi}}}\right),
        \qquad
        \bar A_{\mathrm{te}}
        \sim
        N\!\left(\theta,\frac{\sigma_A^2}{n_{A,\mathrm{te}}}\right),
    \]
    and, for each \(i=1,\ldots,M\),
    \[
        \bar B_{i,\mathrm{pi}}
        \sim
        N\!\left(\theta+\epsilon_i,\frac{\sigma_{B,i}^2}{n_{B,i,\mathrm{pi}}}\right),
        \qquad
        \bar B_{i,\mathrm{te}}
        \sim
        N\!\left(\theta+\epsilon_i,\frac{\sigma_{B,i}^2}{n_{B,i,\mathrm{te}}}\right).
    \]

    For the pilot-sample variance estimators, we adopt the scaled chi-squared approximations
    \[
        \frac{S_{A,\mathrm{pi}}^2}{\sigma_A^2}
        \sim
        \frac{\chi^2_{\nu_{A,\mathrm{pi}}}}{\nu_{A,\mathrm{pi}}},
        \qquad
        \nu_{A,\mathrm{pi}}
        :=
        \frac{2n_{A,\mathrm{pi}}}
        {\kappa_A-\frac{n_{A,\mathrm{pi}}-3}{n_{A,\mathrm{pi}}-1}},
    \]
    and, for each \(i=1,\ldots,M\),
    \[
        \frac{S_{B_i,\mathrm{pi}}^2}{\sigma_{B,i}^2}
        \sim
        \frac{\chi^2_{\nu_{B_i,\mathrm{pi}}}}{\nu_{B_i,\mathrm{pi}}},
        \qquad
        \nu_{B_i,\mathrm{pi}}
        :=
        \frac{2n_{B,i,\mathrm{pi}}}
        {\kappa_{B,i}-\frac{n_{B,i,\mathrm{pi}}-3}{n_{B,i,\mathrm{pi}}-1}}.
    \]
    Under this working approximation, each pilot-sample mean is treated
    as independent of its corresponding pilot-sample variance estimator:
    \[
        \bar A_{\mathrm{pi}}
        \perp\!\!\!\perp
        S_{A,\mathrm{pi}}^2,
        \qquad
        \bar B_{i,\mathrm{pi}}
        \perp\!\!\!\perp
        S_{B_i,\mathrm{pi}}^2,
        \qquad i=1,\ldots,M.
    \]

    Whenever inverse second moments are required, we assume that the relevant effective degrees
    of freedom are large enough for those moments to exist; in particular,
    \[
        \inf_{i\ge 1}\nu_{B_i,\mathrm{pi}}>4.
    \]
\end{enumerate}

Together with Assumption~\ref{ass:largeM}\ref{ass:largeM:reg}, the
scaled chi-squared approximation implies an inverse-moment bounds that will be used in the \(M\)-asymptotic analysis:
\[
    \widehat V_{B,i,\mathrm{te}}
    =
    V_{B,i,\mathrm{te}}
    \frac{\chi^2_{\nu_{B_i,\mathrm{pi}}}}
    {\nu_{B_i,\mathrm{pi}}}.
\]
Hence, for \(\nu_{B_i,\mathrm{pi}}>4\),
\[
    \E\!\left[\widehat V_{B,i,\mathrm{te}}^{-1}\right]
    =
    \frac{1}{V_{B,i,\mathrm{te}}}
    \frac{\nu_{B_i,\mathrm{pi}}}
    {\nu_{B_i,\mathrm{pi}}-2},
    \qquad
    \E\!\left[\widehat V_{B,i,\mathrm{te}}^{-2}\right]
    =
    \frac{1}{V_{B,i,\mathrm{te}}^2}
    \frac{\nu_{B_i,\mathrm{pi}}^2}
    {(\nu_{B_i,\mathrm{pi}}-2)(\nu_{B_i,\mathrm{pi}}-4)}.
\]
Because
\[
    V_{B,i,\mathrm{te}}\ge \underline v>0,
    \qquad
    \inf_{i\ge1}\nu_{B_i,\mathrm{pi}}>4,
\]
it follows that
\[
    \sup_{i\ge1}
    \E\!\left[\widehat V_{B,i,\mathrm{te}}^{-1}\right]
    <\infty,
    \qquad
    \sup_{i\ge1}
    \E\!\left[\widehat V_{B,i,\mathrm{te}}^{-2}\right]
    <\infty.
\]
These bounds are used repeatedly in the \(M\)-asymptotic proofs below.

\subsection{Exact statement of Assumption~\ref{ass:working_full}}\label{app:working-FH}

In addition to Assumption~\ref{ass:model}, suppose the following hold.
\begin{enumerate}[label=\textup{(A\arabic*)},leftmargin=2.5em]

    \item
    The target distribution and each auxiliary distribution have finite
    fourth moments and zero third central moments:
    \[
        \E[(A_j-\theta)^4]<\infty,
        \qquad
        \E[(A_j-\theta)^3]=0,
    \]
    and, for each \(i=1,\ldots,M\),
    \[
        \E[(B_{i,j}-(\theta+\epsilon_i))^4]<\infty,
        \qquad
        \E[(B_{i,j}-(\theta+\epsilon_i))^3]=0.
    \]
    Define
    \[
        \kappa_A
        :=
        \frac{\E[(A_j-\theta)^4]}{\sigma_A^4}.
    \]

    \item
    Suppose \(N_A\ge2\), \(N_{B,i}\ge2\) for every \(i\), and
    \(V_A>0\), \(V_{B,i}>0\). The full-sample means are approximated by
    Gaussian random variables with the corresponding population means
    and variances:
    \[
        \bar A\sim N(\theta,V_A),
        \qquad
        V_A=\frac{\sigma_A^2}{N_A},
    \]
    and, for each \(i=1,\ldots,M\),
    \[
        \bar B_i
        \sim
        N(\theta+\epsilon_i,V_{B,i}),
        \qquad
        V_{B,i}
        =
        \frac{\sigma_{B,i}^2}{N_{B,i}}.
    \]

    For the target full-sample variance estimator, we adopt the
    chi-squared approximation
    \[
        \frac{\widehat V_A}{V_A}
        =
        \frac{S_A^2}{\sigma_A^2}
        \sim
        \frac{\chi_{\nu_A}^2}{\nu_A},
        \qquad
        \nu_A
        :=
        \frac{2N_A}{
            \kappa_A-\frac{N_A-3}{N_A-1}
        }.
    \]

    Under the working approximation, for every \(M\),
    \[
        (\bar A,\bar B_1,\ldots,\bar B_M)
        \perp\!\!\!\perp
        (S_A^2,S_{B,1}^2,\ldots,S_{B,M}^2).
    \]

\end{enumerate}

Assumption~\ref{ass:working_full} imposes distributional approximations on the
full-sample summaries. The Gaussian approximation for the sample means is
motivated by the central limit theorem, while the scaled chi-squared
approximation matches the first two moments of the sample variance. Under Gaussian sampling, these summary-level relations are exact and
\[
\nu_A=N_A-1.
\]

\section{Univariate case: one auxiliary estimator}\label{app:univariate}
This appendix complements the main \(M\)-asymptotic analysis by studying the practical
plug-in estimator with a single auxiliary estimator. The purpose is to provide finite-sample
intuition for how estimated shrinkage weights affect the bias--variance tradeoff.
In the oracle setting, the shrinkage
estimator in Equation~\eqref{eq:oracle-theta1} improves upon the unbiased benchmark because
its weight is computed from the true bias and variance components. In practice, however, those
quantities are unknown and must be estimated from the pilot data. As a result, the shrinkage
weight is random and may deviate from its oracle value, introducing additional variability.

Let \(B\) denote the single auxiliary dataset. The plug-in shrinkage estimator is
\begin{equation}\label{eq:eq:feasible-theta1}
        \hat\theta(\hat w)
    =
    (1-\hat w)\bar A_{\mathrm{te}}
    +
    \hat w\,\bar B_{\mathrm{te}},
\end{equation}
where
\begin{equation}\label{eq:univariate_plugin_weights}
    \hat w
    =
    \frac{\widehat V_{A,\mathrm{te}}}
    {
        \hat\epsilon^2
        +
        \widehat V_{A,\mathrm{te}}
        +
        \widehat V_{B,\mathrm{te}}
    },
    \qquad
    \hat\epsilon
    =
    \bar B_{\mathrm{pi}}-\bar A_{\mathrm{pi}}.
\end{equation}
The estimator \(\hat\theta(\hat w)\) has the same linear-combination form as the oracle
estimator \(\hat\theta(w)\) in \eqref{eq:oracle-theta1}, but uses a random weight estimated from
the pilot data. Our next goal is to determine when the practical plug-in shrinkage estimator improves upon the unbiased benchmark  \(\bar A\) based on the full target sample. That is, we study conditions under which
\[
\MSE\bigl(\hat\theta(\hat w)\bigr) < \MSE(\bar A) = V_A = \frac{\sigma^2_A}{N_A}.
\]
By the independence of the pilot and testing samples, the MSE of \eqref{eq:eq:feasible-theta1}
can be written as
\[
    \MSE\bigl(\hat\theta(\hat w)\bigr)
    =
    V_{A,\mathrm{te}}
    -
    2V_{A,\mathrm{te}}\mu_1
    +
    \left(t+\frac{\sigma_B^2}{n_{B,\mathrm{te}}}\right)\mu_2,
\]
where
\[
    \mu_1=\E(\hat w),
    \qquad
    \mu_2=\E(\hat w^2),
    \qquad
    t
    :=
    \epsilon^2+V_{A,\mathrm{te}}
    =
    \epsilon^2+\frac{\sigma_A^2}{n_{A,\mathrm{te}}}.
\]
The first term \(V_{A,\mathrm{te}}\) is the MSE of the no-shrinkage estimator based only on
the testing sample. The second term captures the variance reduction from shrinkage, while the
third term captures the cost of shrinkage, including both the squared bias and the additional
randomness induced by estimating the weight.

Unfortunately, the exact analytical expression of $\MSE\bigl(\hat\theta(\hat w)\bigr)$ is intractable when $\hat w$ is estimated through Equation~\eqref{eq:univariate_plugin_weights}. More specifically, the MSE depends on \(\mathbb E(\widehat w)\) and \(\mathbb E(\widehat w^2)\). Under the working approximation in Assumption~\ref{ass:working}, the plug-in variance estimators are scaled chi-square random variables, while \(\widehat\epsilon^2\) is the square of a normal approximation. The resulting moments of \(\widehat w\) therefore involve ratios of dependent nonlinear combinations of these random variables, for which closed-form finite-sample expressions are not analytically tractable. We, therefore, apply a Delta-approximation approach to extract key insights.

Formally, with the MSE decomposition above, we approximate \(\mu_1\) by a first-order Delta approximation and bound \(\mu_2\) from above. The details are given in Appendix~\ref{app:cubic}. This yields the following sufficient condition for the practical plug-in shrinkage estimator to improve upon the full-sample benchmark \(\bar A\).

\begin{theorem}[Approximate improvement condition]\label{thm:sufficientimprove}
Under Assumption~\ref{ass:model}, Assumption~\ref{ass:working}, and the approximations developed in
Appendix~\ref{app:cubic}, if
\[
    P(n_{B,\mathrm{te}})<0,
\]
then
\[
    \MSE\bigl(\hat\theta(\hat w)\bigr)<\MSE(\bar A).
\]
Here \(P(n_{B,\mathrm{te}})\) is the cubic polynomial
\[
    P(n_{B,\mathrm{te}})
    =
    t^2n_{B,\mathrm{te}}^3
    +
    2\sigma_B^2t\,n_{B,\mathrm{te}}^2
    +
    \alpha^\dagger n_{B,\mathrm{te}}
    +
    \beta^\dagger,
\]
with
\[
    t
    :=
    \epsilon^2+V_{A,\mathrm{te}}
    =
    \epsilon^2+\frac{\sigma_A^2}{n_{A,\mathrm{te}}},
\]
\[
    \alpha^\dagger
    =
    \sigma_B^4\left(1-\frac{2}{C_A}\right)
    +
    \sigma_B^2t
    \left[
        \frac{n_{A,\mathrm{te}}}{C_A}
        \left(1-\frac{n_{A,\mathrm{te}}}{N_A}\right)
        \frac{\sigma_B^2}{\sigma_A^2}
    \right],
\]
\[
    \beta^\dagger
    =
    \sigma_B^4
    \left[
        \frac{n_{A,\mathrm{te}}}{C_A}
        \left(1-\frac{n_{A,\mathrm{te}}}{N_A}\right)
        \frac{\sigma_B^2}{\sigma_A^2}
    \right],
\]
and
\[
    C_A
    :=
    1+\frac{1}{n_{A,\mathrm{pi}}}
    \left(
        \kappa_A-\frac{n_{A,\mathrm{pi}}-3}{n_{A,\mathrm{pi}}-1}
    \right).
\]
\end{theorem}

Theorem~\ref{thm:sufficientimprove} reduces the practical improvement question to the sign of a cubic polynomial in \(n_{B,\mathrm{te}}\). The condition \(P(n_{B,\mathrm{te}})<0\), however, should be interpreted as a conservative sufficient condition rather than a sharp characterization of all cases in which shrinkage improves upon \(\bar A\). This is because the result combines a first-order Delta approximation for \(\mathbb E(\hat w)\) with an upper bound for \(\mathbb E(\hat w^2)\). Thus, if the condition fails, the plug-in estimator may still improve; the theorem simply does not certify such an improvement. We include simulation to evaluate the realized finite-sample performance of the plug-in estimator.

The result for Theorem~\ref{thm:sufficientimprove}is nevertheless useful as a design criterion. It identifies
pilot/testing splits for which improvement can be certified under the analytic approximation,
even though the exact finite-sample MSE of the plug-in estimator depends on unavailable
moments of the random weight \(\hat w\). This leads to the following question: does there exist
a feasible auxiliary testing size \(n_{B,\mathrm{te}}\) such that \(P(n_{B,\mathrm{te}})<0\)?

\begin{proposition}[Approximated sufficient condition for existence of an improving region]\label{prop:cubic-geometry}
Let
\[
n_{B,\mathrm{te}}^\circ=\frac{\sqrt{4\sigma_B^4-3\alpha^\dagger}-2\sigma_B^2}{3t}.
\]
A sufficient condition for the existence of some \(n_{B,\mathrm{te}}>0\) such that \(\MSE\bigl(\hat\theta(\hat w)\bigr)<\MSE(\bar A)\) is:
\begin{enumerate}[label=\textup{(C\arabic*)},leftmargin=2.5em]
    \item \(n_{B,\mathrm{te}}^\circ\) lies on the positive real line, equivalently, \(\alpha^\dagger<0\);
    \item \(P(n_{B,\mathrm{te}}^\circ)<0\).
\end{enumerate}
When the total auxiliary sample size \(N_B\) is finite, a further feasibility condition is required:
\begin{enumerate}[label=\textup{(C\arabic*)},leftmargin=2.5em,start=3]
    \item \(2\le n_{B,\mathrm{te}}^\circ\le N_B-2\).
\end{enumerate}
\end{proposition}

Proposition~\ref{prop:cubic-geometry} gives an analytically checkable sufficient condition for
the existence of an improving auxiliary testing size. The original question is whether there
exists some \(n_{B,\mathrm{te}}>0\) such that
\[
    P(n_{B,\mathrm{te}})<0.
\]
Directly answering this question would require searching over all possible auxiliary testing
sizes. The proposition reduces this search to the closed-form candidate
\(n_{B,\mathrm{te}}^\circ\), the unique positive minimizer of the cubic whenever such a minimizer
exists. Thus, Conditions (C1)--(C2) should be read together as the sufficient condition:
(C1) ensures that the cubic has a meaningful positive interior minimizer, and (C2) verifies
that the cubic is negative at that minimizer. When both conditions hold, the continuity of
\(P\) implies that \(P(n_{B,\mathrm{te}})<0\) on a nonempty interval around
\(n_{B,\mathrm{te}}^\circ\), so an improving auxiliary testing size exists under the approximation
in Theorem~\ref{thm:sufficientimprove}.

When the total auxiliary sample size \(N_B\) is finite, Condition (C3) adds feasibility: the
closed-form improving candidate must lie within the admissible split range. Therefore, the
purpose of Proposition~\ref{prop:cubic-geometry} is not to avoid evaluating \(P\) altogether,
but to convert the existence question into a simple analytic certificate. This certificate is useful
for deriving insights: it shows how the auxiliary bias, target-side variance, and
pilot/testing split affect whether the plug-in shrinkage estimator can be certified to improve
upon the full-sample benchmark \(\bar A\).

This yield three immediately insights. First, increasing \(\epsilon^2\) makes both (C1) and (C2) harder to satisfy, as the bias cost of borrowing becomes larger. Second, increasing \(\sigma_A^2\) makes (C1) and (C2) easier to satisfy, because a noisier target estimator makes shrinkage more attractive. Third, \(\sigma_B^2\) plays a more limited role: it enters only through the feasibility condition (C3), rather than through the intrinsic possibility conditions (C1)--(C2). Thus, relative to the target-side variance \(\sigma_A^2\), the auxiliary-side variance \(\sigma_B^2\) is less central for determining whether improvement is possible in principle; instead, it mainly affects whether a favorable auxiliary split can be attained with the available sample size. In summary, Proposition~\ref{prop:cubic-geometry} shows the univariate plug-in estimator is most
attractive when the auxiliary bias is not too large and the target estimator is sufficiently noisy
that variance reduction is valuable, echoing similar conclusions found in the James-Stein literature.

\subsection{Simulation for Univariate Plug-in Shrinkage Estimator}\label{subsec:sim_uni}

This subsection complements the one-auxiliary analysis above by evaluating the realized
finite-sample MSE of the practical plug-in estimator. Whereas
Theorem~\ref{thm:sufficientimprove} and Proposition~\ref{prop:cubic-geometry} provide
conservative analytical conditions for improvement, the simulation evaluates the implemented
estimator directly.

In all univariate designs, we set
\[
\theta=0,\qquad
N_A=N_B=200,\qquad
\rho_A=\rho_B=0.5,\qquad
\sigma_B^2=1.
\]

In each Monte Carlo replication, we generate the pilot and test datasets based on the splitting scheme in Section~\ref{sec:practical}, estimate the plug-in weight using the pilot samples, and apply the resulting weight to the independent test-sample means. In the univariate case, the estimator is
\[
\hat\theta(\hat w)
=
(1-\hat w)\bar A_{\mathrm{te}}
+
\hat w\bar B_{\mathrm{te}},
\qquad
\hat w
=
\frac{\widehat V_{A,\mathrm{te}}}
{\hat\epsilon^2+\widehat V_{A,\mathrm{te}}+\widehat V_{B,\mathrm{te}}},
\]
where
\[
\hat\epsilon=\bar B_{\mathrm{pi}}-\bar A_{\mathrm{pi}},
\qquad
\widehat V_{A,\mathrm{te}}=\frac{S^2_{A,\mathrm{pi}}}{n_{A,\mathrm{te}}},
\qquad
\widehat V_{B,\mathrm{te}}=\frac{S^2_{B,\mathrm{pi}}}{n_{B,\mathrm{te}}}.
\]

We consider the following designs. First, we vary the target-side variance \(\sigma_A^2\) over 40 equally spaced grid points in \([5,80]\), with the auxiliary bias from B side using
\[
|\epsilon|\in\{0.1,0.3,1.0\}.
\]
Figure~\ref{fig:uni_sigmaA} reports the resulting MSE reductions.

Second, we vary the auxiliary bias magnitude \(|\epsilon|\) over 50 equally spaced grid points in \([0,5]\), with
\[
\sigma_A^2\in\{5,10,40\}.
\]
Figure~\ref{fig:uni_eps} reports the resulting MSE reductions.


\begin{figure}[H]
    \centering
    \includegraphics[width=0.82\textwidth]{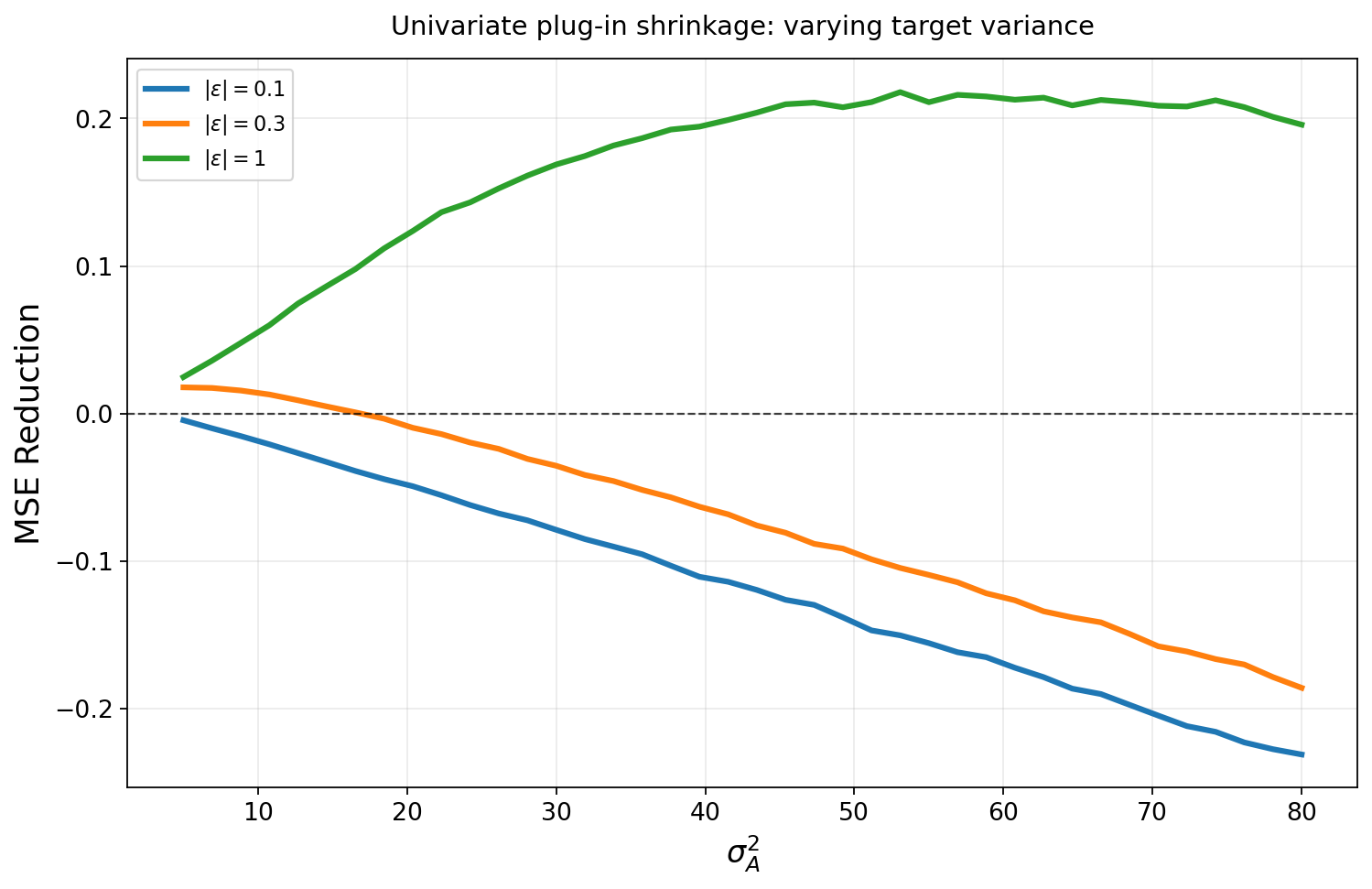}
    \caption{Univariate plug-in shrinkage as the target-side variance varies. The figure reports empirical MSE reduction, \(\widehat{\MSE}\{\hat\theta(\hat w)\}-V_A\), over 40 equally spaced values of \(\sigma_A^2\in[5,80]\), with \(|\epsilon|\in\{0.1,0.3,1.0\}\). The design uses \(\theta=0\), \(N_A=N_B=200\), \(\rho_A=\rho_B=0.5\), and \(\sigma_B^2=1\). Each point is computed from \(50{,}000\) Monte Carlo repetitions. Negative values indicate improvement over the full-sample target benchmark \(\bar A\).}
    \label{fig:uni_sigmaA}
\end{figure}

\begin{figure}[H]
    \centering
    \includegraphics[width=0.82\textwidth]{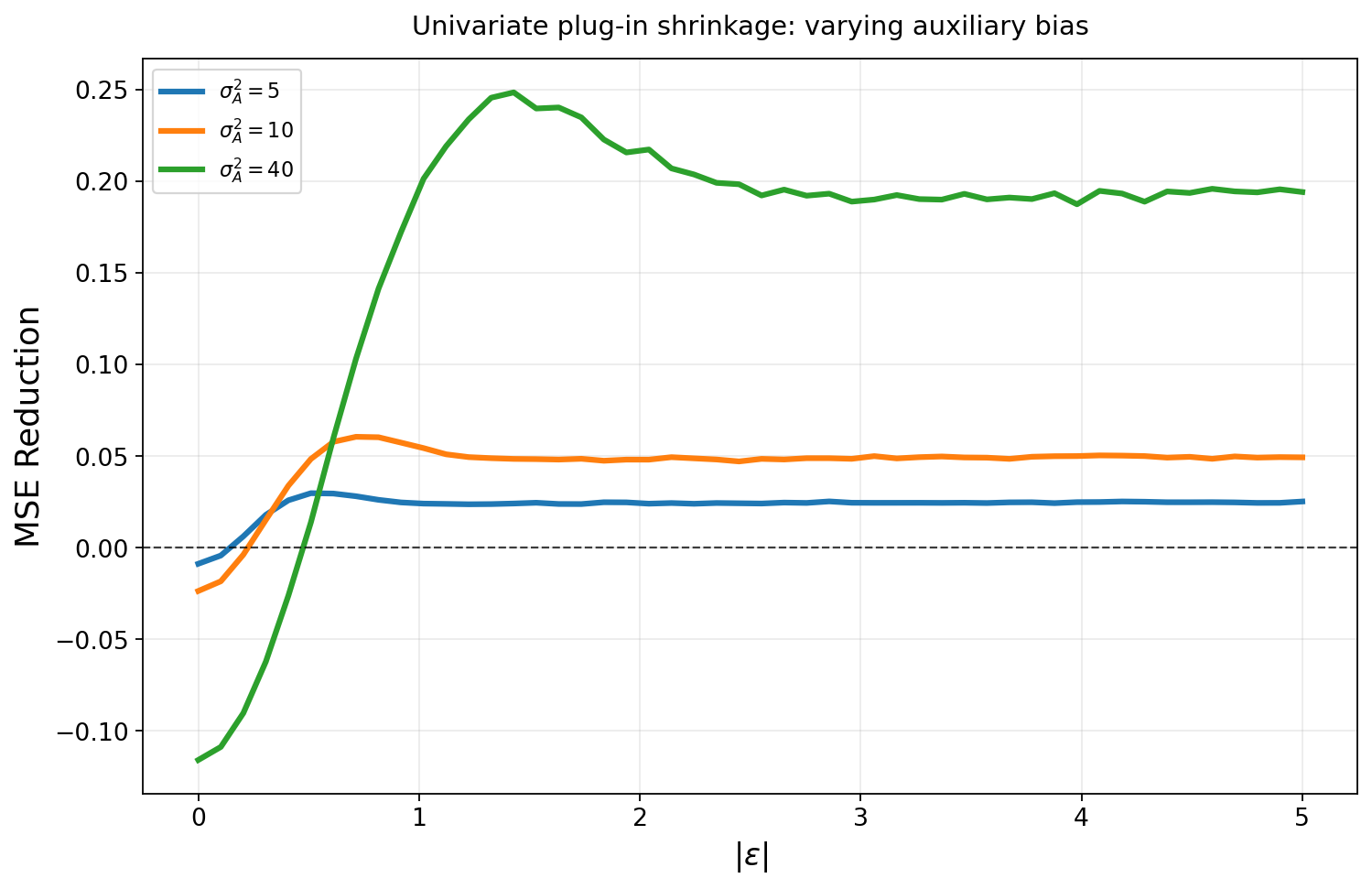}
    \caption{Univariate plug-in shrinkage as the auxiliary bias varies. The figure reports empirical MSE reduction, \(\widehat{\MSE}\{\hat\theta(\hat w)\}-V_A\), over 50 equally spaced values of \(|\epsilon|\in[0,5]\), with \(\sigma_A^2\in\{5,10,40\}\). The design uses \(\theta=0\), \(N_A=N_B=200\), \(\rho_A=\rho_B=0.5\), and \(\sigma_B^2=1\). Each point is computed from \(50{,}000\) Monte Carlo repetitions.}
    \label{fig:uni_eps}
\end{figure}


The univariate simulations largely support the message of Proposition~\ref{prop:cubic-geometry}. Figure~\ref{fig:uni_sigmaA} shows that, when the auxiliary bias is small or moderate, increasing \(\sigma_A^2\) makes the MSE reduction more negative. This pattern is consistent with the theoretical message that a noisier target experiment creates more room for shrinkage to improve upon \(\bar A\). The simulations also reveal an important finite-sample qualification. When \(|\epsilon|\) is large, increasing \(\sigma_A^2\) can make the plug-in estimator perform worse rather than better. The reason is that a noisier target estimator gives the plug-in rule more incentive to borrow from the auxiliary sample. If the auxiliary bias is small, this additional borrowing is beneficial. If \(|\epsilon|\) is large, however, even a moderate amount of mistakenly assigned auxiliary weight can introduce substantial squared bias and increase MSE. Thus, in large-bias regimes, a noisier target experiment does not automatically make shrinkage beneficial; it can instead amplify the cost of imperfect weight estimation.

Figure~\ref{fig:uni_eps} confirms the bias-side message of Proposition~\ref{prop:cubic-geometry}. For all three values of \(\sigma_A^2\), the MSE reduction increases as \(|\epsilon|\) moves away from zero, indicating that the plug-in estimator becomes less attractive as the auxiliary estimator becomes more biased. This pattern is especially pronounced when \(\sigma_A^2\) is large. In that case, the target estimator is noisy, so the plug-in rule tends to borrow more aggressively from the auxiliary sample, and auxiliary bias therefore translates into a larger MSE cost.

\section{Two-Channel Decomposition of Pilot Alignment for the General Plug-in Estimator}
\label{app:pilot-alignment-general}

This appendix provides a formal explanation of the role of cross-study
bias dispersion in the general plug-in estimator. In particular, we
show how study-specific reweighting achieves alignment with the A-side
pilot mean through two channels, and why increasing bias dispersion
reduces the systematic-bias contribution while increasing the influence
of A-side pilot randomness on the final estimator.

Based on the \(M\)-asymptotic behavior of the plug-in weights established
in the proof of Corollary~\ref{cor:simplified}, the shrinkage rule
asymptotically places all weight on the B side and makes the weighted
B-side pilot mean aligned with the A-side pilot mean, that is:
\[
    \sum_{i=1}^M \widehat w_i
    \xrightarrow[M\to\infty]{p}1,
    \qquad
    \sum_{i=1}^M
    \widehat w_i\bar B_{i,\mathrm{pi}}
    \xrightarrow[M\to\infty]{p}
    \bar A_{\mathrm{pi}}.
\]

To see how the weighting rule achieves this required alignment and how bias
dispersion enters the mechanism, write $\eta_i:=\bar B_{i,\mathrm{pi}}-(\theta+\epsilon_i)$
for the auxiliary pilot error, then decompose \(\widehat\epsilon_i\) and rewrite
the second convergence above as
\begin{equation}\label{eq:pilot-align}
    \underbrace{
    \sum_{i=1}^M\widehat w_i\epsilon_i
    }_{\text{true weighted auxiliary bias}}
    +
    \underbrace{
    \sum_{i=1}^M\widehat w_i\eta_i
    }_{\text{B-side pilot randomness}}
    =
    \bar A_{\mathrm{pi}}-\theta
    +o_p(1).
\end{equation}
Hence, the required pilot alignment can be achieved through two channels: by reallocating the weights
\((\widehat w_1,\widehat w_2,\ldots,\widehat w_M)\) across auxiliary studies,
the rule can change the true weighted auxiliary bias
\(\sum_i\widehat w_i\epsilon_i\), while it can also exploit realized B-side
pilot randomness through \(\sum_i\widehat w_i\eta_i\).

The distinction between these two channels is important because only the
first persists directly in the final estimation error. From the large-\(M\)
behavior of the plug-in estimator, the weight on
\(\bar A_{\mathrm{te}}\) vanishes and the B-side testing-sample randomness
averages out, so
\[
    \widehat\theta_M-\theta
    =
    \sum_{i=1}^M\widehat w_i\epsilon_i
    +o_p(1).
\]
The second term in \eqref{eq:pilot-align},
\(\sum_i\widehat w_i\eta_i\), is generated by B-side pilot-sample
randomness. Because the final estimator in
\eqref{eq:feasible-thetaM} is constructed from independent testing sample means,
this realized pilot noise \(\eta_i\)
does not reappear directly in the final
estimation error. Hence, the effect of bias dispersion on the final MSE depends on how it
changes the way the alignment in \eqref{eq:pilot-align} is achieved:
whether the weighting rule relies more on adjusting the first term, the true weighted
auxiliary bias, which persists in the final estimator error, or on the second term, the B-side pilot
randomness, which does not.

When \(\delta=0\), all auxiliary studies have the same true bias
\(\epsilon_i=e\) for all $i$. Since the total auxiliary weight converges to one,
any reallocation of weight across auxiliary studies leaves
\(
    \sum_{i=1}^M\widehat w_i\epsilon_i
    \xrightarrow[M\to\infty]{p} e.
\)
There is no genuine cross-study bias variation for the
study-specific weights to exploit. Therefore, the limiting estimation
error remains \(e\), yielding \(\MSE_\infty=e^2\), while the required
alignment with \(\bar A_{\mathrm{pi}}\) is achieved entirely through
the B-side pilot randomness \(\sum_i\widehat w_i\eta_i\).

When \(\delta>0\), by contrast, the true biases differ across auxiliary
studies. Study-specific reweighting can then change
\(\sum_i\widehat w_i\epsilon_i\) to be shifted from $e$, allowing part of the required pilot
alignment to be achieved by exploiting genuine cross-study bias
differences rather than only B-side pilot noise. This additional flexibility to shift the true weighted auxiliary bias \(\sum_i\widehat w_i\epsilon_i\)   away
from the systematic average bias \(e\) explains why initially increasing
\(\delta\) from zero can reduce MSE.

This flexibility, however, eventually becomes costly. As \(\delta\)
grows large relative to \(\widetilde v'_B\), a larger share of the pilot
alignment to \(\bar A_{\mathrm{pi}}-\theta\) is achieved by changing the true weighted auxiliary bias. In
particular, the true weighted auxiliary bias increasingly tracks the
alignment target \(\bar A_{\mathrm{pi}}-\theta\), hence the final estimator increasingly inherits randomness from the
A-side pilot sample. Accordingly, \eqref{eq:MSE-plug-in} implies
\(
\MSE_\infty(\delta)
\xrightarrow[\delta\to\infty]{}
V_{A,\mathrm{pi}}.\)

\section{Proof of Proposition \ref{prop:oracle_mse_monotone}}

Let
\[
    Q_M
    =
    \veps\veps^\top+\Sigma+V_A\mathbf 1_M\mathbf 1_M^\top,
    \qquad
    \veps=(\epsilon_1,\ldots,\epsilon_M)^\top,
    \qquad
    \Sigma=\diag(V_{B,1},\ldots,V_{B,M}).
\]
After adding the \((M+1)\)-st auxiliary estimator, define
\[
    \widetilde \veps
    =
    \begin{pmatrix}
        \veps\\
        \epsilon_{M+1}
    \end{pmatrix},
    \qquad
    \widetilde \Sigma
    =
    \diag(V_{B,1},\ldots,V_{B,M},V_{B,M+1}),
    \qquad
    \mathbf 1_{M+1}
    =
    \begin{pmatrix}
        \mathbf 1_M\\
        1
    \end{pmatrix}.
\]
Then
\[
    Q_{M+1}
    =
    \widetilde \veps\widetilde \veps^\top
    +
    \widetilde \Sigma
    +
    V_A\mathbf 1_{M+1}\mathbf 1_{M+1}^\top .
\]
Writing \(Q_{M+1}\) in block form gives
\[
    Q_{M+1}
    =
    \begin{pmatrix}
        Q_M & \mathbf b\\
        \mathbf b^\top & d
    \end{pmatrix},
\]
where
\[
    \mathbf b
    =
    \epsilon_{M+1}\veps+V_A\mathbf 1_M,
    \qquad
    d
    =
    \epsilon_{M+1}^2+V_{B,M+1}+V_A.
\]
Since \(Q_{M+1}\) is positive definite, the Schur complement
\[
    S
    :=
    d-\mathbf b^\top Q_M^{-1}\mathbf b
\]
is strictly positive.

By the block inverse formula,
\[
Q_{M+1}^{-1}
=
\begin{pmatrix}
Q_M^{-1}+Q_M^{-1}\mathbf bS^{-1}\mathbf b^\top Q_M^{-1}
&
-Q_M^{-1}\mathbf bS^{-1}\\[4pt]
-S^{-1}\mathbf b^\top Q_M^{-1}
&
S^{-1}
\end{pmatrix}.
\]
Therefore,
\[
\begin{aligned}
\mathbf 1_{M+1}^\top Q_{M+1}^{-1}\mathbf 1_{M+1}
&=
\mathbf 1_M^\top Q_M^{-1}\mathbf 1_M
+
S^{-1}
\left[
\left(\mathbf b^\top Q_M^{-1}\mathbf 1_M\right)^2
-
2\mathbf b^\top Q_M^{-1}\mathbf 1_M
+
1
\right]  \\
&=
\mathbf 1_M^\top Q_M^{-1}\mathbf 1_M
+
\frac{
\left(1-\mathbf b^\top Q_M^{-1}\mathbf 1_M\right)^2
}{
S
}.
\end{aligned}
\]
Using
\[
    \MSE_M^*
    =
    V_A
    -
    V_A^2\mathbf 1_M^\top Q_M^{-1}\mathbf 1_M
\]
and the analogous expression for \(\MSE_{M+1}^*\), we obtain
\[
\begin{aligned}
\MSE_M^*-\MSE_{M+1}^*
&=
V_A^2
\left(
\mathbf 1_{M+1}^\top Q_{M+1}^{-1}\mathbf 1_{M+1}
-
\mathbf 1_M^\top Q_M^{-1}\mathbf 1_M
\right)\\
&=
V_A^2
\frac{
\left(1-\mathbf b^\top Q_M^{-1}\mathbf 1_M\right)^2
}{
S
}\\
&=
V_A^2
\frac{
\left[
1-
(\epsilon_{M+1}\veps+V_A\mathbf 1_M)^\top
Q_M^{-1}\mathbf 1_M
\right]^2
}{
\epsilon_{M+1}^2+V_{B,M+1}+V_A
-
(\epsilon_{M+1}\veps+V_A\mathbf 1_M)^\top
Q_M^{-1}
(\epsilon_{M+1}\veps+V_A\mathbf 1_M)
}.
\end{aligned}
\]
Because \(S>0\), this proves
\[
    \MSE_M^*-\MSE_{M+1}^*\ge 0.
\]

It remains to characterize when equality holds. From the gain expression above, equality holds
if and only if
\[
    1-\mathbf b^\top Q_M^{-1}\mathbf 1_M=0.
\]
The oracle weight on the newly added auxiliary estimator is the last component of
\[
    V_A Q_{M+1}^{-1}\mathbf 1_{M+1}.
\]
Using the block inverse formula, this last component is
\[
    V_A S^{-1}
    \left(
        1-\mathbf b^\top Q_M^{-1}\mathbf 1_M
    \right).
\]
Thus equality holds if and only if the \((M+1)\)-st auxiliary estimator receives zero oracle
weight.

We now show that this zero-weight condition is equivalent to the stated bias condition. Define
the temporary scalar quantities
\[
    a_0:=\mathbf 1_M^\top\Sigma^{-1}\mathbf 1_M,
    \qquad
    a_1:=\mathbf 1_M^\top\Sigma^{-1}\veps
    =
    \sum_{i=1}^M\frac{\epsilon_i}{V_{B,i}},
    \qquad
    a_2:=\veps^\top\Sigma^{-1}\veps
    =
    \sum_{i=1}^M\frac{\epsilon_i^2}{V_{B,i}}.
\]
Let
\[
    \mathbf x:=Q_M^{-1}\mathbf 1_M.
\]
Then
\[
    Q_M\mathbf x=\mathbf 1_M,
\]
or equivalently,
\[
    \Sigma\mathbf x
    +
    \veps(\veps^\top\mathbf x)
    +
    V_A\mathbf 1_M(\mathbf 1_M^\top\mathbf x)
    =
    \mathbf 1_M.
\]
Set
\[
    \alpha:=\veps^\top\mathbf x,
    \qquad
    \beta:=\mathbf 1_M^\top\mathbf x.
\]
Multiplying the display above by \(\veps^\top\Sigma^{-1}\) and
\(\mathbf 1_M^\top\Sigma^{-1}\), respectively, gives the two equations
\[
    (1+a_2)\alpha+V_Aa_1\beta=a_1,
\]
and
\[
    a_1\alpha+(1+V_Aa_0)\beta=a_0.
\]
Solving this two-equation system yields
\[
    \alpha
    =
    \frac{a_1}
    {1+a_2+V_A\{a_0(1+a_2)-a_1^2\}},
\]
and
\[
    \beta
    =
    \frac{a_0(1+a_2)-a_1^2}
    {1+a_2+V_A\{a_0(1+a_2)-a_1^2\}}.
\]
Therefore,
\[
\begin{aligned}
\mathbf b^\top Q_M^{-1}\mathbf 1_M
&=
(\epsilon_{M+1}\veps+V_A\mathbf 1_M)^\top \mathbf x\\
&=
\epsilon_{M+1}\alpha+V_A\beta\\
&=
\frac{
    \epsilon_{M+1}a_1
    +
    V_A\{a_0(1+a_2)-a_1^2\}
}{
    1+a_2+V_A\{a_0(1+a_2)-a_1^2\}
}.
\end{aligned}
\]
Hence
\[
    \mathbf b^\top Q_M^{-1}\mathbf 1_M=1
\]
if and only if
\[
    \epsilon_{M+1}a_1
    +
    V_A\{a_0(1+a_2)-a_1^2\}
    =
    1+a_2+V_A\{a_0(1+a_2)-a_1^2\}.
\]
The \(V_A\)-terms cancel, so the condition reduces to
\[
    \epsilon_{M+1}a_1=1+a_2.
\]
Substituting back the definitions of \(a_1\) and \(a_2\), equality holds if and only if
\[
    \epsilon_{M+1}
    \sum_{i=1}^M\frac{\epsilon_i}{V_{B,i}}
    =
    1+\sum_{i=1}^M\frac{\epsilon_i^2}{V_{B,i}}.
\]
This proves the proposition.

\newpage
\section{Proof of Theorem~\ref{thm:sufficientimprove}}\label{app:cubic}

We prove the sufficient cubic condition for the univariate practical shrinkage estimator in Theorem~\ref{thm:sufficientimprove}.

Recall that
\[
\hat\theta(\hat w)=(1-\hat w)\bar A_{\mathrm{te}}+\hat w \bar B_{\mathrm{te}},
\qquad
\hat w
=
\frac{\widehat V_{A,\mathrm{te}}}
{\hat\epsilon^2+\widehat V_{A,\mathrm{te}}+\widehat V_{B,\mathrm{te}}},
\qquad
\hat\epsilon=\bar B_{\mathrm{pi}}-\bar A_{\mathrm{pi}}.
\]
By the independence of the pilot and test samples, the MSE decomposition in the main text is
\[
\MSE\bigl(\hat\theta(\hat w)\bigr)
=
V_{A,\mathrm{te}}
-
2V_{A,\mathrm{te}}\mu_1
+
\left(t+\frac{\sigma_B^2}{n_{B,\mathrm{te}}}\right)\mu_2,
\]
where
\[
\mu_1=\E(\hat w),\qquad
\mu_2=\E(\hat w^2),\qquad
t:=\epsilon^2+V_{A,\mathrm{te}}
=
\epsilon^2+\frac{\sigma_A^2}{n_{A,\mathrm{te}}}.
\]
Therefore, a sufficient condition for improvement over the unbiased benchmark \(\bar A\) is
\[
V_{A,\mathrm{te}}
-
2V_{A,\mathrm{te}}\mu_1
+
\left(t+\frac{\sigma_B^2}{n_{B,\mathrm{te}}}\right)\mu_2
<
V_A.
\]
We now replace \(\mu_1\) and \(\mu_2\) by the first-order delta approximation and an upper bound.

First, the first-order approximation gives
\[
\mu_1
\approx
\frac{V_{A,\mathrm{te}}}{t+\sigma_B^2/n_{B,\mathrm{te}}}.
\]
Second, for \(\mu_2\),
\[
\mu_2
=
\E\!\left[
\frac{\widehat V_{A,\mathrm{te}}^{\,2}}
{\left(\hat\epsilon^2+\widehat V_{A,\mathrm{te}}+\widehat V_{B,\mathrm{te}}\right)^2}
\right].
\]
Since \(\hat\epsilon^2\ge 0\) and \(\widehat V_{A,\mathrm{te}}\ge 0\), we have
\[
\mu_2
\le
\E\!\left[
\frac{\widehat V_{A,\mathrm{te}}^{\,2}}
{\widehat V_{B,\mathrm{te}}^{\,2}}
\right].
\]
Because the A-side and B-side pilot samples are independent under the sample-splitting scheme,
\[
\mu_2
\le
\E\!\left(\widehat V_{A,\mathrm{te}}^{\,2}\right)
\E\!\left(\frac{1}{\widehat V_{B,\mathrm{te}}^{\,2}}\right).
\]
We now evaluate the two factors on the right-hand side.

For the A-side term,
\[
\widehat V_{A,\mathrm{te}}=\frac{S_{A,\mathrm{pi}}^2}{n_{A,\mathrm{te}}},
\]
so under Assumption~\ref{ass:working},
\[
\E\!\left(\widehat V_{A,\mathrm{te}}^{\,2}\right)
=
\left(\frac{\sigma_A^2}{n_{A,\mathrm{te}}}\right)^2
\left[
1+\frac{1}{n_{A,\mathrm{pi}}}
\left(
\kappa_A-\frac{n_{A,\mathrm{pi}}-3}{n_{A,\mathrm{pi}}-1}
\right)
\right].
\]
Define
\[
C_A
:=
1+\frac{1}{n_{A,\mathrm{pi}}}
\left(
\kappa_A-\frac{n_{A,\mathrm{pi}}-3}{n_{A,\mathrm{pi}}-1}
\right),
\]
so that
\[
\E\!\left(\widehat V_{A,\mathrm{te}}^{\,2}\right)
=
\left(\frac{\sigma_A^2}{n_{A,\mathrm{te}}}\right)^2 C_A.
\]
For the B-side term,
\[
\widehat V_{B,\mathrm{te}}=\frac{S_{B,\mathrm{pi}}^2}{n_{B,\mathrm{te}}}.
\]
Under the moment-matched scaled chi-squared approximation in Assumption~\ref{ass:working},
\[
\frac{S_{B,\mathrm{pi}}^2}{\sigma_B^2}
\approx
\frac{\chi^2_{\nu_{B,\mathrm{pi}} }}{\nu_{B,\mathrm{pi}} },
\qquad
\nu_{B,\mathrm{pi}} 
=
\frac{2n_{B,\mathrm{pi}}}
{\kappa_B-\frac{n_{B,\mathrm{pi}}-3}{n_{B,\mathrm{pi}}-1}}.
\]
Hence, provided \(\nu_{B,\mathrm{pi}} >4\),
\[
\E\!\left(\frac{1}{\widehat V_{B,\mathrm{te}}^{\,2}}\right)
\approx
\left(\frac{n_{B,\mathrm{te}}}{\sigma_B^2}\right)^2
\frac{\nu_{B,\mathrm{pi}} ^2}{(\nu_{B,\mathrm{pi}} -2)(\nu_{B,\mathrm{pi}} -4)}.
\]
To obtain the low-degree polynomial condition used in the main text, we further impose the large-\(n_{B,\mathrm{pi}}\) approximation
\[
\nu_{B,\mathrm{pi}} \to\infty,
\]
which, under fixed kurtosis \(\kappa_B\), is equivalent to \(n_{B,\mathrm{pi}}\to\infty\). Since
\[
\frac{\nu_{B,\mathrm{pi}} ^2}{(\nu_{B,\mathrm{pi}} -2)(\nu_{B,\mathrm{pi}} -4)}
\to 1
\qquad\text{as }\nu_{B,\mathrm{pi}} \to\infty,
\]
we use the approximation
\[
\frac{\nu_{B,\mathrm{pi}} ^2}{(\nu_{B,\mathrm{pi}} -2)(\nu_{B,\mathrm{pi}} -4)}
\approx 1,
\]
which yields the tractable upper bound
\[
\E\!\left(\frac{1}{\widehat V_{B,\mathrm{te}}^{\,2}}\right)
\approx
\left(\frac{n_{B,\mathrm{te}}}{\sigma_B^2}\right)^2.
\]
Therefore,
\[
\mu_2
\lesssim
\left(\frac{n_{B,\mathrm{te}}}{n_{A,\mathrm{te}}}\right)^2
C_A
\frac{\sigma_A^4}{\sigma_B^4}.
\]
Substituting the approximation for \(\mu_1\) and the upper bound for \(\mu_2\) into the sufficient improvement condition gives
\[
V_{A,\mathrm{te}}
-
\frac{2V_{A,\mathrm{te}}^2}{t+\sigma_B^2/n_{B,\mathrm{te}}}
+
\left(t+\frac{\sigma_B^2}{n_{B,\mathrm{te}}}\right)
\left(\frac{n_{B,\mathrm{te}}}{n_{A,\mathrm{te}}}\right)^2
C_A
\frac{\sigma_A^4}{\sigma_B^4}
<
V_A.
\]
Using
\[
V_{A,\mathrm{te}}=\frac{\sigma_A^2}{n_{A,\mathrm{te}}},
\qquad
V_A=\frac{\sigma_A^2}{N_A},
\]
and multiplying both sides by the positive quantity
\[
\frac{n_{A,\mathrm{te}}^2\sigma_B^4}{C_A\sigma_A^4}
\left(t+\frac{\sigma_B^2}{n_{B,\mathrm{te}}}\right)n_{B,\mathrm{te}},
\]
we obtain the sufficient cubic condition
\[
P(n_{B,\mathrm{te}})<0,
\]
where
\[
P(n_{B,\mathrm{te}})
=
t^2 n_{B,\mathrm{te}}^3
+
2\sigma_B^2 t\, n_{B,\mathrm{te}}^2
+
\alpha^\dagger n_{B,\mathrm{te}}
+
\beta^\dagger,
\]
with
\[
\alpha^\dagger
=
\sigma_B^4
\left(
1-\frac{2}{C_A}
\right)
+
\sigma_B^2 t
\left[
\frac{n_{A,\mathrm{te}}}{C_A}
\left(
1-\frac{n_{A,\mathrm{te}}}{N_A}
\right)
\frac{\sigma_B^2}{\sigma_A^2}
\right],
\]
and
\[
\beta^\dagger
=
\sigma_B^4
\left[
\frac{n_{A,\mathrm{te}}}{C_A}
\left(
1-\frac{n_{A,\mathrm{te}}}{N_A}
\right)
\frac{\sigma_B^2}{\sigma_A^2}
\right].
\]
It remains to justify conditions (C1)–(C3). Differentiate \(P\) with respect to \(n_{B,\mathrm{te}}\):
\[
P'(n_{B,\mathrm{te}})
=
3t^2 n_{B,\mathrm{te}}^2
+
4\sigma_B^2 t\, n_{B,\mathrm{te}}
+
\alpha^\dagger,
\]
and
\[
P''(n_{B,\mathrm{te}})
=
6t^2 n_{B,\mathrm{te}}+4\sigma_B^2 t.
\]
Since \(t>0\), we have \(P''(n_{B,\mathrm{te}})>0\) for every \(n_{B,\mathrm{te}}>0\). Thus, any critical point of \(P\) on the positive real line is a strict local minimum. Solving \(P'(n_{B,\mathrm{te}})=0\) gives
\[
n_{B,\mathrm{te}}
=
\frac{-2\sigma_B^2\pm\sqrt{4\sigma_B^4-3\alpha^\dagger}}{3t}.
\]
Hence, whenever the discriminant is nonnegative, the only candidate local minimizer on the positive real line is
\[
n_{B,\mathrm{te}}^\circ
=
\frac{\sqrt{4\sigma_B^4-3\alpha^\dagger}-2\sigma_B^2}{3t}.
\]

Condition (C1) requires \(n_{B,\mathrm{te}}^\circ>0\), so that the cubic has an interior local minimum on the positive axis. Under (C1), the point \(n_{B,\mathrm{te}}^\circ\) is the unique local minimum of \(P\) on the positive real line. If, in addition,
\[
P(n_{B,\mathrm{te}}^\circ)<0,
\]
then by continuity \(P(n_{B,\mathrm{te}})<0\) on a nonempty interval around \(n_{B,\mathrm{te}}^\circ\), and the sufficient condition for improvement holds on that interval. This proves (C2).

Finally, when the full auxiliary sample size is finite, the admissible test-sample sizes must satisfy
\[
2\le n_{B,\mathrm{te}}\le N_B-2.
\]
Thus, even if (C1) and (C2) hold, the sufficient condition can only be realized if
\[
2\le n_{B,\mathrm{te}}^\circ\le N_B-2,
\]
which is condition (C3).

\section{(read through for final check)Proof of Theorem~\ref{thm:MSE_infty1}}

Let \(\mathcal T_M\) denote the \(\sigma\)-field generated by all
pilot-sample quantities. 
By the pilot/testing sample split,
\(\widehat{\boldsymbol w}\) is \(\mathcal T_M\)-measurable and is
independent of the testing-sample means conditional on
\(\mathcal T_M\). Therefore, the MSE admits the conditional
second-moment decomposition
\[
\begin{aligned}
\operatorname{MSE}_M
=\mathbb E\!\left[
    \operatorname{Var}
    \left(
        \widehat\theta_M
        \mid
        \mathcal T_M
    \right)
\right]
+
\mathbb E\!\left[
    \left\{
        \mathbb E
        \left(
            \widehat\theta_M-\theta
            \mid
            \mathcal T_M
        \right)
    \right\}^{2}
\right].
\end{aligned}
\]
Accordingly, the proof reduces to identifying the large-\(M\)
behavior of the conditional variance and conditional mean. We will
show that
\[
\operatorname{Var}
\left(
    \widehat\theta_M
    \mid
    \mathcal T_M
\right)
\xrightarrow[M\to\infty]{p}0,
\]
while
\[
\mathbb E
\left(
    \widehat\theta_M-\theta
    \mid
    \mathcal T_M
\right)
\xrightarrow[M\to\infty]{p}
\frac{
    \widetilde\mu_1\widetilde\nu
    +
    \zeta\widetilde A
}{
    \widetilde\Delta
}.
\]
Under the uniform integrability condition in Assumption~\ref{ass:largeM}\textup{(A3)}, these probability convergences can then be upgraded to convergence of the corresponding expectations, yielding the limiting MSE.

Recall that
\[
    \widehat\theta_M
    =
    \left(
        1-\mathbf 1^\top\widehat w
    \right)\bar A_{\mathrm{te}}
    +
    \sum_{i=1}^M
    \widehat w_i\bar B_{i,\mathrm{te}},
    \qquad
    \widehat w
    =
    \widehat V_{A,\mathrm{te}}
    \widehat Q^{-1}\mathbf 1,
\]
where
\[
    \widehat Q
    =
    \widehat\epsilon\widehat\epsilon^\top
    +
    \widehat\Sigma_{\mathrm{te}}
    +
    \widehat V_{A,\mathrm{te}}
    \mathbf 1\mathbf 1^\top.
\]
Since \(\widehat\Sigma_{\mathrm{te}}\) is diagonal and positive definite,
solving
\[
    \widehat Q\widehat w
    =
    \widehat V_{A,\mathrm{te}}\mathbf 1
\]
gives, for each \(i=1,\dots,M\),
\[
    \widehat w_i
    =
    \frac{\widehat V_{A,\mathrm{te}}}{D_M}
    \left[
        \left(
            1+
            \sum_{j=1}^M
            \frac{\widehat\epsilon_j^2}
            {\widehat V_{B,j,\mathrm{te}}}
        \right)
        \frac{1}{\widehat V_{B,i,\mathrm{te}}}
        -
        \left(
            \sum_{j=1}^M
            \frac{\widehat\epsilon_j}
            {\widehat V_{B,j,\mathrm{te}}}
        \right)
        \frac{\widehat\epsilon_i}
        {\widehat V_{B,i,\mathrm{te}}}
    \right],
\]
where
\[
\begin{aligned}
    D_M
    &:=
    \left(
        1+
        \sum_{j=1}^M
        \frac{\widehat\epsilon_j^2}
        {\widehat V_{B,j,\mathrm{te}}}
    \right)
    \left(
        1+
        \widehat V_{A,\mathrm{te}}
        \sum_{j=1}^M
        \frac{1}{\widehat V_{B,j,\mathrm{te}}}
    \right)
    -
    \widehat V_{A,\mathrm{te}}
    \left(
        \sum_{j=1}^M
        \frac{\widehat\epsilon_j}
        {\widehat V_{B,j,\mathrm{te}}}
    \right)^2.
\end{aligned}
\]

Define
\[
    \zeta
    :=
    \bar A_{\mathrm{pi}}-\theta,
    \qquad
    \eta_i
    :=
    \bar B_{i,\mathrm{pi}}-(\theta+\epsilon_i).
\]
These quantities are the random parts of the pilot-sample means after
subtracting their population means. Then
\[
    \widehat\epsilon_i
    =
    \epsilon_i+\eta_i-\zeta.
\]

Under Assumption~\ref{ass:working}, we adopt the working approximation
\[
    \zeta\sim N(0,V_{A,\mathrm{pi}}),
    \qquad
    \eta_i\sim N(0,V_{B,i,\mathrm{pi}}).
\]
The auxiliary-study collections are independent across \(i\), and
\(\zeta\) is independent of all \(B\)-side quantities. Each
\(\eta_i\) is treated as independent of
\(\widehat V_{B,i,\mathrm{te}}\). Although
\(\widehat V_{B,i,\mathrm{te}}\) estimates the variance of the
testing-sample mean, it is constructed from the pilot-sample variance
estimator:
\[
    \widehat V_{B,i,\mathrm{te}}
    =
    \frac{S_{B_i,\mathrm{pi}}^2}{n_{B,i,\mathrm{te}}}.
\]
Thus, the independence used here is the working independence between
the pilot-sample mean and the pilot-sample variance estimator
\(S_{B_i,\mathrm{pi}}^2\).

We first establish the required weighted-noise limits. By
independence, the cross terms vanish and
\[
\begin{aligned}
    &\mathbb E\left[
        \left(
            \frac1M
            \sum_{i=1}^M
            \frac{\eta_i}
            {\widehat V_{B,i,\mathrm{te}}}
        \right)^2
    \right]
    =
    \frac1{M^2}
    \sum_{i=1}^M
    V_{B,i,\mathrm{pi}}
    \mathbb E\left[
        \widehat V_{B,i,\mathrm{te}}^{-2}
    \right]
    =
    O(M^{-1}).
\end{aligned}
\]
Therefore,
\[
    \frac1M
    \sum_{i=1}^M
    \frac{\eta_i}
    {\widehat V_{B,i,\mathrm{te}}}
    \xrightarrow[M\to\infty]{}0
    \qquad\text{in }L^2,
\]
and hence in probability.

Similarly,
\[
\begin{aligned}
    &\mathbb E\left[
        \left(
            \frac1M
            \sum_{i=1}^M
            \frac{\epsilon_i\eta_i}
            {\widehat V_{B,i,\mathrm{te}}}
        \right)^2
    \right]
    =
    \frac1{M^2}
    \sum_{i=1}^M
    \epsilon_i^2V_{B,i,\mathrm{pi}}
    \mathbb E\left[
        \widehat V_{B,i,\mathrm{te}}^{-2}
    \right].
\end{aligned}
\]
By Assumption~\ref{ass:largeM}\textup{(A1)},
\[
    \sup_{i\geq1}|\epsilon_i|<\infty,
\]
and therefore
\[
    \frac1M\sum_{i=1}^M\epsilon_i^2
    \leq
    \sup_{i\geq1}|\epsilon_i|^2
    =
    O(1).
\]
Together with the uniform bounds on \(V_{B,i,\mathrm{pi}}\) and
\(\mathbb E[\widehat V_{B,i,\mathrm{te}}^{-2}]\), this gives
\[
    \mathbb E\left[
        \left(
            \frac1M
            \sum_{i=1}^M
            \frac{\epsilon_i\eta_i}
            {\widehat V_{B,i,\mathrm{te}}}
        \right)^2
    \right]
    =
    O(M^{-1}),
\]
and consequently
\[
    \frac1M
    \sum_{i=1}^M
    \frac{\epsilon_i\eta_i}
    {\widehat V_{B,i,\mathrm{te}}}
    \xrightarrow[M\to\infty]{p}0.
\]

Under the Gaussian working approximation,
\[
    \operatorname{Var}(\eta_i^2)
    =
    2V_{B,i,\mathrm{pi}}^2.
\]
Therefore,
\[
\begin{aligned}
    &\mathbb E\left[
        \left(
            \frac1M
            \sum_{i=1}^M
            \frac{\eta_i^2-V_{B,i,\mathrm{pi}}}
            {\widehat V_{B,i,\mathrm{te}}}
        \right)^2
    \right]
    =
    \frac2{M^2}
    \sum_{i=1}^M
    V_{B,i,\mathrm{pi}}^2
    \mathbb E\left[
        \widehat V_{B,i,\mathrm{te}}^{-2}
    \right]
    =
    O(M^{-1}),
\end{aligned}
\]
which gives
\[
    \frac1M
    \sum_{i=1}^M
    \frac{\eta_i^2-V_{B,i,\mathrm{pi}}}
    {\widehat V_{B,i,\mathrm{te}}}
    \xrightarrow[M\to\infty]{p}0.
\]

Combining these limits with
Assumption~\ref{ass:largeM}\textup{(A2)}, we obtain
\[
    \frac1M
    \sum_{i=1}^M
    \frac1{\widehat V_{B,i,\mathrm{te}}}
    \xrightarrow[M\to\infty]{p}
    \widetilde\mu_0,
\]
\[
\begin{aligned}
    \frac1M
    \sum_{i=1}^M
    \frac{\widehat\epsilon_i}
    {\widehat V_{B,i,\mathrm{te}}}
    &=
    \frac1M
    \sum_{i=1}^M
    \frac{\epsilon_i}
    {\widehat V_{B,i,\mathrm{te}}}
    +
    \frac1M
    \sum_{i=1}^M
    \frac{\eta_i}
    {\widehat V_{B,i,\mathrm{te}}}
    -
    \zeta
    \frac1M
    \sum_{i=1}^M
    \frac1{\widehat V_{B,i,\mathrm{te}}}
    \xrightarrow[M\to\infty]{p}
    \widetilde\mu_1
    -
    \zeta\widetilde\mu_0,
\end{aligned}
\]
and
\[
\begin{aligned}
    \frac1M
    \sum_{i=1}^M
    \frac{\epsilon_i\widehat\epsilon_i}
    {\widehat V_{B,i,\mathrm{te}}}
    &=
    \frac1M
    \sum_{i=1}^M
    \frac{\epsilon_i^2}
    {\widehat V_{B,i,\mathrm{te}}}
    +
    \frac1M
    \sum_{i=1}^M
    \frac{\epsilon_i\eta_i}
    {\widehat V_{B,i,\mathrm{te}}}
    -
    \zeta
    \frac1M
    \sum_{i=1}^M
    \frac{\epsilon_i}
    {\widehat V_{B,i,\mathrm{te}}}
    \xrightarrow[M\to\infty]{p}
    \widetilde\mu_2
    -
    \zeta\widetilde\mu_1.
\end{aligned}
\]

Expanding
\(\widehat\epsilon_i^2=(\epsilon_i+\eta_i-\zeta)^2\) gives
\[
\begin{aligned}
    &\frac1M
    \sum_{i=1}^M
    \frac{\widehat\epsilon_i^2}
    {\widehat V_{B,i,\mathrm{te}}}
=
    \frac1M
    \sum_{i=1}^M
    \frac{\epsilon_i^2}
    {\widehat V_{B,i,\mathrm{te}}}
    +
    \frac1M
    \sum_{i=1}^M
    \frac{V_{B,i,\mathrm{pi}}}
    {\widehat V_{B,i,\mathrm{te}}}
    -
    2\zeta
    \frac1M
    \sum_{i=1}^M
    \frac{\epsilon_i}
    {\widehat V_{B,i,\mathrm{te}}}
    +
    \zeta^2
    \frac1M
    \sum_{i=1}^M
    \frac1{\widehat V_{B,i,\mathrm{te}}}
    +
    o_p(1)
    \\
    &\xrightarrow[M\to\infty]{p}
    \widetilde\mu_2
    +
    \widetilde\nu
    -
    2\zeta\widetilde\mu_1
    +
    \zeta^2\widetilde\mu_0.
\end{aligned}
\]

We next note explicitly that \(\widetilde\mu_0>0\). Under the
scaled chi-squared approximation in
Assumption~\ref{ass:working},
\[
    \mathbb E\left[
        \widehat V_{B,i,\mathrm{te}}^{-1}
    \right]
    =
    \frac{1}{V_{B,i,\mathrm{te}}}
    \frac{\nu_{B,i,\mathrm{pi}}}
    {\nu_{B,i,\mathrm{pi}}-2}
    \geq
    \frac{1}{\overline v}.
\]
Moreover, independence across \(i\) and the uniform inverse-second-
moment bound imply
\[
    \frac1M
    \sum_{i=1}^M
    \left\{
        \widehat V_{B,i,\mathrm{te}}^{-1}
        -
        \mathbb E[
            \widehat V_{B,i,\mathrm{te}}^{-1}
        ]
    \right\}
    \xrightarrow[M\to\infty]{}0
    \qquad\text{in }L^2.
\]
Combining this with the first limit in
Assumption~\ref{ass:largeM}\textup{(A2)} gives
\[
    \widetilde\mu_0
    =
    \lim_{M\to\infty}
    \frac1M
    \sum_{i=1}^M
    \mathbb E\left[
        \widehat V_{B,i,\mathrm{te}}^{-1}
    \right]
    \geq
    \frac1{\overline v}
    >0.
\]

Define
\[
    \widetilde A
    :=
    \widetilde\mu_0\widetilde\mu_2
    -
    \widetilde\mu_1^2,
    \qquad
    \widetilde\Delta
    :=
    \widetilde A
    +
    \widetilde\mu_0\widetilde\nu.
\]
For every \(M\), weighted Cauchy--Schwarz gives
\[
    \left(
        \frac1M
        \sum_{i=1}^M
        \frac{\epsilon_i}
        {\widehat V_{B,i,\mathrm{te}}}
    \right)^2
    \leq
    \left(
        \frac1M
        \sum_{i=1}^M
        \frac1{\widehat V_{B,i,\mathrm{te}}}
    \right)
    \left(
        \frac1M
        \sum_{i=1}^M
        \frac{\epsilon_i^2}
        {\widehat V_{B,i,\mathrm{te}}}
    \right).
\]
Passing to the limit yields
\[
    \widetilde A\geq0.
\]
Moreover,
\[
    \frac1M
    \sum_{i=1}^M
    \frac{V_{B,i,\mathrm{pi}}}
    {\widehat V_{B,i,\mathrm{te}}}
    \geq
    \underline v
    \frac1M
    \sum_{i=1}^M
    \frac1{\widehat V_{B,i,\mathrm{te}}},
\]
so
\[
    \widetilde\nu
    \geq
    \underline v\widetilde\mu_0
    >0.
\]
Hence
\[
    \widetilde\Delta>0.
\]
Using the limits above,
\[
    \frac{D_M}
    {\widehat V_{A,\mathrm{te}}M^2}
    \xrightarrow[M\to\infty]{p}
    \widetilde\Delta,
\]
and\[
    \frac{\widehat V_{A,\mathrm{te}}}{D_M}
    =
    O_p(M^{-2}).
\]

We next show that the conditional variance vanishes. Define
\[
    T_M:=\mathbf 1^\top\widehat w.
\]
Since
\[
    \widehat Q\widehat w
    =
    \widehat V_{A,\mathrm{te}}\mathbf 1,
\]
premultiplying by \(\widehat w^\top\) gives
\[
    (\widehat w^\top\widehat\epsilon)^2
    +
    \widehat w^\top
    \widehat\Sigma_{\mathrm{te}}
    \widehat w
    +
    \widehat V_{A,\mathrm{te}}T_M^2
    =
    \widehat V_{A,\mathrm{te}}T_M.
\]
Hence
\[
    (\widehat w^\top\widehat\epsilon)^2
    +
    \widehat w^\top
    \widehat\Sigma_{\mathrm{te}}
    \widehat w
    =
    \widehat V_{A,\mathrm{te}}T_M(1-T_M).
\]
The left-hand side is nonnegative, so \(0\le T_M\le1\). Therefore,
\[
    \widehat w^\top
    \widehat\Sigma_{\mathrm{te}}
    \widehat w
    \le
    \widehat V_{A,\mathrm{te}}T_M(1-T_M)
    \le
    \widehat V_{A,\mathrm{te}}(1-T_M).
\]
From the explicit total-weight expression,
\[
    1-T_M
    =
    \frac{
        1+
        \displaystyle\sum_{i=1}^M
        \widehat\epsilon_i^2/
        \widehat V_{B,i,\mathrm{te}}
    }{
        D_M
    }.
\]
The preceding limits imply
\[
    1-T_M=O_p(M^{-1}),
\]
and since the A-side design is fixed,
\(\widehat V_{A,\mathrm{te}}=O_p(1)\). Consequently,
\[
    \sum_{i=1}^M
    \widehat V_{B,i,\mathrm{te}}\widehat w_i^2
    =
    \widehat w^\top
    \widehat\Sigma_{\mathrm{te}}
    \widehat w
    =
    O_p(M^{-1}).
\]

It remains to replace the estimated auxiliary variances $\widehat V_{B,i,\mathrm{te}}$ by the true
testing-sample variances $V_{B,i,\mathrm{te}}$. By Assumption~\ref{ass:largeM}\textup{(A1)},
\(V_{B,i,\mathrm{te}}\le\overline v\), and hence
\[
\begin{aligned}
    \sum_{i=1}^M
    V_{B,i,\mathrm{te}}\widehat w_i^2
    =
    \sum_{i=1}^M
    \frac{
    V_{B,i,\mathrm{te}}
    }{
    \widehat V_{B,i,\mathrm{te}}
    }
    \widehat V_{B,i,\mathrm{te}}
    \widehat w_i^2
    &\le
    \overline v
    \left(
        \max_{1\le i\le M}
        \widehat V_{B,i,\mathrm{te}}^{-1}
    \right)
    \sum_{i=1}^M
    \widehat V_{B,i,\mathrm{te}}\widehat w_i^2.
\end{aligned}
\]
The uniform inverse-second-moment bound and Markov implies
\[
\begin{aligned}
    \Pr\left(
        \max_{1\le i\le M}
        \widehat V_{B,i,\mathrm{te}}^{-1}
        >
        K\sqrt M
    \right)
    &\le
    \frac{1}{K^2M}
    \sum_{i=1}^M
    \mathbb E
    \left[
        \widehat V_{B,i,\mathrm{te}}^{-2}
    \right]
    \\
    &\le
    \frac{C}{K^2}
\end{aligned}
\]
 for any \(K>0\) and some finite constant \(C\) independent of \(M\).
Hence
\[
    \max_{1\le i\le M}
    \widehat V_{B,i,\mathrm{te}}^{-1}
    =
    O_p(M^{1/2}).
\]

Thus,
\[
    \sum_{i=1}^M
    V_{B,i,\mathrm{te}}\widehat w_i^2
    =
    O_p(M^{-1/2})
    =
    o_p(1).
\]

Finally, conditional on \(\mathcal T_M\),
\[
\begin{aligned}
    \operatorname{Var}
    \left(
        \widehat\theta_M
        \mid
        \mathcal T_M
    \right)
    &=
    V_{A,\mathrm{te}}(1-T_M)^2
    +
    \sum_{i=1}^M
    V_{B,i,\mathrm{te}}\widehat w_i^2.
\end{aligned}
\]
The first term is \(O_p(M^{-2})\), whereas the second is
\(o_p(1)\). Therefore,
\[
    \operatorname{Var}
    \left(
        \widehat\theta_M
        \mid
        \mathcal T_M
    \right)
    \xrightarrow[M\to\infty]{p}0.
\]Conditional on \(\mathcal T_M\),
\[
    \mathbb E
    \left(
        \widehat\theta_M-\theta
        \mid
        \mathcal T_M
    \right)
    =
    \sum_{i=1}^M\widehat w_i\epsilon_i.
\]
Substitution of the explicit weights gives
\[
\begin{aligned}
    \sum_{i=1}^M\widehat w_i\epsilon_i
    &=
    \frac{\widehat V_{A,\mathrm{te}}}{D_M}
    \Bigg[
        \left(
            1+
            \sum_{i=1}^M
            \frac{\widehat\epsilon_i^2}
            {\widehat V_{B,i,\mathrm{te}}}
        \right)
        \left(
            \sum_{i=1}^M
            \frac{\epsilon_i}
            {\widehat V_{B,i,\mathrm{te}}}
        \right)
        -
        \left(
            \sum_{i=1}^M
            \frac{\widehat\epsilon_i}
            {\widehat V_{B,i,\mathrm{te}}}
        \right)
        \left(
            \sum_{i=1}^M
            \frac{\epsilon_i\widehat\epsilon_i}
            {\widehat V_{B,i,\mathrm{te}}}
        \right)
    \Bigg].
\end{aligned}
\]
Dividing the numerator by \(M^2\) and the denominator by
\(\widehat V_{A,\mathrm{te}}M^2\) yields
\begin{equation}
    \mathbb E
    \left(
        \widehat\theta_M-\theta
        \mid
        \mathcal T_M
    \right)
    \xrightarrow[M\to\infty]{p}
    \frac{
        \widetilde\mu_1\widetilde\nu
        +
        \zeta\widetilde A
    }{
        \widetilde\Delta
    }.
    \label{eq:general-cond-mean-limit}
\end{equation}

By the continuous mapping theorem,
\[
    \left[
        \mathbb E
        \left(
            \widehat\theta_M-\theta
            \mid
            \mathcal T_M
        \right)
    \right]^2
    \xrightarrow[M\to\infty]{p}
    \left[
        \frac{
            \widetilde\mu_1\widetilde\nu
            +
            \zeta\widetilde A
        }{
            \widetilde\Delta
        }
    \right]^2.
\]
By Assumption~\ref{ass:largeM}\textup{(A3)} and Vitali's convergence
theorem,  we have
\[
    \mathbb E\left[
        \operatorname{Var}
        \left(
            \widehat\theta_M
            \mid
            \mathcal T_M
        \right)
    \right]
    \xrightarrow[M\to\infty]{}0,
\]
while \eqref{eq:general-cond-mean-limit} and the continuous mapping
theorem imply
\[
\begin{aligned}
    &\mathbb E\left[
        \left\{
            \mathbb E
            \left(
                \widehat\theta_M-\theta
                \mid
                \mathcal T_M
            \right)
        \right\}^2
    \right]
    \xrightarrow[M\to\infty]{}
    \mathbb E\left[
        \left(
            \frac{
                \widetilde\mu_1\widetilde\nu
                +
                \zeta\widetilde A
            }{
                \widetilde\Delta
            }
        \right)^2
    \right].
\end{aligned}
\]

With conditional second-moment decomposition,
\[
\begin{aligned}
    \mathbb E\left[
        (\widehat\theta_M-\theta)^2
    \right]
    &=
    \mathbb E\left[
        \operatorname{Var}
        \left(
            \widehat\theta_M
            \mid
            \mathcal T_M
        \right)
    \right]
    +
    \mathbb E\left[
        \left\{
            \mathbb E
            \left(
                \widehat\theta_M-\theta
                \mid
                \mathcal T_M
            \right)
        \right\}^2
    \right].
\end{aligned}
\]
Hence
\[
    \operatorname{MSE}_M
    \xrightarrow[M\to\infty]{}
    \mathbb E\left[
        \left(
            \frac{
                \widetilde\mu_1\widetilde\nu
                +
                \zeta\widetilde A
            }{
                \widetilde\Delta
            }
        \right)^2
    \right].
\]
Finally, Assumption~\ref{ass:model} gives
\[
    \mathbb E[\zeta]=0,
    \qquad
    \mathbb E[\zeta^2]=V_{A,\mathrm{pi}}.
\]
Hence
\[
\mathbb E\left[\left(\frac{\widetilde\mu_1\widetilde\nu+\zeta\widetilde A}{\widetilde\Delta}\right)^2\right]
=
\frac{(\widetilde\mu_1\widetilde\nu)^2+V_{A,\mathrm{pi}}\widetilde A^2}{\widetilde\Delta^2}.
\]
This proves Theorem~\ref{thm:MSE_infty1}.

\section{Proof of Proposition \ref{prop:pooled_bias_opt}}
Fix a scalar \(\epsilon_{\mathrm{pool}}\), and consider the oracle pooled-bias version of the plug-in matrix:
\[
Q_{\mathrm{pool}}(\epsilon_{\mathrm{pool}})
=
\diag\!\bigl(V_{B,1,\mathrm{te}},\dots,V_{B,M,\mathrm{te}}\bigr)
+
\epsilon_{\mathrm{pool}}^2\mathbf{1}\mathbf{1}^\top
+
V_{A,\mathrm{te}}\mathbf{1}\mathbf{1}^\top.
\]
The corresponding weight vector is
\[
w(\epsilon_{\mathrm{pool}})
=
V_{A,\mathrm{te}}\,Q_{\mathrm{pool}}(\epsilon_{\mathrm{pool}})^{-1}\mathbf{1}.
\]

Applying the Sherman--Morrison formula gives
\[
Q_{\mathrm{pool}}(\epsilon_{\mathrm{pool}})^{-1}\mathbf{1}
=
\frac{1}{
1+\bigl(V_{A,\mathrm{te}}+\epsilon_{\mathrm{pool}}^2\bigr)\sum_{i=1}^M 1/V_{B,i,\mathrm{te}}
}
\begin{pmatrix}
1/V_{B,1,\mathrm{te}}\\
\vdots\\
1/V_{B,M,\mathrm{te}}
\end{pmatrix}.
\]
Hence, for each \(i=1,\dots,M\),
\[
w_i(\epsilon_{\mathrm{pool}})
=
\frac{V_{A,\mathrm{te}}}{
1+\bigl(V_{A,\mathrm{te}}+\epsilon_{\mathrm{pool}}^2\bigr)\sum_{j=1}^M 1/V_{B,j,\mathrm{te}}
}
\cdot
\frac{1}{V_{B,i,\mathrm{te}}},
\]
and therefore
\[
\mathbf{1}^\top w(\epsilon_{\mathrm{pool}})
=
\frac{
V_{A,\mathrm{te}}\sum_{i=1}^M 1/V_{B,i,\mathrm{te}}
}{
1+\bigl(V_{A,\mathrm{te}}+\epsilon_{\mathrm{pool}}^2\bigr)\sum_{i=1}^M 1/V_{B,i,\mathrm{te}}
}.
\]

The corresponding pooled-bias estimator based on the test means is
\[
\hat\theta_{\mathrm{pool}}(\epsilon_{\mathrm{pool}})
=
\bigl(1-\mathbf{1}^\top w(\epsilon_{\mathrm{pool}})\bigr)\bar A_{\mathrm{te}}
+
\sum_{i=1}^M w_i(\epsilon_{\mathrm{pool}})\bar B_{i,\mathrm{te}}.
\]
Conditioning on the test variances and biases, its conditional mean squared error is
\[
\MSE_{\mathrm{pool}}(\epsilon_{\mathrm{pool}})
=
\left(
\sum_{i=1}^M w_i(\epsilon_{\mathrm{pool}})\epsilon_i
\right)^2
+
V_{A,\mathrm{te}}\bigl(1-\mathbf{1}^\top w(\epsilon_{\mathrm{pool}})\bigr)^2
+
\sum_{i=1}^M V_{B,i,\mathrm{te}}\,w_i(\epsilon_{\mathrm{pool}})^2.
\]

Substituting the explicit form of \(w_i(\epsilon_{\mathrm{pool}})\) yields
\[
\sum_{i=1}^M w_i(\epsilon_{\mathrm{pool}})\epsilon_i
=
\frac{
V_{A,\mathrm{te}}\sum_{i=1}^M \epsilon_i/V_{B,i,\mathrm{te}}
}{
1+\bigl(V_{A,\mathrm{te}}+\epsilon_{\mathrm{pool}}^2\bigr)\sum_{i=1}^M 1/V_{B,i,\mathrm{te}}
},
\]
\[
1-\mathbf{1}^\top w(\epsilon_{\mathrm{pool}})
=
\frac{
1+\epsilon_{\mathrm{pool}}^2\sum_{i=1}^M 1/V_{B,i,\mathrm{te}}
}{
1+\bigl(V_{A,\mathrm{te}}+\epsilon_{\mathrm{pool}}^2\bigr)\sum_{i=1}^M 1/V_{B,i,\mathrm{te}}
},
\]
and
\[
\sum_{i=1}^M V_{B,i,\mathrm{te}}\,w_i(\epsilon_{\mathrm{pool}})^2
=
\frac{
V_{A,\mathrm{te}}^2\sum_{i=1}^M 1/V_{B,i,\mathrm{te}}
}{
\left[
1+\bigl(V_{A,\mathrm{te}}+\epsilon_{\mathrm{pool}}^2\bigr)\sum_{i=1}^M 1/V_{B,i,\mathrm{te}}
\right]^2 }.
\]

Therefore,
\[
\MSE_{\mathrm{pool}}(\epsilon_{\mathrm{pool}})
=
\frac{
V_{A,\mathrm{te}}^2\left(\sum_{i=1}^M \epsilon_i/V_{B,i,\mathrm{te}}\right)^2
+
V_{A,\mathrm{te}}\left(1+\epsilon_{\mathrm{pool}}^2\sum_{i=1}^M 1/V_{B,i,\mathrm{te}}\right)^2
+
V_{A,\mathrm{te}}^2\sum_{i=1}^M 1/V_{B,i,\mathrm{te}}
}{
\left[
1+\bigl(V_{A,\mathrm{te}}+\epsilon_{\mathrm{pool}}^2\bigr)\sum_{i=1}^M 1/V_{B,i,\mathrm{te}}
\right]^2 }.
\]

Now view this as a function of \(\epsilon_{\mathrm{pool}}^2\). Differentiating with respect to \(\epsilon_{\mathrm{pool}}^2\) and setting the derivative equal to zero yields
\[
\epsilon_{\mathrm{pool}}^2
=
\left(
\frac{\sum_{i=1}^M \epsilon_i/V_{B,i,\mathrm{te}}}
{\sum_{i=1}^M 1/V_{B,i,\mathrm{te}}}
\right)^2.
\]
Hence the MSE-optimal pooled scalar is
\[
\epsilon_{\mathrm{pool},\mathrm{opt}}
=
\frac{\sum_{i=1}^M \epsilon_i/V_{B,i,\mathrm{te}}}
{\sum_{i=1}^M 1/V_{B,i,\mathrm{te}}}.
\]

Replacing the unknown quantities \(\epsilon_i\) and \(V_{B,i,\mathrm{te}}\) by their plug-in analogues \(\hat\epsilon_i\) and \(\widehat V_{B,i,\mathrm{te}}\) gives the feasible implementation
\[
\hat\epsilon_{\mathrm{pool}}
=
\frac{\sum_{i=1}^M \hat\epsilon_i/\widehat V_{B,i,\mathrm{te}}}
{\sum_{i=1}^M 1/\widehat V_{B,i,\mathrm{te}}}.
\]
This proves the proposition.

\section{(do a final check)Proof of Theorem~\ref{thm:pooled_bias_limit}}
\label{app:proof_thm_pool}

All probability statements and expectations are taken under the
summary-level working model in Assumption~\ref{ass:working}.

Let \(\mathcal T_M\) denote the \(\sigma\)-field generated by all
pilot-sample quantities. Define
\[
    \zeta
    :=
    \bar A_{\mathrm{pi}}-\theta,
    \qquad
    \eta_i
    :=
    \bar B_{i,\mathrm{pi}}-(\theta+\epsilon_i),
    \quad i=1,\ldots,M.
\]
Thus, \(\zeta\) and \(\eta_i\) are the centered random components
of the target and auxiliary pilot-sample means, respectively.
In particular,
\[
    \widehat\epsilon_i
    =
    \bar B_{i,\mathrm{pi}}-\bar A_{\mathrm{pi}}
    =
    \epsilon_i+\eta_i-\zeta.
\]

Under Assumption~\ref{ass:working}, we use the working approximation
\[
    \zeta\sim N(0,V_{A,\mathrm{pi}}),
    \qquad
    \eta_i\sim N(0,V_{B,i,\mathrm{pi}}).
\]

The proof proceeds by first identifying the large-\(M\) limits of
the pooled bias estimate, the total weight assigned to the auxiliary
side, and the precision-weighted auxiliary testing mean. These limits
yield the probability limit of the pooled plug-in estimator. We then establish convergence of the MSE using the conditional
second-moment decomposition. Conditional on \(\mathcal T_M\), the
estimated weights are fixed and the testing-sample means are
independent of the pilot-sample quantities. We will show that
\[
    \mathbb E\left[
        \left.
        \widehat\theta_{\mathrm{pool},M}-\theta
        \right|
        \mathcal T_M
    \right]
    \xrightarrow[M\to\infty]{p}
    e\widehat T_\infty,
\]
and
\[
    \operatorname{Var}\left(
        \left.
        \widehat\theta_{\mathrm{pool},M}
        \right|
        \mathcal T_M
    \right)
    \xrightarrow[M\to\infty]{p}
    V_{A,\mathrm{te}}(1-\widehat T_\infty)^2.
\]
With the  uniform integrability conditions verified, these probability limits can be upgraded to
convergence of the corresponding expectations and hence get the limiting MSE.

Define the normalized estimated precision weights
\[
    a_{i,M}
    :=
    \frac{\widehat V_{B,i,\mathrm{te}}^{-1}}
    {\displaystyle\sum_{j=1}^M
        \widehat V_{B,j,\mathrm{te}}^{-1}},
    \qquad i=1,\ldots,M.
\]
Then
\[
    \widehat\epsilon_{\mathrm{pool}}
    =
    \sum_{i=1}^M a_{i,M}\widehat\epsilon_i.
\]
Recall that
\[
    \widehat Q_{\mathrm{pool}}
    =
    \widehat\Sigma_{\mathrm{te}}
    +
    \left(
        \widehat V_{A,\mathrm{te}}
        +
        \widehat\epsilon_{\mathrm{pool}}^{\,2}
    \right)
    \mathbf 1\mathbf 1^\top,
\]
and
\[
    \widehat{\boldsymbol w}_{\mathrm{pool}}
    =
    \widehat V_{A,\mathrm{te}}
    \widehat Q_{\mathrm{pool}}^{-1}\mathbf 1.
\]
By the Sherman--Morrison formula,
\[
    \widehat Q_{\mathrm{pool}}^{-1}\mathbf 1
    =
    \frac{
        \widehat\Sigma_{\mathrm{te}}^{-1}\mathbf 1
    }{
        1+
        \left(
            \widehat V_{A,\mathrm{te}}
            +
            \widehat\epsilon_{\mathrm{pool}}^{\,2}
        \right)
        \displaystyle\sum_{j=1}^M
        \widehat V_{B,j,\mathrm{te}}^{-1}
    }.
\]
Therefore, for each \(i=1,\ldots,M\),
\[
    \widehat w_{\mathrm{pool},i}
    =
    \frac{
        \widehat V_{A,\mathrm{te}}
    }{
        1+
        \left(
            \widehat V_{A,\mathrm{te}}
            +
            \widehat\epsilon_{\mathrm{pool}}^{\,2}
        \right)
        \displaystyle\sum_{j=1}^M
        \widehat V_{B,j,\mathrm{te}}^{-1}
    }
    \widehat V_{B,i,\mathrm{te}}^{-1}.
\]

Define the total weight assigned to the auxiliary side by
\begin{equation}
    \widehat T_M
    :=
    \mathbf 1^\top\widehat{\boldsymbol w}_{\mathrm{pool}}
    =
    \frac{
        \widehat V_{A,\mathrm{te}}
    }{
        \widehat V_{A,\mathrm{te}}
        +
        \widehat\epsilon_{\mathrm{pool}}^{\,2}
        +
        \left(
            \displaystyle\sum_{i=1}^M
            \widehat V_{B,i,\mathrm{te}}^{-1}
        \right)^{-1}
    }.
    \label{eq:pooled_total_weight}
\end{equation}
In particular,
\[
    0\leq \widehat T_M\leq 1 
\]
and we have
\begin{equation}
    \widehat w_{\mathrm{pool},i}
    =
    \widehat T_M a_{i,M}.
    \label{eq:pooled_weight_decomposition}
\end{equation}

We first establish the limit
\begin{equation}
    \sum_{i=1}^M a_{i,M}\epsilon_i
    \xrightarrow[M\to\infty]{p} e.
    \label{eq:weighted_true_bias_limit}
\end{equation}
To combine the stochastic averaging limits in
Assumption~\ref{ass:largeM}(A2) with the deterministic
no-alignment condition in Assumption~\ref{ass:noalign}, we first
show that the empirical estimated precisions concentrate around
their expectations.

Define
\[
    q_i
    :=
    \mathbb E\left[
        \widehat V_{B,i,\mathrm{te}}^{-1}
    \right].
\]
By independence across auxiliary studies,
\[
\begin{aligned}
    &\mathbb E\left[
        \left\{
            \frac{1}{M}
            \sum_{i=1}^M
            \left(
                \widehat V_{B,i,\mathrm{te}}^{-1}-q_i
            \right)
        \right\}^{2}
    \right]
    =
    \frac{1}{M^2}
    \sum_{i=1}^M
    \operatorname{Var}
    \left(
        \widehat V_{B,i,\mathrm{te}}^{-1}
    \right).
\end{aligned}
\]
Assumption~\ref{ass:largeM}(A1), together with the
effective-degree-of-freedom condition in
Assumption~\ref{ass:working}, implies
\[
    \sup_{i\geq1}
    \mathbb E\left[
        \widehat V_{B,i,\mathrm{te}}^{-2}
    \right]
    <\infty.
\]
Consequently,
\[
    \mathbb E\left[
        \left\{
            \frac{1}{M}
            \sum_{i=1}^M
            \left(
                \widehat V_{B,i,\mathrm{te}}^{-1}-q_i
            \right)
        \right\}^{2}
    \right]
    =
    O(M^{-1}),
\]
and hence
\[
    \frac{1}{M}
    \sum_{i=1}^M
    \widehat V_{B,i,\mathrm{te}}^{-1}
    -
    \frac{1}{M}
    \sum_{i=1}^M q_i
    \xrightarrow[M\to\infty]{L^2}0.
\]

Similarly,
\[
\begin{aligned}
    &\mathbb E\left[
        \left\{
            \frac{1}{M}
            \sum_{i=1}^M
            \epsilon_i
            \left(
                \widehat V_{B,i,\mathrm{te}}^{-1}-q_i
            \right)
        \right\}^{2}
    \right]
    =
    \frac{1}{M^2}
    \sum_{i=1}^M
    \epsilon_i^2
    \operatorname{Var}
    \left(
        \widehat V_{B,i,\mathrm{te}}^{-1}
    \right)
    \leq
    \frac{C}{M^2}
    \sum_{i=1}^M\epsilon_i^2
\end{aligned}
\]
for some finite constant \(C\) independent of \(M\).
By Assumption~\ref{ass:largeM}(A1),
\[
    \sup_{i\geq1}|\epsilon_i|<\infty,
\]
and therefore
\[
    \frac{1}{M}\sum_{i=1}^M\epsilon_i^2
    \leq
    \left(
        \sup_{i\geq1}|\epsilon_i|
    \right)^2
    <\infty.
\]
It follows that
\[
    \mathbb E\left[
        \left\{
            \frac{1}{M}
            \sum_{i=1}^M
            \epsilon_i
            \left(
                \widehat V_{B,i,\mathrm{te}}^{-1}-q_i
            \right)
        \right\}^{2}
    \right]
    =
    O(M^{-1}),
\]
and therefore
\[
    \frac{1}{M}
    \sum_{i=1}^M
    \epsilon_i\widehat V_{B,i,\mathrm{te}}^{-1}
    -
    \frac{1}{M}
    \sum_{i=1}^M
    \epsilon_iq_i
    \xrightarrow[M\to\infty]{L^2}0.
\]

Combining these two \(L^2\) limits with
Assumption~\ref{ass:largeM}(A2) gives
\[
    \frac{1}{M}\sum_{i=1}^M q_i
    \xrightarrow[M\to\infty]{}
    \widetilde\mu_0,
    \qquad
    \frac{1}{M}\sum_{i=1}^M\epsilon_iq_i
    \xrightarrow[M\to\infty]{}
    \widetilde\mu_1.
\]
The sequences on the left-hand sides are deterministic.

Let
\[
    \bar\epsilon_M
    :=
    \frac{1}{M}\sum_{i=1}^M\epsilon_i,
    \qquad
    \bar q_M
    :=
    \frac{1}{M}\sum_{i=1}^M q_i.
\]
The first no-alignment condition in
Assumption~\ref{ass:noalign} gives
\[
    \frac{1}{M}
    \sum_{i=1}^M
    (\epsilon_i-\bar\epsilon_M)
    (q_i-\bar q_M)
    \xrightarrow[M\to\infty]{}0.
\]
Expanding the left-hand side,
\[
\begin{aligned}
    &\frac{1}{M}
    \sum_{i=1}^M
    (\epsilon_i-\bar\epsilon_M)
    (q_i-\bar q_M)
    =
    \frac{1}{M}\sum_{i=1}^M\epsilon_iq_i
    -
    \bar\epsilon_M\bar q_M.
\end{aligned}
\]
By Assumption~\ref{ass:noalign},
\[
    \bar\epsilon_M\xrightarrow[M\to\infty]{} e.
\]
Together with the preceding deterministic limits, the no-alignment
condition therefore implies
\[
    \widetilde\mu_1-e\widetilde\mu_0=0,
\]
and hence
\begin{equation}
    \widetilde\mu_1=e\widetilde\mu_0.
    \label{eq:mu1_mu0_relation}
\end{equation}

We also have \(\widetilde\mu_0>0\). Under the scaled chi-squared
approximation in Assumption~\ref{ass:working},
\[
    \widehat V_{B,i,\mathrm{te}}
    =
    V_{B,i,\mathrm{te}}
    \frac{\chi^2_{\nu_{B,i,\mathrm{pi}}}}
    {\nu_{B,i,\mathrm{pi}}},
\]
and therefore
\[
    q_i
    =
    \frac{1}{V_{B,i,\mathrm{te}}}
    \frac{\nu_{B,i,\mathrm{pi}}}
    {\nu_{B,i,\mathrm{pi}}-2}.
\]
Since Assumption~\ref{ass:largeM}(A1) gives
\(V_{B,i,\mathrm{te}}\leq\overline v\), it follows that
\[
    q_i\geq\frac{1}{\overline v}.
\]
Thus,
\[
    \widetilde\mu_0
    =
    \lim_{M\to\infty}
    \frac{1}{M}\sum_{i=1}^M q_i
    \geq
    \frac{1}{\overline v}
    >0.
\]

It now follows from Assumption~\ref{ass:largeM}(A2),
\eqref{eq:mu1_mu0_relation}, and the continuous mapping theorem that
\[
\begin{aligned}
    \sum_{i=1}^M a_{i,M}\epsilon_i
    &=
    \frac{
        \displaystyle
        M^{-1}\sum_{i=1}^M
        \epsilon_i\widehat V_{B,i,\mathrm{te}}^{-1}
    }{
        \displaystyle
        M^{-1}\sum_{i=1}^M
        \widehat V_{B,i,\mathrm{te}}^{-1}
    }
    &\xrightarrow[M\to\infty]{p}
    \frac{\widetilde\mu_1}{\widetilde\mu_0}
    =
    e,
\end{aligned}
\]
which proves \eqref{eq:weighted_true_bias_limit}.

Using
\(\widehat\epsilon_i=\epsilon_i+\eta_i-\zeta\), we may write
\[
    \widehat\epsilon_{\mathrm{pool}}
    =
    \sum_{i=1}^M a_{i,M}\epsilon_i
    +
    \sum_{i=1}^M a_{i,M}\eta_i
    -
    \zeta.
\]
We next show that the weighted pilot-sample noise vanishes. By
independence across \(i\), the working independence between
\(\eta_i\) and \(\widehat V_{B,i,\mathrm{te}}\), and
\(\mathbb E[\eta_i]=0\),
\[
\begin{aligned}
    &\mathbb E\left[
        \left(
            \frac{1}{M}
            \sum_{i=1}^M
            \frac{\eta_i}
            {\widehat V_{B,i,\mathrm{te}}}
        \right)^2
    \right]
    =
    \frac{1}{M^2}
    \sum_{i=1}^M
    V_{B,i,\mathrm{pi}}
    \mathbb E\left[
        \widehat V_{B,i,\mathrm{te}}^{-2}
    \right]
    =
    O(M^{-1}).
\end{aligned}
\]
Therefore,
\[
    \frac{1}{M}
    \sum_{i=1}^M
    \frac{\eta_i}
    {\widehat V_{B,i,\mathrm{te}}}
    \xrightarrow[M\to\infty]{L^2}0.
\]
Since
\[
    \frac{1}{M}
    \sum_{i=1}^M
    \widehat V_{B,i,\mathrm{te}}^{-1}
    \xrightarrow[M\to\infty]{p}
    \widetilde\mu_0>0,
\]
we obtain
\[
\begin{aligned}
    \sum_{i=1}^M a_{i,M}\eta_i
    &=
    \frac{
        \displaystyle
        M^{-1}\sum_{i=1}^M
        \eta_i\widehat V_{B,i,\mathrm{te}}^{-1}
    }{
        \displaystyle
        M^{-1}\sum_{i=1}^M
        \widehat V_{B,i,\mathrm{te}}^{-1}
    }
    \xrightarrow[M\to\infty]{p}0.
\end{aligned}
\]
Combining this result with
\eqref{eq:weighted_true_bias_limit} yields
\begin{equation}
    \widehat\epsilon_{\mathrm{pool}}
    \xrightarrow[M\to\infty]{p}
    e-\zeta
    =
    e-(\bar A_{\mathrm{pi}}-\theta).
    \label{eq:pooled_bias_probability_limit}
\end{equation}

Moreover,
\[
    \sum_{i=1}^M
    \widehat V_{B,i,\mathrm{te}}^{-1}
    =
    M
    \left(
        \frac{1}{M}
        \sum_{i=1}^M
        \widehat V_{B,i,\mathrm{te}}^{-1}
    \right)
    \xrightarrow[M\to\infty]{p}\infty,
\]
because the average in parentheses converges in probability to
\(\widetilde\mu_0>0\). Hence
\[
    \left(
        \sum_{i=1}^M
        \widehat V_{B,i,\mathrm{te}}^{-1}
    \right)^{-1}
    \xrightarrow[M\to\infty]{p}0.
\]
Substituting this result and
\eqref{eq:pooled_bias_probability_limit} into
\eqref{eq:pooled_total_weight}, and applying the continuous mapping
theorem, gives
\begin{equation}
    \widehat T_M
    \xrightarrow[M\to\infty]{p}
    \widehat T_\infty,
    \qquad
    \widehat T_\infty
    :=
    \frac{
        \widehat V_{A,\mathrm{te}}
    }{
        \widehat V_{A,\mathrm{te}}
        +
        [e-(\bar A_{\mathrm{pi}}-\theta)]^2
    }.
    \label{eq:pooled_total_weight_limit}
\end{equation}

We next consider the precision-weighted auxiliary testing
mean. We may write
\[
\begin{aligned}
    \sum_{i=1}^M a_{i,M}\bar B_{i,\mathrm{te}}
    &=
    \theta
    +
    \sum_{i=1}^M a_{i,M}\epsilon_i
    +
        \sum_{i=1}^M
    a_{i,M}
    \left\{
        \bar B_{i,\mathrm{te}}
        -
        (\theta+\epsilon_i)
    \right\}.
\end{aligned}
\]
The second term converges in probability to \(e\) by
\eqref{eq:weighted_true_bias_limit}.

Conditional on \(\mathcal T_M\), the weights \(a_{i,M}\) are fixed.
By the pilot/testing sample split, the testing-sample means are
mutually independent across auxiliary studies and independent of
\(\mathcal T_M\), with
\[
    \mathbb E\left[
        \left.
        \bar B_{i,\mathrm{te}}
        -
        (\theta+\epsilon_i)
        \right|
        \mathcal T_M
    \right]
    =0.
\]
Therefore,
\[
    \mathbb E\left[
        \left.
        \sum_{i=1}^M
        a_{i,M}
        \left\{
            \bar B_{i,\mathrm{te}}
            -
            (\theta+\epsilon_i)
        \right\}
        \right|
        \mathcal T_M
    \right]
    =0,
\]
and
\[
\begin{aligned}
    &\operatorname{Var}\left(
        \left.
        \sum_{i=1}^M
        a_{i,M}
        \left\{
            \bar B_{i,\mathrm{te}}
            -
            (\theta+\epsilon_i)
        \right\}
        \right|
        \mathcal T_M
    \right)
    =
    \sum_{i=1}^M a_{i,M}^2V_{B,i,\mathrm{te}}
    =
    \frac{
        \displaystyle
        \sum_{i=1}^M
        V_{B,i,\mathrm{te}}
        \widehat V_{B,i,\mathrm{te}}^{-2}
    }{
        \left(
            \displaystyle
            \sum_{j=1}^M
            \widehat V_{B,j,\mathrm{te}}^{-1}
        \right)^2
    }.
\end{aligned}
\]
By Markov's inequality, Assumption~\ref{ass:largeM}(A1), and the
uniform inverse-second-moment bound,
\[
    \frac{1}{M}
    \sum_{i=1}^M
    V_{B,i,\mathrm{te}}
    \widehat V_{B,i,\mathrm{te}}^{-2}
    =
    O_{\mathrm p}(1).
\]
Also,
\[
    \left(
        \sum_{j=1}^M
        \widehat V_{B,j,\mathrm{te}}^{-1}
    \right)^2
    =
    M^2\left(\widetilde\mu_0+o_{\mathrm p}(1)\right)^2.
    \]
It follows that
\begin{equation}
    \sum_{i=1}^M a_{i,M}^2V_{B,i,\mathrm{te}}
    =
    O_{\mathrm p}(M^{-1})
    \xrightarrow[M\to\infty]{p}0.
    \label{eq:B_testing_variance_vanishes}
\end{equation}
Moreover,
\[
    0
    \leq
    \sum_{i=1}^M a_{i,M}^2V_{B,i,\mathrm{te}}
    \leq
    \overline v
    \sum_{i=1}^M a_{i,M}^2
    \leq
    \overline v.
\]
Thus, conditional Chebyshev's inequality, together with
\eqref{eq:B_testing_variance_vanishes}, implies
\[
    \sum_{i=1}^M
    a_{i,M}
    \left\{
        \bar B_{i,\mathrm{te}}
        -
        (\theta+\epsilon_i)
    \right\}
    \xrightarrow[M\to\infty]{p}0.
\]
Consequently,
\begin{equation}
    \sum_{i=1}^M a_{i,M}\bar B_{i,\mathrm{te}}
    \xrightarrow[M\to\infty]{p}
    \theta+e.
    \label{eq:weighted_B_testing_mean_limit}
\end{equation}

By \eqref{eq:pooled_weight_decomposition}, the pooled estimator can
be written as
\[
    \widehat\theta_{\mathrm{pool},M}
    =
    (1-\widehat T_M)\bar A_{\mathrm{te}}
    +
    \widehat T_M
    \sum_{i=1}^M a_{i,M}\bar B_{i,\mathrm{te}}.
\]
Combining \eqref{eq:pooled_total_weight_limit} and
\eqref{eq:weighted_B_testing_mean_limit}, and applying Slutsky's
theorem, gives
\[
    \widehat\theta_{\mathrm{pool},M}
    \xrightarrow[M\to\infty]{p}
    (1-\widehat T_\infty)\bar A_{\mathrm{te}}
    +
    \widehat T_\infty(\theta+e).
\]
This shows
Theorem~\ref{thm:pooled_bias_limit}.

It remains to prove convergence of the MSE. Conditional on
\(\mathcal T_M\), the estimated weights are fixed and independent of
all testing-sample means. Therefore,
\[
\begin{aligned}
    &\mathbb E\left[
        \left.
        \widehat\theta_{\mathrm{pool},M}-\theta
        \right|
        \mathcal T_M
    \right]
    =
    (1-\widehat T_M)\theta
    +
    \widehat T_M
    \sum_{i=1}^M
    a_{i,M}(\theta+\epsilon_i)
    -
    \theta
    =
    \widehat T_M
    \sum_{i=1}^M a_{i,M}\epsilon_i.
\end{aligned}
\]
By \eqref{eq:weighted_true_bias_limit} and
\eqref{eq:pooled_total_weight_limit},
\begin{equation}
    \left\{
        \mathbb E\left[
            \left.
            \widehat\theta_{\mathrm{pool},M}-\theta
            \right|
            \mathcal T_M
        \right]
    \right\}^{2}
    \xrightarrow[M\to\infty]{p}
    e^2\widehat T_\infty^2.
    \label{eq:squared_conditional_mean_limit}
\end{equation}

The conditional variance is
\[
\begin{aligned}
    \operatorname{Var}\left(
        \left.
        \widehat\theta_{\mathrm{pool},M}
        \right|
        \mathcal T_M
    \right)
    &=
    V_{A,\mathrm{te}}(1-\widehat T_M)^2
    +
    \widehat T_M^2
    \sum_{i=1}^M
    a_{i,M}^2V_{B,i,\mathrm{te}}.
\end{aligned}
\]
By \eqref{eq:pooled_total_weight_limit} and
\eqref{eq:B_testing_variance_vanishes},
\begin{equation}
    \operatorname{Var}\left(
        \left.
        \widehat\theta_{\mathrm{pool},M}
        \right|
        \mathcal T_M
    \right)
    \xrightarrow[M\to\infty]{p}
    V_{A,\mathrm{te}}(1-\widehat T_\infty)^2.
    \label{eq:conditional_variance_limit}
\end{equation}

We now verify uniform integrability of the two conditional-moment
families appearing in the conditional second-moment decomposition.
Since \(0\leq\widehat T_M\leq1\),
\[
\begin{aligned}
    \operatorname{Var}\left(
        \left.
        \widehat\theta_{\mathrm{pool},M}
        \right|
        \mathcal T_M
    \right)
    &\leq
    V_{A,\mathrm{te}}
    +
    \sum_{i=1}^M a_{i,M}^2V_{B,i,\mathrm{te}}
    \\
    &\leq
    V_{A,\mathrm{te}}
    +
    \overline v
    \sum_{i=1}^M a_{i,M}^2
    \\
    &\leq
    V_{A,\mathrm{te}}+\overline v.
\end{aligned}
\]
Thus, the conditional-variance family is uniformly bounded and
therefore uniformly integrable.

For the squared conditional means, since \(0\leq\widehat T_M\leq1\),
\(a_{i,M}\geq0\), and
\(\sum_{i=1}^M a_{i,M}=1\), Assumption~\ref{ass:largeM}(A1) gives
\[
\begin{aligned}
    \left|
        \mathbb E\left[
            \left.
            \widehat\theta_{\mathrm{pool},M}-\theta
            \right|
            \mathcal T_M
        \right]
    \right|
    &=
    \left|
        \widehat T_M
        \sum_{i=1}^M a_{i,M}\epsilon_i
    \right|
    \\
    &\leq
    \sum_{i=1}^M a_{i,M}|\epsilon_i|
    \\
    &\leq
    \sup_{i\geq1}|\epsilon_i|
    <\infty.
\end{aligned}
\]
Thus,
\[
    \left\{
        \left[
            \mathbb E\left(
                \left.
                \widehat\theta_{\mathrm{pool},M}-\theta
                \right|
                \mathcal T_M
            \right)
        \right]^2
        :
        M\geq1
    \right\}
\]
is uniformly bounded and therefore uniformly integrable.

By Vitali's convergence theorem, convergence in probability together
with uniform integrability implies convergence in \(L^1\). Applying
this result to \eqref{eq:squared_conditional_mean_limit} and
\eqref{eq:conditional_variance_limit} gives
\[
\begin{aligned}
    &\mathbb E\left[
        \operatorname{Var}\left(
            \left.
            \widehat\theta_{\mathrm{pool},M}
            \right|
            \mathcal T_M
        \right)
    \right]
    \xrightarrow[M\to\infty]{}
    \mathbb E\left[
        V_{A,\mathrm{te}}(1-\widehat T_\infty)^2
    \right],
\end{aligned}
\]
and
\[
\begin{aligned}
    &\mathbb E\left[
        \left\{
            \mathbb E\left[
                \left.
                \widehat\theta_{\mathrm{pool},M}-\theta
                \right|
                \mathcal T_M
            \right]
        \right\}^{2}
    \right]
    \xrightarrow[M\to\infty]{}
    \mathbb E\left[
        e^2\widehat T_\infty^2
    \right].
\end{aligned}
\]

Therefore we conclude that
\[
    \operatorname{MSE}_{\mathrm{pool},M}
    \xrightarrow[M\to\infty]{}
    \mathbb E\left[
        V_{A,\mathrm{te}}(1-\widehat T_\infty)^2
        +
        e^2\widehat T_\infty^2
    \right].
\]

Finally, by \eqref{eq:pooled_total_weight_limit},
we have
\[
\operatorname{MSE}_{\mathrm{pool},\infty}
=
\mathbb E_{\bar A_{\mathrm{pi}},\widehat V_{A,\mathrm{te}}}
\left[
    \frac{
        e^2\widehat V_{A,\mathrm{te}}^{\,2}
        +
        V_{A,\mathrm{te}}
        \left[e-(\bar A_{\mathrm{pi}}-\theta)\right]^4
    }{
        \left\{
            \widehat V_{A,\mathrm{te}}
            +
            \left[e-(\bar A_{\mathrm{pi}}-\theta)\right]^2
        \right\}^{2}
    }
\right].
\]
This proves Theorem~\ref{thm:pooled_bias_limit}.

\section{Proof of Theorem~\ref{thm:fixed_weight}}

All probability statements and expectations are evaluated under the
summary-level working model in Assumption~\ref{ass:working_full}. We first establish the large-\(M\) behavior of the aggregated auxiliary
mean. Recall that
\[
    \widetilde B_M
    =
    \sum_{i=1}^M u_{i,M}\bar B_i.
\]
Under Assumption~\ref{ass:model},
\[
    \E[\widetilde B_M]
    =
    \theta+\bar\epsilon_M^w,
\]
and
\[
    \Var(\widetilde B_M)
    =
    U_M
    :=
    \sum_{i=1}^M u_{i,M}^2V_{B,i}.
\]
By Assumption~\ref{ass:fw}\textup{(FW1)},
\[
\begin{aligned}
    U_M
    &\le
    \left(\sup_{i\ge1}V_{B,i}\right)
    \sum_{i=1}^M u_{i,M}^2
    \le
    \left(\sup_{i\ge1}V_{B,i}\right)
    \max_{1\le i\le M}u_{i,M}
    \xrightarrow[M\to\infty]{}0.
\end{aligned}
\]
Together with Assumption~\ref{ass:fw}\textup{(FW2)}, this gives
\[
\begin{aligned}
    \E\!\left[
        \{\widetilde B_M-(\theta+e)\}^2
    \right]
    &=
    \Var(\widetilde B_M)
    +
    \{\E[\widetilde B_M]-(\theta+e)\}^2
    =    U_M+(\bar\epsilon_M^w-e)^2
    \xrightarrow[M\to\infty]{}0.
\end{aligned}
\]
Hence,
\[
    \widetilde B_M
    \xrightarrow[M\to\infty]{}
    \theta+e
    \qquad\text{in }L^2,
\]
and therefore also in probability.

We next consider the estimated B-side contribution to the discrepancy
variance,
\[
    \widehat U_M
    =
    \sum_{i=1}^M u_{i,M}^2\widehat V_{B,i}.
\]
Since each \(\widehat V_{B,i}\) is nonnegative and unbiased for
\(V_{B,i}\),
\[
\begin{aligned}
    \E[\widehat U_M]
    &=
    \sum_{i=1}^M
    u_{i,M}^2\E[\widehat V_{B,i}]
    =
    \sum_{i=1}^M
    u_{i,M}^2V_{B,i}
    =
    U_M
    \xrightarrow[M\to\infty]{}0.
\end{aligned}
\]
Because \(\widehat U_M\ge0\),
\[
    \E|\widehat U_M|
    =
    \E[\widehat U_M]
    \xrightarrow[M\to\infty]{}0.
\]
Thus,
\[
    \widehat U_M\xrightarrow[M\to\infty]{}0
    \qquad\text{in }L^1,
\]
and hence in probability.

Moreover, 
\[
    \lambda_M
    =
    \frac{\widetilde N_{B,M}}{N_A+\widetilde N_{B,M}}
    \xrightarrow[M\to\infty]{}1.
\]

Recall that
\[
    D_M
    =
    \widetilde B_M-\bar A.
\]
Since the A-side design is held fixed as \(M\to\infty\),
\(\bar A\) and \(\widehat V_A\) retain their randomness. The preceding
limits imply
\[
    D_M
    \xrightarrow[M\to\infty]{p}
    e-(\bar A-\theta),
\]
and
\[
    \widehat V_A+\widehat U_M
    \xrightarrow[M\to\infty]{p}
    \widehat V_A.
\]

The practical testing indicator is
\[
    \widehat I_M
    =
    \mathbf 1\!\left\{
        |D_M|
        \le
        t_{1-\alpha/2}
        \sqrt{\widehat V_A+\widehat U_M}
    \right\}.
\]
Under Assumption~\ref{ass:working_full},
\(\bar A-\theta\) has a continuous Gaussian distribution and is
independent of \(\widehat V_A\). Hence
\[
    \Pr\!\left(
        |e-(\bar A-\theta)|
        =
        t_{1-\alpha/2}\sqrt{\widehat V_A}
    \right)
    =
    0.
\]
Therefore, the discontinuity boundary of the indicator has probability
zero, and the continuous mapping theorem yields
\[
    \widehat I_M
    \xrightarrow[M\to\infty]{p}
    \mathbf 1\!\left\{
        |e-(\bar A-\theta)|
        \le
        t_{1-\alpha/2}\sqrt{\widehat V_A}
    \right\}.
\]

Using
\[
    \widehat\theta_M^{\mathrm{FH}}
    =
    \bar A+\lambda_MD_M\widehat I_M,
\]
we obtain
\[
\begin{aligned}
    \widehat\theta_M^{\mathrm{FH}}-\theta
    \xrightarrow[M\to\infty]{p}\;&
    e\,
    \mathbf 1\!\left\{
        |e-(\bar A-\theta)|
        \le
        t_{1-\alpha/2}\sqrt{\widehat V_A}
    \right\}
    +
    (\bar A-\theta)
    \mathbf 1\!\left\{
        |e-(\bar A-\theta)|
        >
        t_{1-\alpha/2}\sqrt{\widehat V_A}
    \right\}.
\end{aligned}
\]

It remains to justify convergence of the corresponding second moments.
The definition of the testing gate gives
\[
    D_M^2\widehat I_M
    \le
    t_{1-\alpha/2}^{\,2}
    (\widehat V_A+\widehat U_M).
\]
Since \(0\le\lambda_M\le1\),
\[
\begin{aligned}
    (\widehat\theta_M^{\mathrm{FH}}-\theta)^2
    &=
    \{(\bar A-\theta)+\lambda_MD_M\widehat I_M\}^2
    \le
    2(\bar A-\theta)^2
    +
    2\lambda_M^2D_M^2\widehat I_M
    \le
    2(\bar A-\theta)^2
    +
    2t_{1-\alpha/2}^{\,2}\widehat V_A
    +
    2t_{1-\alpha/2}^{\,2}\widehat U_M.
\end{aligned}
\]
Since \(\widehat U_M\to0\) in \(L^1\), so
\(\{\widehat U_M:M\ge1\}\) is uniformly integrable. Hence the
right-hand side above is a uniformly integrable family, which implies
that
\[
    \left\{
        (\widehat\theta_M^{\mathrm{FH}}-\theta)^2
        :M\ge1
    \right\}
\]
is uniformly integrable.

It follows from convergence in probability and uniform integrability
that
\[
\begin{aligned}
    \MSE_M^{\mathrm{FH}}
    \xrightarrow[M\to\infty]{}
    \E\Bigg[
        &e^2
        \mathbf 1\!\left\{
            |e-(\bar A-\theta)|
            \le
            t_{1-\alpha/2}\sqrt{\widehat V_A}
        \right\}
        +
        (\bar A-\theta)^2
        \mathbf 1\!\left\{
            |e-(\bar A-\theta)|
            >
            t_{1-\alpha/2}\sqrt{\widehat V_A}
        \right\}
    \Bigg].
\end{aligned}
\]

We now evaluate this expectation. Recall that
\[
    R_A
    :=
    \frac{\widehat V_A}{V_A}
    \sim
    \frac{\chi_{\nu_A}^2}{\nu_A},
\]
and, under Assumption~\ref{ass:working_full},
\[
    \frac{\bar A-\theta}{\sqrt{V_A}}
    \sim N(0,1),
    \qquad
    \frac{\bar A-\theta}{\sqrt{V_A}}
    \perp\!\!\!\perp R_A.
\]
Conditional on \(R_A=r\), the acceptance event is
\[
    a_F(r)
    \le
    \frac{\bar A-\theta}{\sqrt{V_A}}
    \le
    b_F(r),
\]
where
\[
    a_F(r)
    =
    \frac{e}{\sqrt{V_A}}
    -
    t_{1-\alpha/2}\sqrt r,
\]
and
\[
    b_F(r)
    =
    \frac{e}{\sqrt{V_A}}
    +
    t_{1-\alpha/2}\sqrt r.
\]
Therefore,
\[
    p_F(r)
    =
    \Phi(b_F(r))-\Phi(a_F(r))
\]
is the conditional borrowing probability.

Conditional on \(R_A=r\), the limiting MSE contribution is
\[
    e^2p_F(r)
    +
    V_A
    \left[
        1-
        \int_{a_F(r)}^{b_F(r)}
        z^2\phi(z)\,dz
    \right].
\]
Using
\[
    \int_a^b z^2\phi(z)\,dz
    =
    \Phi(b)-\Phi(a)
    +
    a\phi(a)-b\phi(b),
\]
we obtain
\[
\begin{aligned}
    &e^2p_F(r)
    +
    V_A
    \left[
        1-p_F(r)
        +
        b_F(r)\phi(b_F(r))
        -
        a_F(r)\phi(a_F(r))
    \right].
\end{aligned}
\]
Finally, taking expectation over \(R_A\) gives
\[
\begin{aligned}
    \MSE_\infty^{\mathrm{FH}}
    =
    \E_{R_A}\Big[
        &e^2p_F(R_A)
        +
        V_A
        \big\{
            1-p_F(R_A)
            +
            b_F(R_A)\phi(b_F(R_A))
            -
            a_F(R_A)\phi(a_F(R_A))
        \big\}
    \Big].
\end{aligned}
\]
This proves the theorem.

\section{Proof of Proposition~\ref{prop:fh-endpoint-comparison}}

From the proof of Theorem~\ref{thm:fixed_weight}, the limiting MSE of the
FH estimator can be written as
\[
\MSE_\infty^{\mathrm{FH}}(e)=\E\!\left[e^2\mathbf 1\!\left\{|e-(\bar A-\theta)|\le t_{1-\alpha/2}\sqrt{\widehat V_A}\right\}+(\bar A-\theta)^2\mathbf 1\!\left\{|e-(\bar A-\theta)|>t_{1-\alpha/2}\sqrt{\widehat V_A}\right\}\right].
\]

Under Assumption~\ref{ass:working_full}, \(\bar A-\theta\sim N(0,V_A)\)
and \(\bar A-\theta\) is independent of \(\widehat V_A\). We first
condition on \(\widehat V_A\). For a fixed finite half-width \(c>0\),
define, for the purpose of this proof,
\[
G_e(c):=e^2\Pr\!\left\{|e-(\bar A-\theta)|\le c\right\}+\E\!\left[(\bar A-\theta)^2\mathbf 1\!\left\{|e-(\bar A-\theta)|>c\right\}\right].
\]

Thus, conditional on \(\widehat V_A\), the limiting FH MSE is obtained by
evaluating \(G_e(c)\) at
\[
c=t_{1-\alpha/2}\sqrt{\widehat V_A}.
\]

Because the distribution of \(\bar A-\theta\) is symmetric around zero,
\(G_e(c)=G_{-e}(c)\). It is therefore sufficient to consider \(e\ge0\).

For fixed \(e\ge0\), differentiation with respect to \(c\) gives
\[
G_e'(c)=\frac{1}{\sqrt{V_A}}\left[\{e^2-(e+c)^2\}\phi\!\left(\frac{e+c}{\sqrt{V_A}}\right)+\{e^2-(e-c)^2\}\phi\!\left(\frac{e-c}{\sqrt{V_A}}\right)\right].
\]

Using
\[
e^2-(e+c)^2=-c^2-2ec
\]
and
\[
e^2-(e-c)^2=-c^2+2ec,
\]
we obtain
\[
G_e'(c)=\frac{c}{\sqrt{V_A}}\left[\phi\!\left(\frac{e-c}{\sqrt{V_A}}\right)+\phi\!\left(\frac{e+c}{\sqrt{V_A}}\right)\right]\left[2e\tanh\!\left(\frac{ec}{V_A}\right)-c\right].
\]

The first two factors are strictly positive for every \(c>0\). Hence the
sign of \(G_e'(c)\) is determined by
\[
2e\tanh\!\left(\frac{ec}{V_A}\right)-c.
\]

When \(e=0\), this quantity equals \(-c<0\), so \(G_0(c)\) is strictly
decreasing for \(c>0\).

Now suppose \(e>0\). Dividing the final factor by \(c>0\) gives
\[
\frac{2e\tanh(ec/V_A)-c}{c}=\frac{2e^2}{V_A}\frac{\tanh(ec/V_A)}{ec/V_A}-1.
\]

The function \(\tanh(x)/x\) is strictly decreasing on \(x>0\). Indeed,
\[
\frac{\mathrm d}{\mathrm dx}\left\{\frac{\tanh(x)}{x}\right\}=\frac{x-\sinh(x)\cosh(x)}{x^2\cosh^2(x)}<0,
\]
because
\[
\sinh(x)\cosh(x)=\frac{1}{2}\sinh(2x)>x
\]
for every \(x>0\). Moreover,
\[
\lim_{x\downarrow0}\frac{\tanh(x)}{x}=1
\]
and
\[
\lim_{x\to\infty}\frac{\tanh(x)}{x}=0.
\]

It follows that, as a function of \(c>0\),
\[
\frac{2e\tanh(ec/V_A)-c}{c}
\]
is strictly decreasing from \(2e^2/V_A-1\) to \(-1\).

If \(e^2\le V_A/2\), then \(G_e'(c)<0\) for every \(c>0\). Hence
\(G_e(c)\) is strictly decreasing.

If \(e^2>V_A/2\), there exists a unique positive value of \(c\) at which
\(G_e'(c)=0\). The derivative is positive before this point and negative
after it. Thus, \(G_e(c)\) first increases and then decreases, and its
unique interior stationary point is a maximum rather than a minimum.

It remains to examine the two endpoint limits. Because
\(\bar A-\theta\) has a continuous distribution,
\[
\lim_{c\downarrow0}G_e(c)=\E[(\bar A-\theta)^2]=V_A.
\]
Also, as \(c\to\infty\), the acceptance event occurs with probability
approaching one, so
\[
\lim_{c\to\infty}G_e(c)=e^2.
\]

If \(e^2\le V_A/2\), \(G_e(c)\) decreases strictly from \(V_A\) toward
\(e^2\), and therefore
\[
G_e(c)>e^2=\min\{e^2,V_A\}
\]
for every finite \(c>0\).

If \(e^2>V_A/2\), \(G_e(c)\) first increases above its left endpoint
\(V_A\) and then decreases strictly toward its right endpoint \(e^2\).
Consequently, for every finite \(c>0\),
\[
G_e(c)>\min\{e^2,V_A\}.
\]

We now return to the estimated-variance FH gate. Under
Assumption~\ref{ass:working_full},
\[
\frac{\widehat V_A}{V_A}\sim\frac{\chi_{\nu_A}^2}{\nu_A},
\]
so \(\widehat V_A>0\) almost surely. For fixed \(\alpha\in(0,1)\),
\(t_{1-\alpha/2}\) is finite and strictly positive. Hence
\[
0<t_{1-\alpha/2}\sqrt{\widehat V_A}<\infty
\]
almost surely.

Therefore, conditional on \(\widehat V_A\),
\[
G_e\!\left(t_{1-\alpha/2}\sqrt{\widehat V_A}\right)>\min\{e^2,V_A\}
\]
almost surely. Taking expectation over \(\widehat V_A\) gives
\[
\MSE_\infty^{\mathrm{FH}}(e)>\min\{e^2,V_A\}.
\]
This proves the proposition.

\section{(not yet)Proof of Proposition \ref{prop:adaptive-ci}}

Recall the general plug-in shrinkage estimator in \eqref{eq:feasible-thetaM}:
\[
\hat\theta(\hat w)
=
\bigl(1-\mathbf{1}^\top \hat w\bigr)\bar A_{\mathrm{te}}
+
\sum_{i=1}^M \hat w_i \bar B_{i,\mathrm{te}},
\qquad
\hat w
=
\widehat V_{A,\mathrm{te}}\widehat Q^{-1}\mathbf{1}.
\]

Let $\mathcal T_M$ denote the $\sigma$-field generated by all pilot-sample quantities. By sample splitting scheme in Section~\ref{sec:practical}, conditional on $\mathcal T_M$, the random weight vector $\hat w$ is fixed and independent of the test-sample means. Therefore, conditional on $\mathcal T_M$,
\[
\hat\theta(\hat w)-\theta \mid \mathcal T_M
\sim
N\!\bigl(b_M,v_M\bigr),
\]
where
\[
b_M:=\sum_{i=1}^M \hat w_i\epsilon_i,
\qquad
v_M:=V_{A,\mathrm{te}}\bigl(1-\mathbf{1}^\top \hat w\bigr)^2+\sum_{i=1}^M V_{B,i,\mathrm{te}}\hat w_i^2.
\]
From the proof of Theorem~\ref{thm:MSE_infty1}, we already have
\[
\hat w_i=O_P(M^{-1}), \qquad 1-\mathbf{1}^\top \hat w=O_P(M^{-1}),
\]
which implies
\[
v_M\xrightarrow[M\to\infty]{p}0.
\]
The same proof also established that
\[
b_M
\xrightarrow[M\to\infty]{p}
\frac{\tilde\mu_1\tilde\nu+(\bar A_{\mathrm{pi}}-\theta)\tilde A}{\tilde\Delta},
\]
where
\[
\tilde A=\tilde\mu_0\tilde\mu_2-\tilde\mu_1^2,
\qquad
\tilde\Delta=\tilde A+\tilde\mu_0\tilde\nu.
\]
Since
\[
\bar A_{\mathrm{pi}}-\theta\sim N(0,V_{A,\mathrm{pi}}),
\]
it follows that
\[
b_M
\xrightarrow[M\to\infty]{d}
N\!\left(
\frac{\tilde\mu_1\tilde\nu}{\tilde\Delta},
\frac{\tilde A^2}{\tilde\Delta^2}V_{A,\mathrm{pi}}
\right).
\]
Recall we have the conditional distribution
\[
\hat\theta(\hat w)-\theta \mid \mathcal T_M
\sim N(b_M,v_M).
\]
Since variance \(v_M\xrightarrow[M\to\infty]{p}0\), the conditional normal distribution collapses around its mean \(b_M\). Therefore
\[
\hat\theta(\hat w)-\theta-b_M \xrightarrow[M\to\infty]{p} 0.
\]
On the other hand, we have already shown that
\[
b_M
\xrightarrow[M\to\infty]{d}
N\!\left(
\frac{\tilde\mu_1\tilde\nu}{\tilde\Delta},
\frac{\tilde A^2}{\tilde\Delta^2}V_{A,\mathrm{pi}}
\right).
\]
It then follows from Slutsky's theorem that
\[
\hat\theta(\hat w)-\theta
\xrightarrow[M\to\infty]{d}
N\!\left(
\frac{\tilde\mu_1\tilde\nu}{\tilde\Delta},
\frac{\tilde A^2}{\tilde\Delta^2}V_{A,\mathrm{pi}}
\right).
\]

\section{(not yet)Proof of Proposition~\ref{prop:fh-conditional-ci}}

Recall that
\[
    D_M
    =
    \widetilde B_M-\bar A,
    \qquad
    \widetilde B_M
    =
    \sum_{i=1}^M u_{i,M}\bar B_i.
\]
Under Assumption~\ref{ass:working_full},
\[
    \bar A-\theta
    \sim
    N(0,V_A).
\]
Because the auxiliary sample means are mutually independent under
Assumption~\ref{ass:model},
\begin{equation}
    \widetilde B_M-\theta
    \sim
    N\!\left(
        \bar\epsilon_M^{\,w},
        U_M
    \right),
    \label{eq:fh-ci-auxiliary-mean}
\end{equation}
where
\[
    \bar\epsilon_M^{\,w}
    =
    \sum_{i=1}^M u_{i,M}\epsilon_i
\]
and
\[
    U_M
    =
    \sum_{i=1}^M u_{i,M}^2V_{B,i}.
\]
Moreover, \(\bar A\) and \(\widetilde B_M\) are independent under
Assumption~\ref{ass:model}. It follows that
\[
    \left(
        \bar A-\theta,\,
        D_M
    \right)
\]
is jointly Gaussian.

The mean of \(D_M\) is
\begin{equation}
    \E(D_M)
    =
    \bar\epsilon_M^{\,w}.
    \label{eq:fh-ci-discrepancy-mean}
\end{equation}
Its variance is
\begin{equation}
    \Var(D_M)
    =
    V_A+U_M.
    \label{eq:fh-ci-discrepancy-variance}
\end{equation}
Furthermore,
\begin{equation}
    \Cov(\bar A-\theta,D_M)
    =
    -V_A.
    \label{eq:fh-ci-covariance}
\end{equation}
and therefore
\[
    \Cov(\bar A-\theta,D_M)
    =
    \Cov(\bar A-\theta,\widetilde B_M-\theta)
    -
    \Var(\bar A-\theta)=0-V_A.
\]

By the conditional distribution formula for jointly Gaussian random variables,
Equations~\eqref{eq:fh-ci-discrepancy-mean}--\eqref{eq:fh-ci-covariance} give
\begin{equation}
    \left(
        \bar A-\theta
        \,\middle|\,
        D_M=d
    \right)
    \sim
    N\!\left(
        -\frac{V_A}{V_A+U_M}
        \left(
            d-\bar\epsilon_M^{\,w}
        \right),
        \,
        V_A-\frac{V_A^2}{V_A+U_M}
    \right).
    \label{eq:fh-ci-target-conditional-initial}
\end{equation}
The conditional variance in
Equation~\eqref{eq:fh-ci-target-conditional-initial} simplifies as
\[
    V_A-\frac{V_A^2}{V_A+U_M}
    =
    \frac{V_AU_M}{V_A+U_M}.
\]
Hence,
\begin{equation}
    \left(
        \bar A-\theta
        \,\middle|\,
        D_M=d
    \right)
    \sim
    N\!\left(
        -\frac{V_A}{V_A+U_M}
        \left(
            d-\bar\epsilon_M^{\,w}
        \right),
        \,
        \frac{V_AU_M}{V_A+U_M}
    \right).
    \label{eq:fh-ci-target-conditional}
\end{equation}

We next account for conditioning on the estimated discrepancy variance.
Under Assumption~\ref{ass:model}, the target dataset and all auxiliary
datasets are mutually independent. Under
Assumption~\ref{ass:working_full}, each full-sample mean is treated as
independent of its corresponding sample variance. Together, these conditions
imply
\begin{equation}
    \left(
        \bar A,\bar B_1,\ldots,\bar B_M
    \right)
    \perp\!\!\!\perp
    \left(
        S_A^2,S_{B_1}^2,\ldots,S_{B_M}^2
    \right).
    \label{eq:fh-ci-mean-variance-independence}
\end{equation}
Since \(D_M\) is a function of the sample means, whereas
\(\widehat V_A+\widehat U_M\) is a function of the sample variances,
Equation~\eqref{eq:fh-ci-mean-variance-independence} implies
\[
    \left(
        \bar A-\theta,D_M
    \right)
    \perp\!\!\!\perp
    \left(
        \widehat V_A+\widehat U_M
    \right).
\]
Consequently, conditioning additionally on
\(\widehat V_A+\widehat U_M=q\) does not change the conditional distribution
in Equation~\eqref{eq:fh-ci-target-conditional}. Thus,
\begin{equation}
    \left(
        \bar A-\theta
        \,\middle|\,
        D_M=d,\,
        \widehat V_A+\widehat U_M=q
    \right)
    \sim
    N\!\left(
        -\frac{V_A}{V_A+U_M}
        \left(
            d-\bar\epsilon_M^{\,w}
        \right),
        \,
        \frac{V_AU_M}{V_A+U_M}
    \right).
    \label{eq:fh-ci-target-conditional-full}
\end{equation}

It remains to incorporate the hypothesis-testing gate. Recall that
\[
    \hat\theta_M^{\mathrm{FH}}
    =
    \bar A+\lambda_MD_M\widehat I_M.
\]
Conditional on
\[
    D_M=d,
    \qquad
    \widehat V_A+\widehat U_M=q,
\]
the testing indicator becomes the deterministic quantity
\begin{equation}
    \widehat I_M
    =
    \mathbf 1
    \left\{
        |d|
        \le
        t_{1-\alpha/2}\sqrt q
    \right\}.
    \label{eq:fh-ci-conditional-indicator}
\end{equation}
It follows that, under the same conditioning event,
\begin{equation}
    \hat\theta_M^{\mathrm{FH}}-\theta
    =
    \bar A-\theta
    +
    \lambda_Md
    \mathbf 1
    \left\{
        |d|
        \le
        t_{1-\alpha/2}\sqrt q
    \right\}.
    \label{eq:fh-ci-conditional-estimator}
\end{equation}

The second term on the right-hand side of
Equation~\eqref{eq:fh-ci-conditional-estimator} is deterministic after
conditioning. Therefore, adding this term to the conditional Gaussian
distribution in Equation~\eqref{eq:fh-ci-target-conditional-full} gives
\begin{equation}
    \left(
        \hat\theta_M^{\mathrm{FH}}-\theta
        \,\middle|\,
        D_M=d,\,
        \widehat V_A+\widehat U_M=q
    \right)
    \sim
    N\!\left(
        m_M(d,q),
        s_M^2
    \right),
    \label{eq:fh-ci-final-conditional-distribution}
\end{equation}
where
\[
    m_M(d,q)
    =
    -\frac{V_A}{V_A+U_M}
    \left(
        d-\bar\epsilon_M^{\,w}
    \right)
    +
    \lambda_Md
    \mathbf 1
    \left\{
        |d|
        \le
        t_{1-\alpha/2}\sqrt q
    \right\},
\]
and
\[
    s_M^2
    =
    \frac{V_AU_M}{V_A+U_M}.
\]

Equation~\eqref{eq:fh-ci-final-conditional-distribution} implies
\begin{equation}
    \left.
    \frac{
        \hat\theta_M^{\mathrm{FH}}
        -
        \theta
        -
        m_M(d,q)
    }{
        s_M
    }
    \,\right|\,
    D_M=d,\,
    \widehat V_A+\widehat U_M=q
    \sim
    N(0,1).
    \label{eq:fh-ci-standardized}
\end{equation}
Therefore,
\begin{equation}
\begin{split}
    &\Pr\!\left(
        -z_{1-\xi/2}
        \le
        \frac{
            \hat\theta_M^{\mathrm{FH}}
            -
            \theta
            -
            m_M(d,q)
        }{
            s_M
        }
        \le
        z_{1-\xi/2}
        \,\middle|\,
        D_M=d,\,
        \widehat V_A+\widehat U_M=q
    \right)
    \\
    &\qquad
    =
    1-\xi.
\end{split}
    \label{eq:fh-ci-conditional-coverage}
\end{equation}
Rearranging the inequalities inside
Equation~\eqref{eq:fh-ci-conditional-coverage} gives the conditional confidence
interval
\[
\left[
    \hat\theta_M^{\mathrm{FH}}
    -
    m_M(d,q)
    -
    z_{1-\xi/2}s_M,
    \;
    \hat\theta_M^{\mathrm{FH}}
    -
    m_M(d,q)
    +
    z_{1-\xi/2}s_M
\right].
\]

Rearranging, we have
\[
\left[
    \bar A
    +
    \frac{V_A}{V_A+U_M}
    \left(
        d-\bar\epsilon_M^{\,w}
    \right)
    -
    z_{1-\xi/2}
    \sqrt{
        \frac{V_AU_M}{V_A+U_M}
    },
    \;
    \bar A
    +
    \frac{V_A}{V_A+U_M}
    \left(
        d-\bar\epsilon_M^{\,w}
    \right)
    +
    z_{1-\xi/2}
    \sqrt{
        \frac{V_AU_M}{V_A+U_M}
    }
\right].
\]

This proves Proposition~\ref{prop:fh-conditional-ci}.

\end{APPENDICES}

\end{document}